\documentclass{emulateapj}

\usepackage{txfonts}
\usepackage{float}
\usepackage[usenames,dvipsnames]{color}
\usepackage{ulem}

\slugcomment{}

\shorttitle{Dynamical Evolution of Viscous Disks around Be Stars}
\shortauthors{Haubois et al.}

\begin{document}


\title{Dynamical Evolution of Viscous Disks around Be Stars. I: photometry}


\author{X. Haubois and A.C. Carciofi}
\affil{Instituto de Astronomia, Geof\'{i}sica e Ci\^{e}ncias Atmosf\'{e}ricas, Universidade de S\~{a}o Paulo, Rua do Mat\~{a}o 1226, Cidade Universit\'{a}ria, S\~{a}o Paulo, SP 05508-900, Brazil}
\email{xhaubois@astro.iag.usp.br}
\author{Th. Rivinius}
\affil{European Organisation for Astronomical Research in the Southern Hemisphere, Casilla 19001, Santiago 19, Chile}
\author{A.T. Okazaki}
\affil{Faculty of Engineering, Hokkai-Gakuen University, Toyohira-ku, Sapporo 062-8605, Japan}
\and
\author{J. E. Bjorkman}
\affil{Ritter Observatory, Department of Physics \& Astronomy, University of Toledo,Toledo, OH 43606, USA}


%\altaffiltext{1}{}
%\altaffiltext{2}{}
%\altaffiltext{3}{}
%\altaffiltext{4}{}


\begin{abstract}
Be stars possess gaseous circumstellar disks that modify in many ways the spectrum of the central B star. Furthermore, they exhibit variability at several timescales and for a large number of observables. Putting the pieces together of this dynamical behavior is not an easy task and requires a detailed understanding of the physical processes that control the temporal evolution of the observables.
There is an increasing body of evidence that suggests that Be disks are well described by standard $\alpha$-disk theory. This paper is the first of a series that aims at studying the possibility of inferring several disk and stellar parameters through the follow-up of various observables. Here we study the temporal evolution of the disk density for different dynamical scenarios, including the disk build-up as a result of a long and steady mass injection from the star, the disk dissipation that occurs after mass injection is turned off, as well as scenarios in which active periods are  followed by periods of quiescence.
For those scenarios, we investigate the temporal evolution of continuum photometric observables using a 3-D non-LTE radiative transfer code. We show that lightcurves for different wavelengths are specific of a mass loss history, inclination angle and $\alpha$ viscosity parameter. The diagnostic potential of those lightcurves is also discussed.

%Which band, niveau de polarization attendu, quels diagrammes sont caracteristiques de quels scenarios, alpha ?? 
%   comparaisons aux vrais donnees : ca se rapporte a des combinaisons de cas classiques ?
\end{abstract}


\keywords{circumstellar matter Ñ radiative transfer Ñ stars: emission-line, Be}

\section{Introduction}
\label{sec:introduction}

%Be stars are non-supergiant, early-type stars that have been known for a long time to possess circumstellar with a circumstellar disk that forms from matter ejected from the star. The way that the disk forms and hence the origin of the Be phenomenon is still unclear. A review of ifferent mechanisms  is presented in \cite{2003PASP..115.1153P} and \citet{carciofi11}.

%General stuff 

Line emission in most stellar spectra arises from ionized gas beyond the
photospheric level. In particular, emission-line spectra of B-type stars
(e.g., Be, B[e], Herbig Ae/Be) are due to extended circumstellar (CS)
envelopes. Additionally, the ionized CS gas produces continuum radiation due
to free-free and free-bound transitions \citep{geh74}. The envelope
contribution to the object's brightness depends on the distribution of the
density, temperature, and ionization degree throughout the CS envelope.
Classical Be stars are a large class of objects in which CS contribution to
the stellar continuum can be significant.  They can display strong IR
excesses, which can be up to $\approx 40$ times larger than the photospheric
flux of the central star at $12\;\mu m$ \citep{1988iras....1.....B}.

In the past decade or so, a consensus has emerged that the CS matter around Be
stars is distributed in a disk. The presence of rotating, flattened material
(i.e.\ a disk) was initially inferred from the typical double-peaked emission
line profiles seen in many Be stars and has since been confirmed by modern
high-angular resolution techniques which have resolved the disks around
several nearby Be stars \citep[see e.g.][]{1997ApJ...479..477Q}.

%Viscous model

The origin of those disks has been the subject of much debate. Clearly, a
necessary ingredient for disk formation is the star's rapid rotation rate,
which allows material to be more easily lifted off the surface of the
star. Nevertheless, since most Be stars seem not to be rotating at their
break-up speed \citep[e.g,][]{2005ApJ...634..585C}, other mechanism(s) are likely
necessary \citep[for an overview see][]{2006ASPC..355..219O}. Regardless of how
material is ejected into orbit, another mechanism is required to distribute
the material throughout the disk. The viscous decretion disk model (VDDM),
first suggested by \cite{1991MNRAS.250..432L} and further developed by
\citet{1997LNP...497..239B}, \citet{2002MNRAS.337..967O} and \citet{2005ASPC..337...75B}, among others, uses the angular momentum transport
by turbulent viscosity to lift material into higher orbits, thereby causing
the disk to grow in size. A theoretical prediction of this model is that
material is in Keplerian rotation throughout the disk.  Spectrointerferometry
and spectroastrometry are starting to provide clear evidence that, in most
observed and so far analyzed systems \citep[see e.g.][]{2011arXivMeilland},
the disks rotate in a Keplerian fashion \citep{2007A&A...464...73M,2011IAUS..272..430S,2011arXiv1109.3447K,2011A&A...529A..87D}. This
important fact, together with other observational facts outlined in Carciofi 2011, are properties that only the VDDM can reproduce.

%These disks operate like accretion disks, except that in the case of viscous decretion, the matter is driven outward under the effect of turbulent viscosity. This process was first suggested by \cite{1991MNRAS.250..432L} and further developed by Bjorkman (1997), Okazaki (2001) and Bjorkman \& Carciofi (2005), among others. It requires that the central star ejects material at Keplerian or super-Keplerian speeds at its surface. Although the driving and the details of the ejection mechanism is not fully understood yet, pulsation and internal or magnetic instabilities coupled with high stellar rotation velocities represent good candidates to cause this feeding of the inner disk by the star (See Owocki 2006, for a review).

%Variability
%An intriguing aspect of Be stars is their variability. 

Photometric observations offer a possibility to study different regions of Be disks \citep[e.g.][]{car06b}. At short wavelengths, say, $V$-band, excess continuum radiation arises from a relatively small area near the star, whereas for long wavelengths the emission area increases with the wavelength approximately as $\lambda^{8/11}$ (see Eq.~A15 of \citealt{car06}).
Some stars are known to have had a nearly stable continuum emission for decades, $\zeta$ Tauri being a notable example \citep{ste09}. This suggests that these stars possess  a decretion disk that has been fed at a roughly steady rate.
However, there are also cases that a star that has been stable for a long time suddenly looses its visible and IR excesses \citep[e.g., $\pi$ Aqr, ][]{2010ApJ...709.1306W}. This is interpreted as the dissipation of the disk by some mechanism, as a result of the mass loss from the star being turned off.
Similarly, there are cases where the disk has been rebuilt after dissipation, and is now stable again.
During the time of disk growth/dissipation there is often short term, small
scale photometric variability, a result of transitory outbursts with timescales of days or weeks \citep[sometimes called ``flickering activity'', as in $\mu$\,Cen, ][]{1993A&A...274..356H}. Finally, a good fraction of Be stars are intrinsically variable at different timescales. Their photometric variability can be found periodic, quasi-periodic or irregular \citep{1994A&AS..108..237M,1996A&A...311..579S,2008A&A...478..659S} or can exhibit episodic outbursts of variable duration \citep{2000ASPC..214..348H, 2002A&A...393..887M}. Recently, a time-dependent model akin to the one employed in this work was successfully applied for modeling the dissipation of the disk of 28\,CMa \citep{car11b}. From that analysis, the authors were able to determine the viscosity parameter of the disk, having found $\alpha=1.0\pm0.2$.

The above suggests the existence of at least two timescales controlling the disk photometric properties: 1) $\tau_{\rm in}$, timescale for the variability of the mass \textit{injection} into the disk, related to the rate of stellar mass ejection events and the length of these events, and, 2) $\tau_{\rm d}$, timescale for the disk to redistribute the injected material.

To critically test the VDDM against observations, the structure of the disk must be determined from hydrodynamical equations.
Using constant mass injection rates\footnote{In this paper, we distinguish the mass loss rate and the mass injection rate which is the normalized rate at which mass is injected into the disk.} and non-LTE codes based on different approaches \cite[e.g.][]{car06,2007ApJ...668..481S}, the first quantitative tests of the near steady state solution have been successfully carried out on \object{$\delta$ Sco} \citep{car06b}, \object{$\chi$ Oph} \citep{2008ApJ...689..461T}, and \object{$\zeta$ Tau} \citep{car09}. However, those models can only be applied to objects that went through a sufficiently long and stable decretion phase.  

The purpose of this paper is to provide a framework to understand the effects
of time variable mass loss rates on the structure of the disk and its
observational consequences. To accomplish this we will first describe the model in \S~\ref{models}, then in
\S~\ref{dynmod} we study two limiting cases, for which the timescale of
mass loss rate variability is either much longer or comparable to the disk
timescale introduced above. In particular, we study the growth of a disk where
none has been previously present and the dissipation of a pre-existing
disk. In the intermediate regime where those timescales are comparable we
study single outburst-like events and periodic ones. In \S~\ref{prediction}
the resulting photometric observables are described, which are then compared to
previous work and observations in \S~\ref{discussion}. We finally summarize this work in \S~\ref{conclusion}.

%\rmACC{In order to model the evolution of active Be stars, whose disk are perturbed under the action of frequent mass ejection from the star, it is crucial to investigate the effect of non-constant injection rates in Be disks. The goal of this work is to present the signatures of simple, yet realistic, decretion scenarios upon the disk structure and the emergent spectrum. This paper deals with photometry in the continuum, i.e., broad-band diagnostics. 
% Plan : ici juste polarim et photom
%In \S~\ref{models}, we describe the modeling approach. Then we present the dynamical models we investigated (\S~\ref{dynmod}) and the corresponding predictions for the temporal behaviour of photometric observables in the \S~\ref{prediction} before concluding. 
%A discussion and a comparison to observed data of these synthetic observables are presented in \S~\ref{comp} before concluding.}

\section{Model Description}
\label{models}

The temporal evolution of Be disks was studied by \cite{2002MNRAS.337..967O},
\cite{2007ASPC..361..230O}, \cite{2008MNRAS.386.1922J}, \cite{2007ApJ...668..481S} and \citet{car11b}. With the exception of
the former, these studies focused on the evolution of the surface density of a
disk fed by a constant mass injection rate and constrained the power-law index
$n$ describing the radial dependency of the density ($\rho (r) \propto r^{-n}
$). In isothermal disks, \cite{2007ASPC..361..230O} found that $n$ is always
larger than 7/2, but approaches this value as time goes to infinity.
In order to analyze the observational signatures of variable mass injection
rates, we use the time-dependent hydrodynamic code {\sc singlebe}
\citep{2007ASPC..361..230O,2002MNRAS.337..967O}.  This code performs
one-dimensional (1-D) simulations of the structure and evolution of an
isothermal viscous decretion disk by solving the time-dependent fluid
equations \citep{1974MNRAS.168..603L} in the thin disk
approximation.
 %For this purpose, it assumes that the disk around an isolated Be star is axisymmetric, in Keplerian rotation and in hydrostatic equilibrium in the vertical direction.   
The evolution of such a disk is described by the following one-dimensional, diffusion-type equation of the surface density $\Sigma$ \citep[see e.g.,][]{pri81},
\begin{equation}
\frac{\partial \Sigma}{\partial t} =  \frac{1}{r}
  \frac{\partial}{\partial r} \left[  \frac {\frac{\partial}{\partial
  r} (r^{2}\Sigma \alpha c_{s}^{2} ) }{\frac{\partial}{\partial r} (r^{2} \Omega ) }
  \right]\,,
\label{eq:density}
\end{equation}
%no new line here.
 where $\alpha$ is the Shakura-Sunyaev viscosity parameter, $c_{s}$ is the isothermal sound speed, and $\Omega$ is the angular frequency of the disk rotation. In this equation, we take the angular frequency of disk rotation to be circularly Keplerian, i.e., $\Omega (r) = \Omega_{K}$, where $\Omega_{K} = (GM/r^{3})^{1/2}$. 

We adopt the (torque-free) outflow boundary condition, $\Sigma = 0$, for both the inner and outer edges of the disk. We further assume that mass is injected at a point located just above the photosphere, $r_{in} = 1 R_\star$.  We place the inner boundary inside the surface of the star, while the outer boundary is placed at $r_{out}=1000\,R_\star$, which is roughly the location of the trans-sonic point for photo-evaporation of the disk. Note that for steady-state outflow, the surface density of the disk scales as $\Sigma \propto ({\dot M} / \alpha) \sqrt{r_{out}/R_\star}$ \citep[see eq.~37][]{1997LNP...497..239B}.  Consequently, if a different physical mechanism is responsible for truncating the disk (for example, a binary companion), our results for the asymptotic values of the surface density should be scaled by $\sqrt{r_{out}/1000 R_\star}$.

 %We adopt an absorbing inner boundary condition where $\Sigma = 0$ close to the stellar surface. Furthermore, we assume that the mass is injected into the disk at $r = 1 R_\star$, i.e. at the photospheric level.
 % As an outer boundary condition, we chose $\Sigma = 0$ at $r=1000\,R_\star$. This large maximum radius value has been chosen so that the disk structure of the first hundreds of stellar radii is not affected by this truncation.  

The output of {\sc singlebe} is the surface density, $\Sigma(r,t)$, as a function of radius and time for a given stellar mass loss history and $\alpha$ viscosity parameter. This surface density is converted to volume density using the usual vertical hydrostatic equilibrium solution (a Gaussian) with a power law scale height, $H=H_0(r/R_{\star})^{1.5}$.
The volume density is then used as input for the three-dimensional non-LTE Monte Carlo radiative transfer code {\sc hdust} \citep{car06,2008ApJ...684.1374C}. This code performs a full spectral synthesis by simultaneously solving the NLTE statistical equilibrium equations and radiative equilibrium equation to obtain the hydrogen level populations and electron temperature throughout the disk. The code's output consists of the SED, polarization spectrum and other observables of interest.

%teff   :
% radius delta sco
% rotaion speed : zeta tauri
% temperature : sigma ori ?
%The stellar temperature is typical of a B2 , spectral type at which the distribution peak of Be stars is located \cite{1997A&A...318..443Z}.

In this work, we adopt a rotationally deformed and gravity darkened star whose parameters are typical of a B2Ve star (Table~\ref{tab_param}). To account for the effects of rotation, the star is divided in a number of latitude bins (typically 100), each with its effective temperature, gravity and a spectral shape given by the appropriate Kurucz model atmosphere \citep{Kur94}. 
%After emission by the star, each photon is followed as it travels through the disk (where it may be scattered, or absorbed and reemitted, many times) until it escapes. 

To specify the disk structure, we need, as input for {\sc singlebe}, the mass loss rate as a function of time, $\dot{M}(t)$, and $\alpha$. Here we define the stellar mass loss rate as the mass injected per unit of time at the stellar radius. In order to be able to compare different mass loss rate scenarios, we introduce the parameter $\Sigma_0$, which is the density at the base of the disk after a sufficient long decretion phase. In this case, the surface density approaches the near steady state solution \citep{2005ASPC..337...75B},
\begin{equation}
\Sigma(r)=\Sigma_0(r/R_\star)^{-2}\,.
\label{eq:sigma_steady_state}
\end{equation}
Because $\Sigma_0$ depends on the ratio $\dot{M}/\alpha$, these quantities are not independent parameters in our models. Hence,  given a value for $\Sigma_0$, a value of $\alpha$ is uniquely associated with a value of $\dot{M}$ and vice-versa.

 Table~\ref{tab_param2} summarizes the disk parameters adopted in this work.
We chose the value $\Sigma_0 = 0.85\;\rm g\;cm^{-2}$, which corresponds to a volume density of $3 \times 10^{-11}\;\rm g\; cm^{-3}$, a typical value for dense Be disks \citep{car06b,1988A&A...198..200W}.
Several values of $\alpha$ were investigated in the range  0.1~---~1.0, whose corresponding mass loss rates lie in the range 1.6~---~$16\times 10^{-9}\;\rm M_{\sun}\;yr^{-1}$. We specify here that $\alpha$ is constant through all the radii. The possibility of a radial dependence of $\alpha$ is mentioned in Sect.~\ref{kphot}.
   
%In addition, we explored four disk sizes in the range 5~---~$100\,R$, to study the effect of disk truncation by a secondary star. 

 % {\sc hdust} includes both continuum processes and spectral lines in the opacity and emissivity of the gas. Even if the viscous diffusion is treated in the isothermal case by {\sc singlebe}, our models are not isothermal because the full NLTE radiative transfer performed by {\sc hdust} calculates the temperature along the disk. The interested reader is referred to \cite{car06} for details of the Monte Carlo NLTE solution. To summarize, {\sc hdust} turns the structural information provided by {\sc singlebe} into astrophysical observables. By repeating the coupling of {\sc singlebe} and {\sc hdust} for different epochs of the disk evolution, we can follow the evolution of different observables (SED, polarization, images, etc). To summarize, we aim at investigating a range of dynamic behaviours of a typical decretion Be disk. 

 \begin{deluxetable}{cc}
% \tableline\tableline
\tablecolumns{2}
\tablecaption{Stellar main parameters used in the simulations. \label{tab_param}}
\tablehead{\colhead{Parameter} & \colhead{Value}}
\startdata
Mass &   9.0 $M_{\odot}$ \\
Polar radius $R_{\rm pole} $& 5.7 $R_{\odot}$\\
Equatorial radius & $R_{\star} = 6.5 R_{\odot}$ \\ %delta sco
Rotation speed & 273 km/s \\
Keplerian speed at equator& $V_{K}=$ 514 km/s \\
Breakup speed$\sqrt{\frac{2}{3}GM/R_{pol}}$ & $V_{c}=$ 448 km/s \\
$\Omega$ / $\Omega_{c}$ &  0.8 \\ %273*(1.5/1.14)/448.
Oblateness & 1.14 \\
Polar temperature & 22000 K \\ % sigma Ori ?
%Ratio between the polar and equatorial temperature & 1.16 \\ 
Luminosity & 5980 $L_{\odot}$\\ 
%\tableline
\enddata
\end{deluxetable}

  \begin{deluxetable}{cc}
\tablecolumns{2}
\tablecaption{Disk parameters used in the simulations.\label{tab_param2}}
\tablehead{\colhead{Parameter} & \colhead{Value}}
\startdata
%Maximum radius & 5, 10, 20 and 100 $R_{\star}$ \\
$\Sigma_{0}$, surface density at the base of the disk &   0.85 g.cm$^{-2}$ \\
% ? =sqrt(2¹)H?0 = sqrt(2¹) * 9.09219e9 *3.0 e-11 = 0.68 en theorie mais dans le code d Atsuo, H est calcule autrement.
 $\dot{M}$ [10$^{-9}$ M$_{\odot}$/year] & 1.64, 4.90, 8.15, 11.41 and 16.29  \\ % le mettre car valeur ok pour tout sauf dm05 ?
 
 % sigma=0.85 
% rho= sigma/(sqrt((6.5*0.6955e8)^3*(299792458.)^2/(6.67384e-11*9*1.98892e30)))
 $\alpha$ & 0.1, 0.3, 0.5, 0.7 and 1.0 \\ 
Flaring parameter & 1.5 \\
\enddata
\end{deluxetable}

% \section{Density behavior in the dynamical scenarios }
 \section{Temporal evolution of the surface density }
\label{dynmod}

Our purpose is to guide the analysis of Be disk observations by exploring simple dynamical models and their repercussions on the photometric observables. 
As discussed in the introduction, observations indicate that the state of the disk, and hence its photometric properties, is controlled by two competing timescales, $\tau_{\rm in}$ and $\tau_{\rm d}$. This tells us that there will be three regimes:
\begin{enumerate}
\item $\tau_{\rm in} \ll \tau_{\rm d}$. This situation is irrelevant for the purpose of this paper because it does not produce significant observable effects.
\item $\tau_{\rm in} \gg \tau_{\rm d}$. Here, as limiting cases we study the creation of a new disk fed at a constant mass injection rate and the dissipation of a pre-existing disk.
\item $\tau_{\rm in} \sim \tau_{\rm d}$. In this case there will be an interesting interaction between the two competing timescales. In this work we explore periodic mass injection scenarios and episodic ones.
\end{enumerate}

%Typically, one dynamical scenario is characterized by a mass loss history and an $\alpha$ viscosity parameter value. We also investigated different maximum extensions for the disk in order to take into account tidal truncation of the disk by a secondary star. We thus created three different kinds of dynamical models that include different types of mass loss histories (Figure~\ref{scenarios}). The first kind allows investigating the growth of a disk fed by a constant mass injection rate (disk build-up phase) and its subsequent dissipation. The second kind of models explores periodic mass loss scenarios, i.e. the mass loss is successively turned on and off every period. The last kind takes into account irregular alternative mass loss scenarios. In those models, the mass loss is turned on and off with no periodicity and with different amplitudes.

%A fourth family ? With inconstant irregular mass loss scenarios (dm05) dm05_0. : mass loss rate factor 10 -> 2.

\subsection{Reference cases: disk growth and dissipation}
\label{refcase}

\subsubsection{Disk growth}
\label{growth}

A decretion disk never experiences steady state: it either grows or decays \citep{2007ASPC..361..230O}. However, it is useful to study the solution of a disk subject to a constant mass injection rate when time goes to infinity. 
Even though steady state is never physically realized, a limit value for the surface density of an ideally grown, unperturbed and isothermal disk can be determined analytically, in which case the surface density takes the simple form of Eq.~(\ref{eq:sigma_steady_state}) \citep{2005ASPC..337...75B}.
%\rmACC{This density limit solution is described by a power law  $\rho (r) \propto r^{-3.5} $.  The disk is also found to be in keplerian rotation and geometrically thin with a scaleheight H varying as  $H (r) \propto r^{1.5} $.}
%\rmACC{Here, we computed a quasi-steady state solution in one dimension, in this paper we chose not to explore the vertical stratification for the density, in order to have a reference limit value to compare the evolution of the surface density with. }

\begin{figure*}[t!]
% \vspace*{-2.0 cm}
\begin{center}
 \includegraphics[width=3.4in]{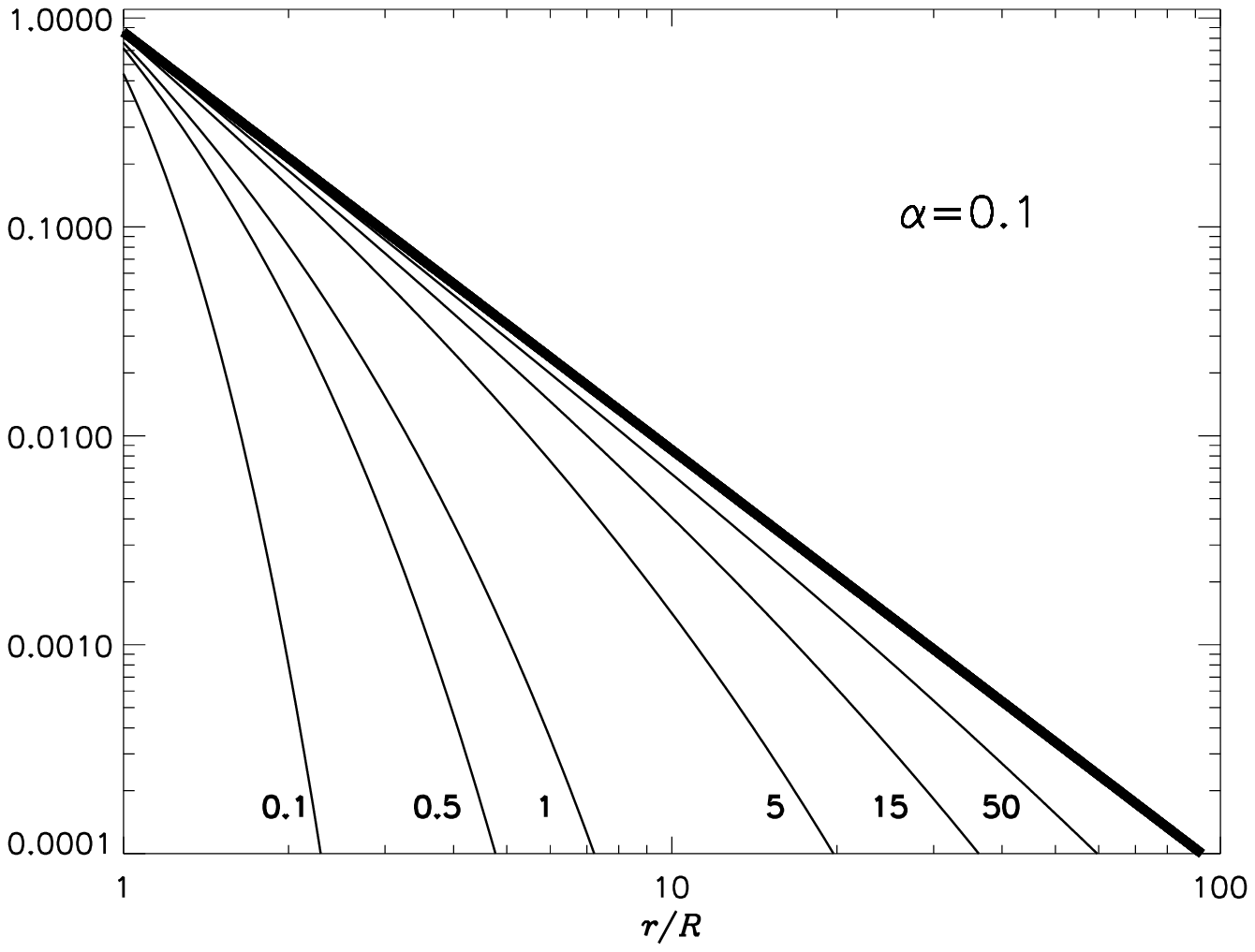}
 \includegraphics[width=3.4in]{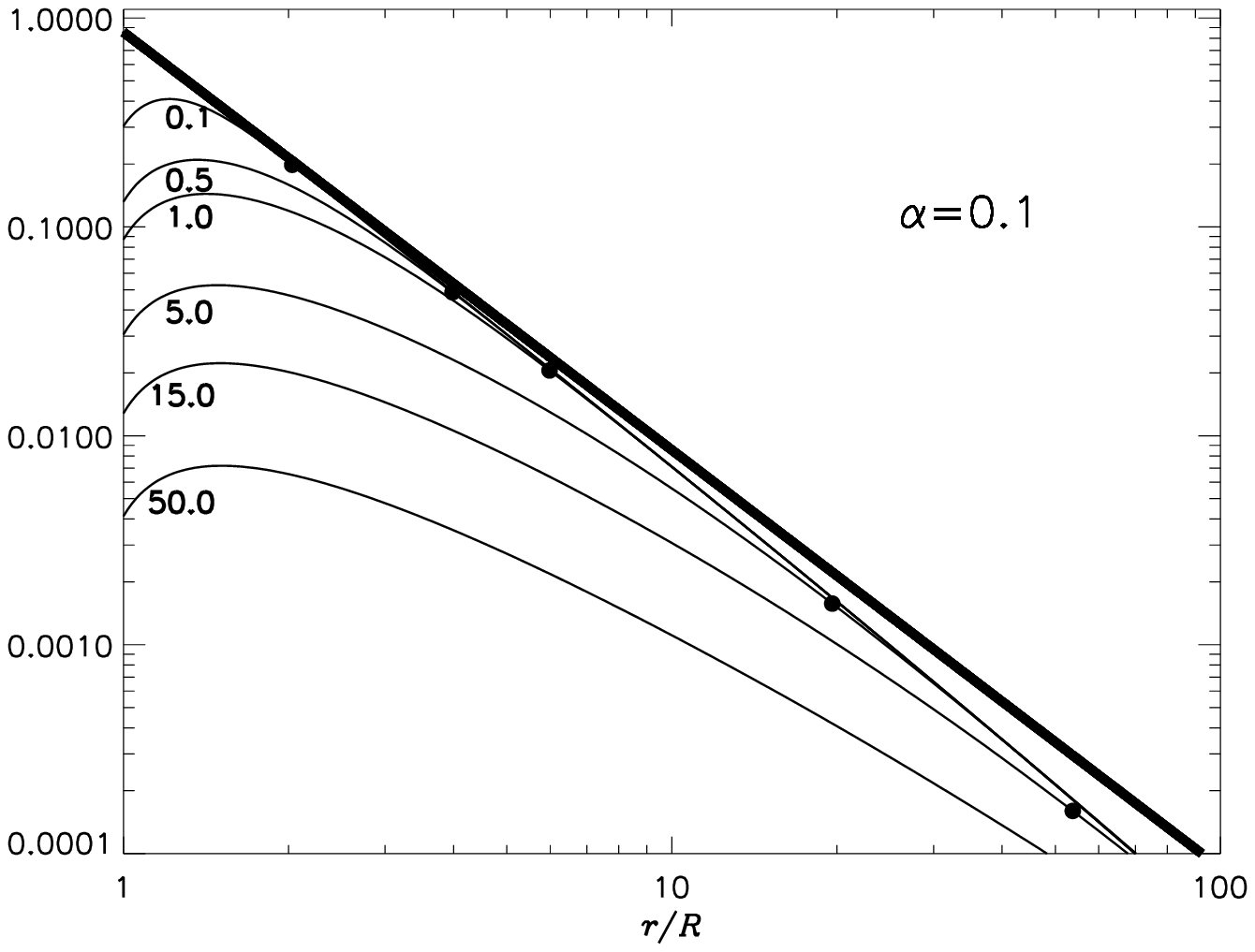}
 \includegraphics[width=3.4in]{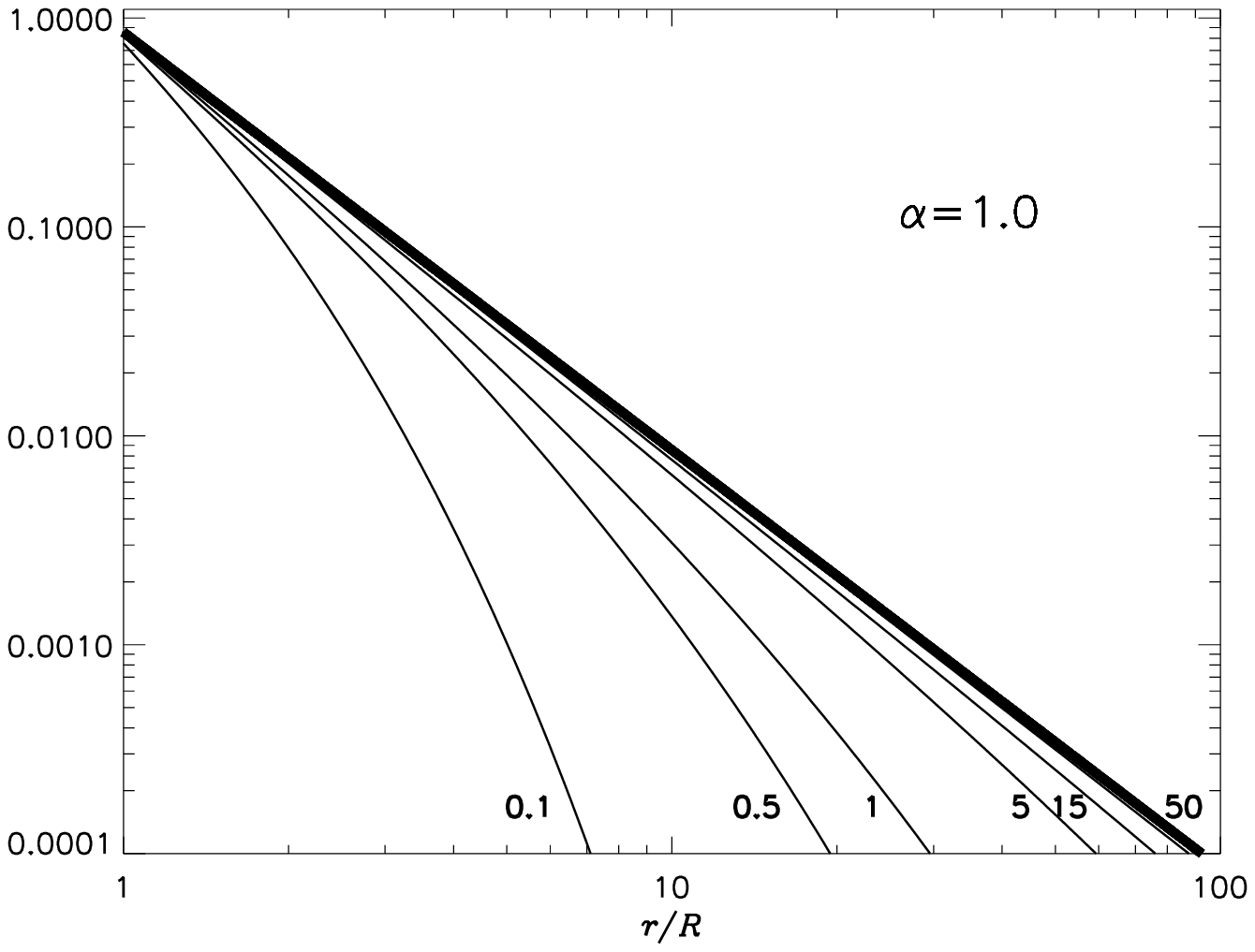} 
 \includegraphics[width=3.4in]{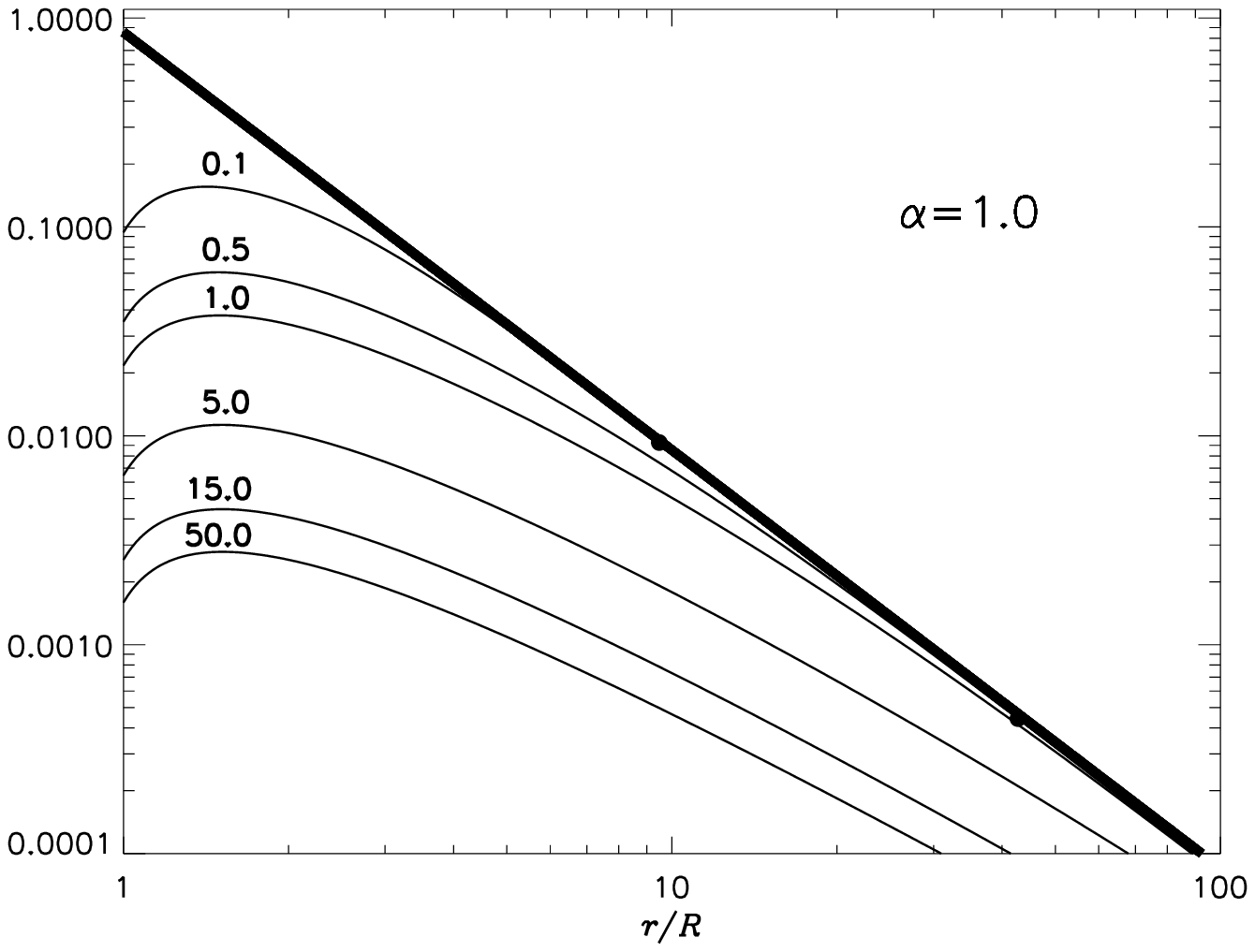} 
% \includegraphics[width=3.5in]{Images/decay.eps} 
% \vspace*{-1.0 cm}
 \caption{Temporal evolution of the surface density during the disk build-up (left) and dissipation (right) phases for $\alpha = 0.1$ (top) and $\alpha = 1.0$ (bottom). The dissipation phase is described in \S~\ref{dissip}. Epochs are counted in years and the thick line represents the quasi-steady state density profile of  Eq.~(\ref{eq:sigma_steady_state}). Black dots indicate the outer radius of the accretion region where the radial velocity is zero (stagnation point).}
   \label{density}
\end{center}
\end{figure*}

Figure~\ref{density}, left panels, shows the temporal evolution of the surface density of a newly formed disk subject to a constant injection rate for 50 years. The steady-state limit of  Eq.~(\ref{eq:sigma_steady_state}) is shown as the thick line.
In the entire disk the surface density grows with time, albeit with very
different rates depending on the distance from the star. Therefore, the time
required to reach quasi-steady state depends strongly on the location in the
disk as well as on the viscosity parameter $\alpha$.
%\rmACC{From the inner part of disk to the outermost borders, surface density progressively increases to reach its limit value.  On  Figure~\ref{time_QS}, we can see the $\alpha$ dependence of the time required for the density to reach $95\%$ of the limit value. }
Also, the slope of the surface density profile is very steep at the early stages of disk formation and then slowly decreases with time, asymptotically reaching the steady-state slope.

In this simple situation of a disk fed at a constant rate, the influence of the viscosity parameter $\alpha$ on the disk is simply to scale time up and down. It is immediately evident from Eq.~(\ref{eq:density}) that changing $\alpha$ is equivalent to changing the time, as long as the ratio between the orbital speed and the sound speed is kept constant.
Therefore, the higher the $\alpha$ the faster the disk will grow. For instance,  a disk with $\alpha = 1$  will evolve strictly 10 times faster than  a disk with  $\alpha=0.1$. 

%\rmACC{The influence of the viscosity parameter $\alpha$ on the surface density can be therefore understood as a time scaling factor. As defined in \citet{1973A&A....24..337S}, the higher the $\alpha$ value, the higher the turbulent velocity in  the disk and/or the size of the turbulent eddies. Consequently, the associated viscous torque will get higher and the disk will decrete and grow faster. Under a constant mass injection rate, high $\alpha$ values mean that the disk gets dense and big sooner than for low $\alpha$ values.  From the equations presented in \citet{carciofi11}  we can see directly that proportional link between radial velocity $v$ and the $\alpha$ parameter:}

%\beginequation}
%v \propto 1/ \Sigma ,\hspace*{0.3 cm}
%\Sigma \propto  1/\alpha ,\hspace*{0.3 cm}
%thus \hspace*{0.2 cm} v \propto \alpha
%\label{eq:prop}
%\end{equation}

%\rmACC{Moreover, $\alpha$ controlling directly the turbulence speed in the disk, it sets the time in a proportional way as we can see in  Figure~\ref{time_QS}, e.g. a disk described by an $\alpha$ = 1.0 value will evolve strictly 10 times faster than for a disk with  $\alpha$=0.1.}

 \begin{figure}[t!]
% \vspace*{-2.0 cm}
\begin{center}
 \includegraphics[width=3.5in]{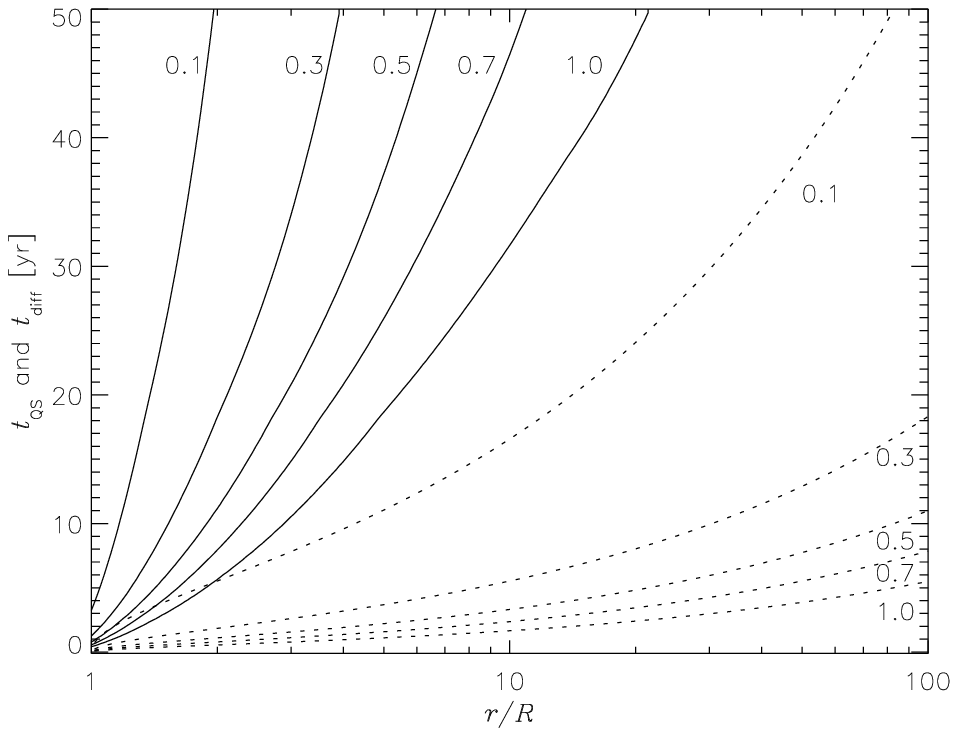} 
% \includegraphics[width=3.5in]{Images/superpop_time.eps} 
% \vspace*{-1.0 cm}
 \caption{Time required for the disk density to reach $95\%$ of the quasi-steady state value ($t_{\rm QS}$ (solid lines) and viscous timescale, $t_{\rm diff}$  (dashed lines), as functions of the distance from the star and computed for different values of $\alpha$: 0.1, 0.3, 0.5, 0.7 and 1.0.}
 %Right: Same figure as left but with $t_{\rm QS}$ and  $t_{\rm diff}$ multiplied by $\alpha$.}
   \label{time_QS}
\end{center}
\end{figure}

% Normalization problem : why the inner part never reach more than 90% of the steady state ? it should quite quiclky ...
% Comparison Diffusion time and gaz time.

Naturally, one could expect that the disk formation occurs in the viscous diffusion timescale. However, the timescale for the disk to get close to the steady state is much longer than the viscous timescale. In order to explicitly show this, we plot, in Figure~\ref{time_QS},  $t_{\rm QS}(r)$, defined as the time the surface density at radius $r$ takes to reach $95\%$ of its limit value (in solid line). Here, QS stands for quasi-steady state. We compare those timescales to the viscous timescale,
\begin{equation}
     t_{\rm diff} = r^2 / \nu \enspace,
\end{equation}
which is a timescale for how quickly diffusion can change the density at a particular radius. Here, $\nu$ is the kinematic viscosity of the gas, for which we adopt the turbulent (or eddy) viscosity of \citet{1973A&A....24..337S}, $\nu$ = $\alpha c_s H$, with $c_s$ being the sound speed and $H$ the disk scale height at a given radius. % \rmACC{This results in $T_{diff} \propto 1/\alpha$.}
The viscous diffusion timescales are plotted in  Figure~\ref{time_QS}  as the dashed lines. The viscous diffusion timescales are much shorter than $t_{\rm QS}$ and are not, therefore, representative of the time the disk takes to settle in a quasi-steady steady after a long period of constant injection rate. For instance, a disk with $\alpha= 1.0$ takes 30 years to reach 95\% of the limit density at $10\,R_\star$, whereas the diffusion time for this radius is only 1.7 years. Although most of the changes occur at a viscous diffusion timescale, it follows from the above that the disk requires a longer time to be filled up and stabilize in its outermost reaches. It is probably difficult to find a Be system in which the entire disk is close to steady state, since this would require an exceptionally long and stable period of mass loss from the central star. However, $\zeta$ Tauri and 1 Del could probably be emblematic cases of this behaviour and more generally, late-type Be stars, since they are supposedly more stable than early-type ones as seen from photometric studies \citep{1989A&AS...81..151C,1992A&AS...92..533B,1987MNRAS.227..213S,2011AJ....141..150J}. We conclude that the timescales for the disk to fully redistribute the material ejected from the star, $\tau_{\rm d}$ (\S~\ref{sec:introduction}), are much larger than the viscous diffusion timescales.

From observations \citep[e.g.,][]{1986A&A...162..121W} and theoretical considerations \citep[e.g.,][]{2005ASPC..337...75B}, Be disk models often use a power-law to describe the radial mass density profile of the disk, i.e.,
$$\rho(r) \propto r^{-n} \enspace .$$ 
 Let us now discuss what are the model predictions for the value of $n$.
%both in the observational and theoretical literature (sometimes $n$ has different definitions. In the following, we discuss our model predictions about the m index values.
%Let us express the mass density $\rho(r,z)$, in cylindrical coordinates, in terms of the surface density.
Assuming vertical hydrostatic equilibrium and an isothermal gas, the mass density is related to the surface density by
%\begin{equation}
%\rho(r,z) = \rho_0(r) \exp[-0.5(z/H)^2] \enspace,
%\end{equation}
%where $\rho_0$ describes the radial dependence of the density. The surface density is calculated by integrating $\rho$ in the vertical direction
%\begin{equation}
%    \Sigma(r)=\int_{-\infty}^\infty \rho(r,z) \, dz = \sqrt{2\pi}H\rho_0 \enspace .
%    \label{eq:sigmadef}
%\end{equation}
%Thus
\begin{equation}
\rho(r,z) = \frac{\Sigma(r)}{\sqrt{2\pi}H(r)} \exp\left[ -\frac{z^2}{2H^2(r)} \right] \enspace,
\label{eq:rho}
\end{equation}
where $r$ and $z$ are the usual cylindrical coordinates. Since the isothermal
scale height increases as $r^{1.5}$ and the steady-state surface density
decreases as $r^{-2}$, we obtain the familiar result that the steady state
volume density falls as $r^{-3.5}$. \cite{2008ApJ...684.1374C} investigated
how the disk temperature structure affects the viscous diffusion. The
combination of the radial temperature structure, disc scale height, and
viscous transport produces a complex radial dependence for the disk density
that departs very much from the simple $n=3.5$ power-law. This is certainly a source of explanation to the wide range of $n$ values reported in the literature. But in the present paper, our results suggest that there is an additional, alternative explanation for this.

%The index $n$ of this power law is equal to 3.5 for steady state isothermal outflow \citep{1999A&A...348..512P,1991MNRAS.250..432L}. However, as pointed out by \cite{2008ApJ...684.1374C} , the combination of the radial temperature structure, disk scaleheight, and viscous transport produces a complex radial dependence for the disk density that departs very much from the simple n = 3.5 power-law. This may account for the large scatter of the index n reported in the literature together with the differences in the modeling.

%\rouge{Atsuo veut enlever cette figure ou au moins pour un alpha , car on a deja dit que alpha = 1.0 ca va 10 fois plus vite que alpha=0.1}

 \begin{figure*}[t!]
% \vspace*{-2.0 cm}
\begin{center}
\includegraphics[width=3.5in]{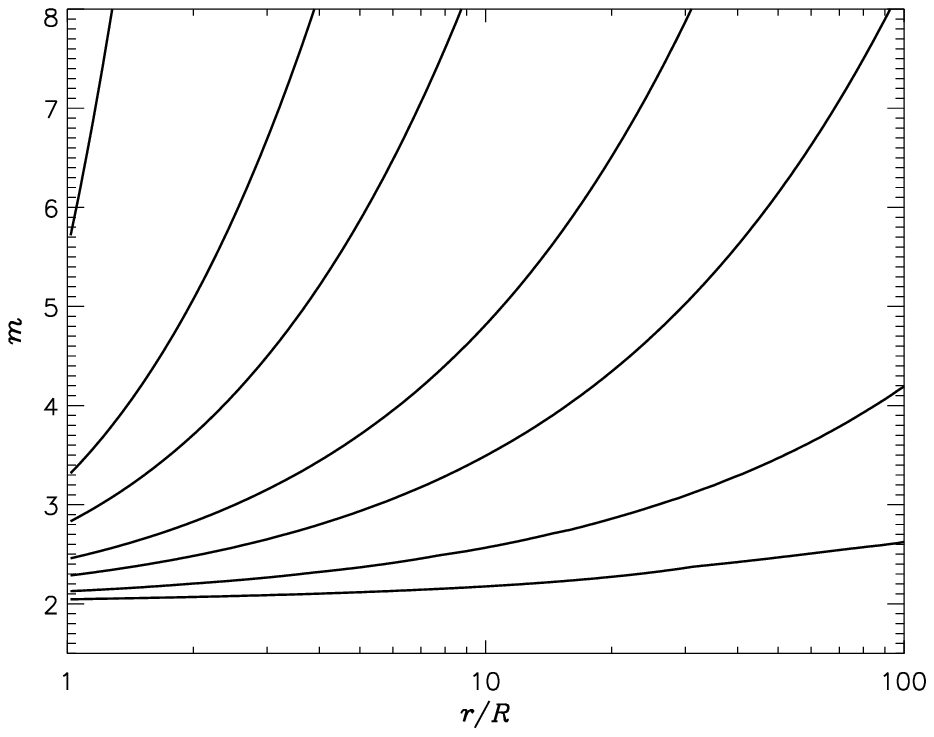} 
\includegraphics[width=3.5in]{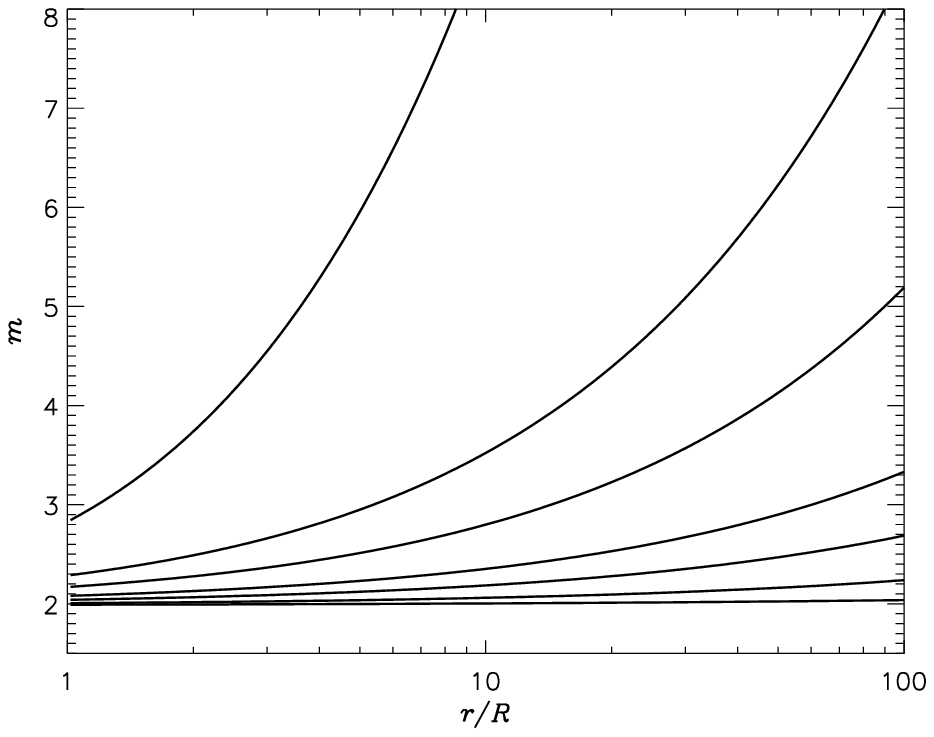} 
% \vspace*{-1.0 cm}
 \caption{Index $m$ of the surface density, $\Sigma \propto r^{-m}$, as a function of the radius during disk build-up phase (left:  $\alpha$ = 0.1, right: $\alpha$ = 1.0). Epochs from top to bottom are:   0.1, 0.5, 1.0 , 2.5, 5.0, 15.0 and 50 years. See Figure~1 and text.}
   \label{index1}
\end{center}
\end{figure*}

%the large range observed in the radial density slope.
In Figure~\ref{index1} we show the local index of the surface density profile defined as $m= - d \ln(\Sigma)/d \ln(r)$. 
%In order to compare $m$ with the predictions from our modeling of a build-up phase, we derived the behaviour of the m index from the radial surface density profile which is defined as: $\Sigma(r) \propto r^{-m}$, so m = n - 1.5 (in the isothermal case).
From Eq.~(\ref{eq:rho}) we see that local index of the volume density, $n$, is related to $m$ by $n = m + 1.5$, for isothermal disks.
A large value of $m$ at a particular location indicates that the disk is far from steady state at that location. %For disks far from steady-state, the $n$ is much larger than $2$. 
As the disk builds up, the slope progressively reaches the limit value of $2$, beginning at the inner disk and spreading out to larger radii. The time required for the outer disk to reach the $m=2$ slope can be quite long (decades), even for large values of $\alpha$. Therefore, if one estimates the slope of the density profile (for instance, via the IR SED), one is likely to find values of $m > 2$ (or, equivalently, $n > 3.5$). This could therefore account for the wide range of $n$ values in the literature. Note that this is only true for the disk build-up phase since, as presented below, the density slope for the dissipation phase is quite different. 

% from the inner to the outer disk. At 10 $R_\star$ the time required to do so is around 80 years for $\alpha= 1.0$. 
%As an alternative to non-thermal effects in the disk, these results also show that the temporal evolution of the surface density profiles of a growing disk can account for the wide range observed in the density radial exponent.

%On this figure, we can see that n' is an exponential function of the radius for one given epoch and thus far from being constant.  We can see that the previously quoted quasi-steady state n value of 3.5 corresponding to n'= 2  is reached only when the disk is stabilized, after more than 10 years.

\subsubsection{Disk dissipation}
\label{dissip}

When mass injection from the star ceases, the angular momentum supply to the disk stops and the inner disk starts to re-accrete back onto the star. Reaccretion of material from the inner disk occurs because the turbulent viscosity transports angular momentum outward, transferring angular momentum from the inner to the outer parts of the disk.

%I moved the steady-state accretion down below.

\begin{figure}[t!]
\begin{center}
 \includegraphics[width=3.5in]{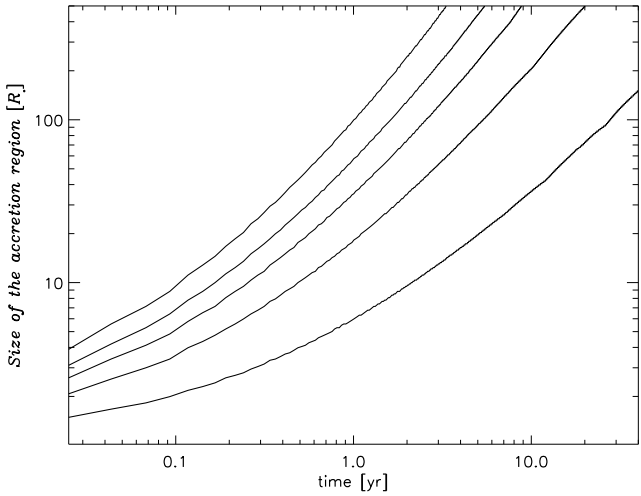} 
 \caption{Position of the stagnation point, i.e. size of the accretion region, as a function of time for a dissipating disk for different $\alpha$ (bottom to top: 0.1, 0.3, 0.5, 0.7 and 1.0).}
   \label{inflex}
\end{center}
\end{figure}

To study the disk dissipation we start from a pre-existing disk and turn off  mass input from the star. The temporal evolution of the surface density is shown in the right panels of Figure~\ref{density} for two values of $\alpha$ (0.1 and 1).
As soon as the mass input is turned off, the inner disk re-accretes back onto the star, giving rise to a  \emph{stagnation point}, i.e.\  a region in the disk where the radial velocity is zero. The stagnation point slowly propagates outwards and the density of the inner part decreases in a homologous way with time, i.e. the density structure within the stagnation point remains self-similar during this phase. The speed in which the stagnation point propagates outwards is shown in Figure~\ref{inflex}; it is, as expected, strictly proportional to $\alpha$. For instance, for $\alpha = 1$ the stagnation point reaches about 100 $R_\star$ in 1 year. This time, at which accretion starts at 100 $R_\star$, is actually shorter than the viscous time ($\sim$ 5.5 years).

Using the formalism presented in \cite{2005ASPC..337...75B}, it is possible to demonstrate that the surface density of an accretion disk in the steady-state regime is given by
\begin{equation}
 \Sigma(r) = \frac{\dot{M} V_{\rm crit} R_\star^{1/2}} {3 \pi \alpha c_s^{2} r^{3/2}} \left(1-\sqrt{\frac{R_{0}}{r}}\right)\,,
\label{eq:accretion}
\end{equation}
where $c_s$ is the sound speed.
 
\begin{figure}[t!]
\begin{center}
 \includegraphics[width=3.5in]{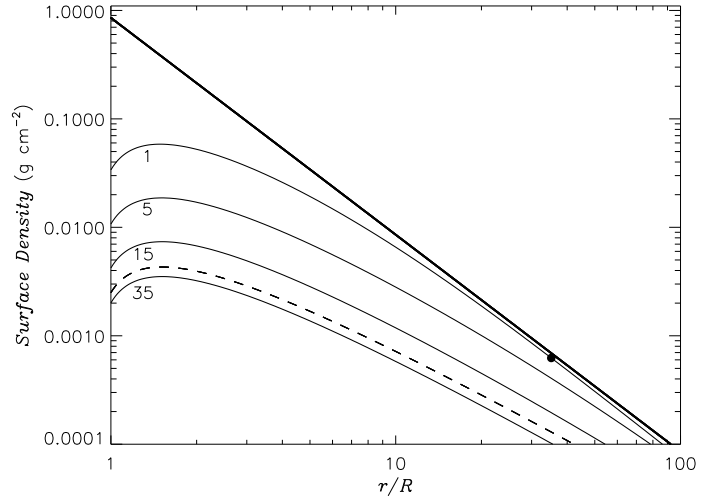} 
 \caption{Surface density profiles from {\sc singlebe} simulations (solid lines) and from the analytical expression of Equation~\ref{eq:accretion} (dashed line) in the disk dissipation scenario. The thick line represents the quasi-steady state density profile of Eq. (2) and the black dot indicates the stagnation point. $\alpha = 0.5$ and the four epochs shown are from top to bottom: 1, 5, 15 and 35 years. }
   \label{accretionth}
\end{center}
\end{figure}

In Figure~\ref{accretionth}, the shape of the simulated density profiles for a decaying disk is perfectly matched by the analytical curve of the surface density for a steady accretion state using $R_{0} = 0.85 R_\star$. This parameter is particularly important because it determines the shape of the density in the inner part of the disk.  As time goes on, the stagnation point goes outwards, the accretion region grows and a larger fraction of the disk assumes a density distribution similar to Eq.~\ref{eq:accretion}. For example, 15 years after accretion starts, the first 50 $R_\star$ are already in a steady accretion state.

% and $R_{0}$ is the outer radius of the disk. -> no, R_0 is the same as above.
%This predicts a 

%We should state here that during the dissipation phase, the extended accreting region has a density distribution that depends directly on the boundary condition at $r=R_{0}$. Our choice for this condition ($R_{0}= 0.56 R_\star$) is debatable but as the surface density at 1 $R_\star$ is non-zero (as it is expected when the particles are accreted back onto the star), the value we assume is consistent. Quantitatively, however, there isa degeneracy in the shape of surface density at the inner part of the disk as a result of the indetermination of the inner boundary condition.

%Hence, the disk assumes a dual behavior in which the internal part is accreting back to the star while the external part still continues to decrete outwards. % \rmACC{in the steady state regime and therefore keeps the same limit density as if the injection rate was still on. Thus the disk  both vanishes inwards and outwards. During this dissipation phase, the position of the inflection point (frontier between accretion and decretion parts) grows in time (Figure~\ref{inflex}), i.e., the accretion region size increases with a temporal dependence on $\alpha$.} 

%\rmACC{It is interesting to realize that, during this phase, the decretion speed sets the accretion speed and the faster decretion, the faster the accretion.}

% Fit pas evident en fait, pas possible de trouver une loi facile.
% Discussion sur la dependence de l accretion en fonction de \alpha

\begin{figure*}[t]
\begin{center}
 \includegraphics[width=3.in]{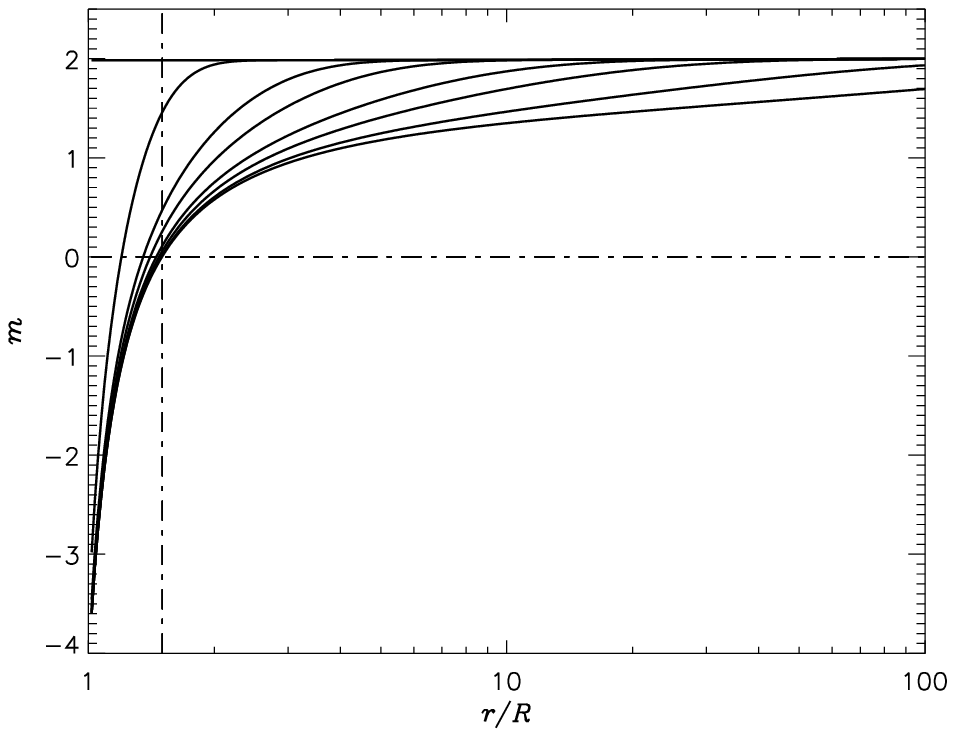} 
 \includegraphics[width=3.in]{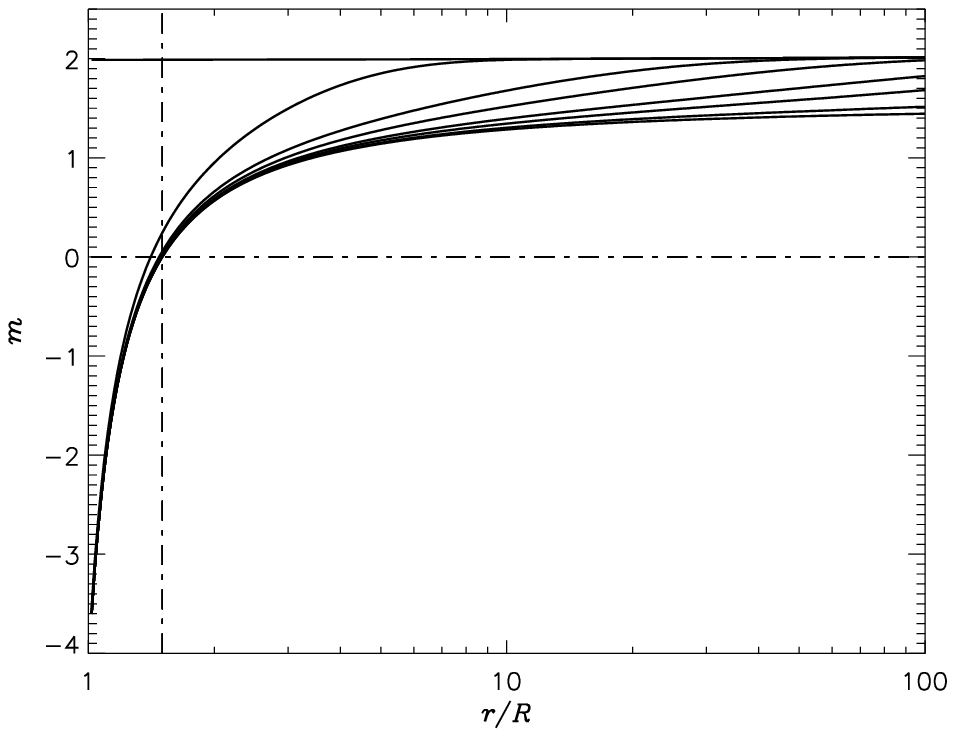}  
 \caption{Same as Figure ~\ref{index1} for a dissipating disk. Shown are results for two values of $\alpha$ (left: $\alpha$= 0.1, right: $\alpha$  =1.0) and for the following epochs (top to bottom): 0, 0.1, 0.5, 1., 2.5, 5, 15 and 50 years. The horizontal and vertical dashed lines mark the final position of the maximum surface density.}
   \label{index_dec}
\end{center}
\end{figure*}

We conclude that during the disk dissipation the disk assumes a dual behavior, with two distinct regions (inner accretion and outer decretion) separated by the stagnation point. In the decreting part, the index $m$ of the surface density is about $2$, whereas in the accreting part the slope goes from 1.5, which is the value for a steady-state accreting disk, to negative values closer to the star (see Figure~\ref{index_dec}).

%Because of accretion, the evolution of the surface density for the dissipation phase has not the same timescale as the build-up phase. For $\alpha$ = 1.0, only 0.1 year is needed for the accretion region to reach 4 $R_\star$, this is much faster than the diffusion time at this radius which is about 1 year.} 

%\rmACC{Such a decaying disk would actually look like a ring with a decreasing density. It is noteworthy that at this stage, the position for the density maximum is always located around 1.5 stellar radius and it doesn't depend on $\alpha$. However, the higher the $\alpha$ value, the faster the ring will form and stabilize  at its definitive radius: from 6 to 0.5 year for 0.1 to 0.9  $\alpha$ values. {This stabilization of the peak density can be explained by a limit value for the accreted mass from the outer part of the disk}.}

%When the disk decays, we see that the surface density doesn't dissipate symmetrically compared to the build-up phase because of the characteristic outward motion of mass decretion.  We can see at 5 $R_\star$ that it requires a bit less than 4 years for the surface density to  get to 0.01 g.cm$^{-2}$ in the bulding phase. In the dissipation phase, it takes about twice that time to get back to that value. Because of the radial dependency of the diffusion time, the closer to the central star, the faster the surface density will react to mass decretion.

\subsection{Periodic injection rate}
\label{period}

\begin{figure*}[t]
%\placefigure{}
% \vspace*{-2.0 cm}
\begin{center}
\includegraphics[width=3.4in]{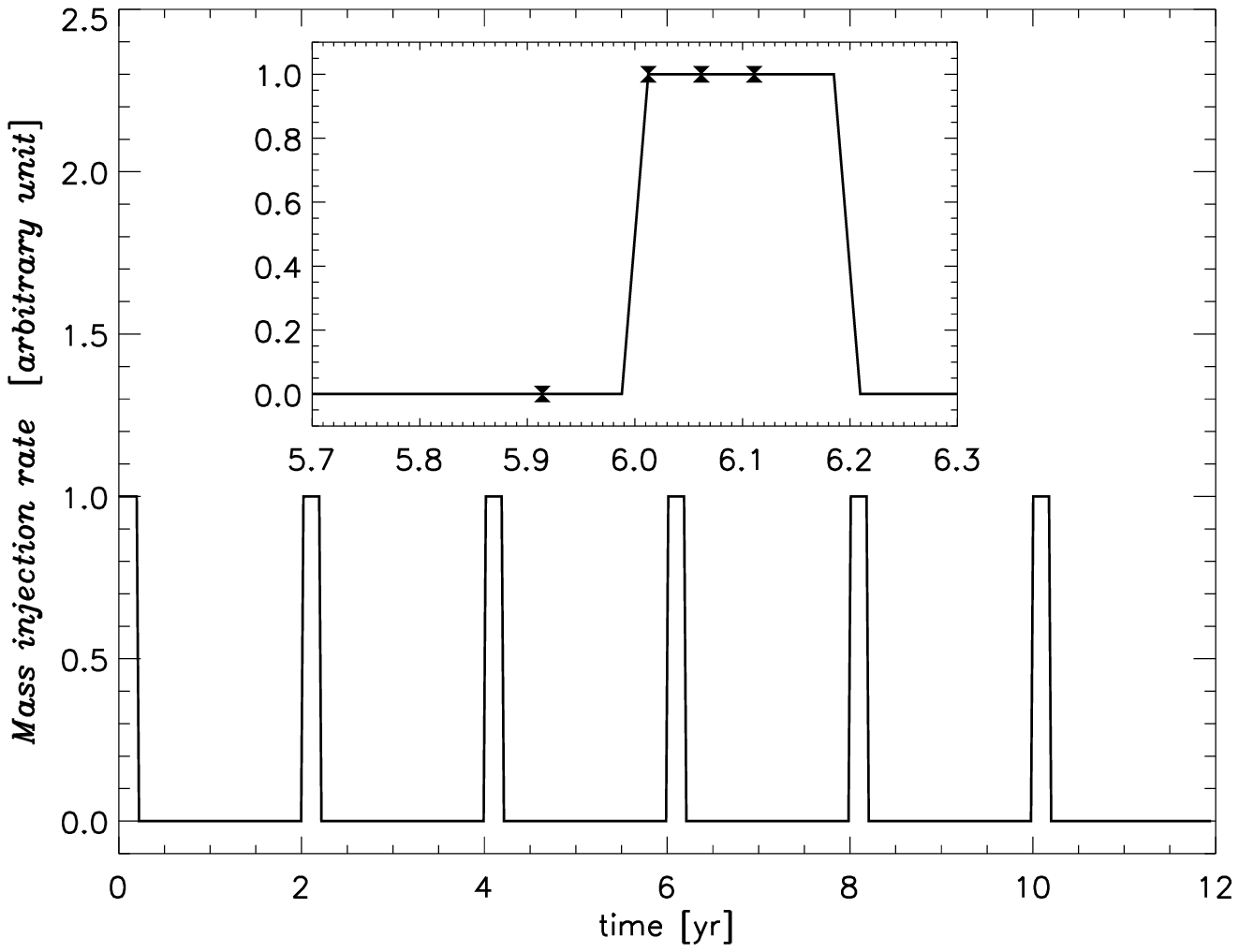} 
\hspace*{0.5 cm}
\includegraphics[width=3.4in]{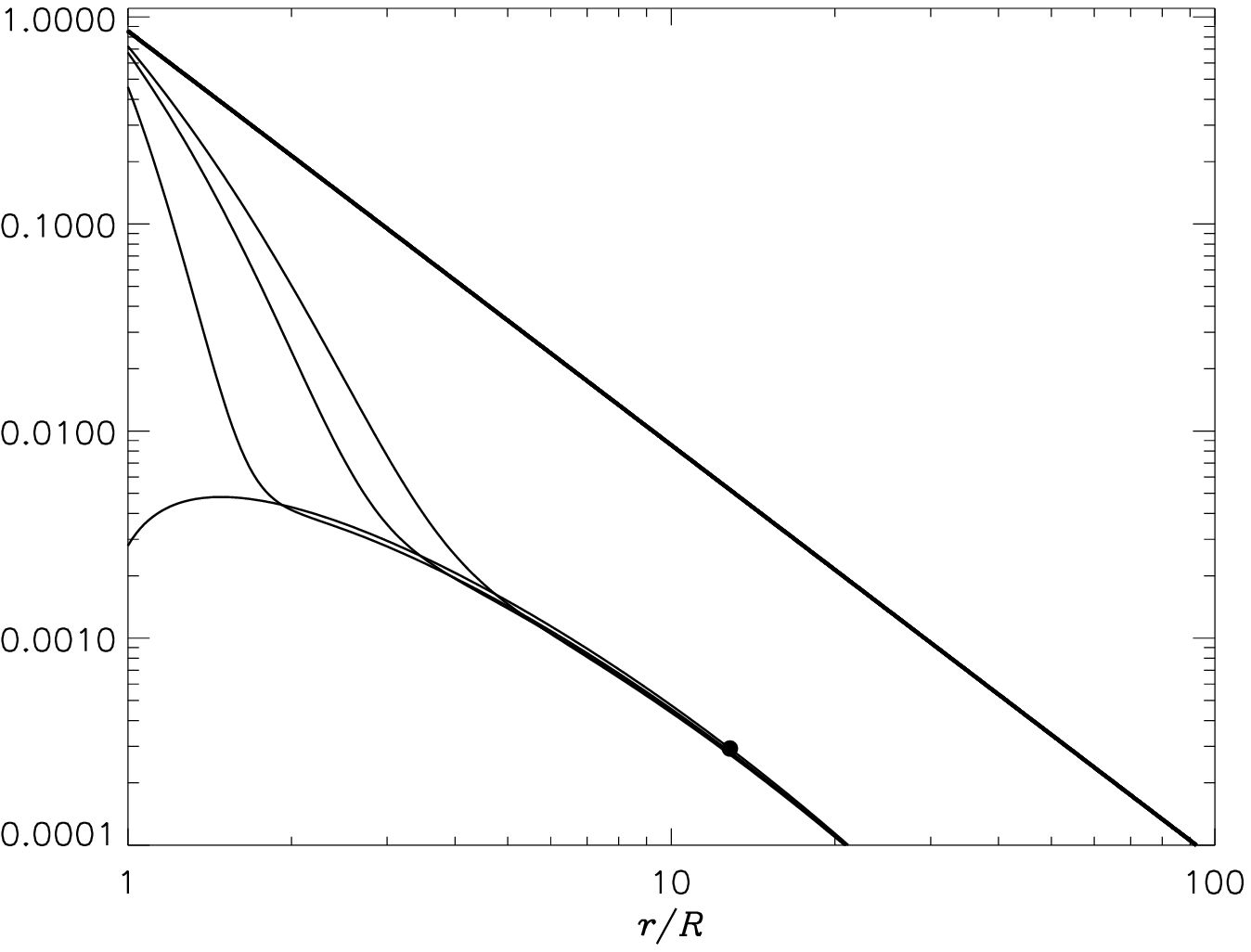} 
\includegraphics[width=3.4in]{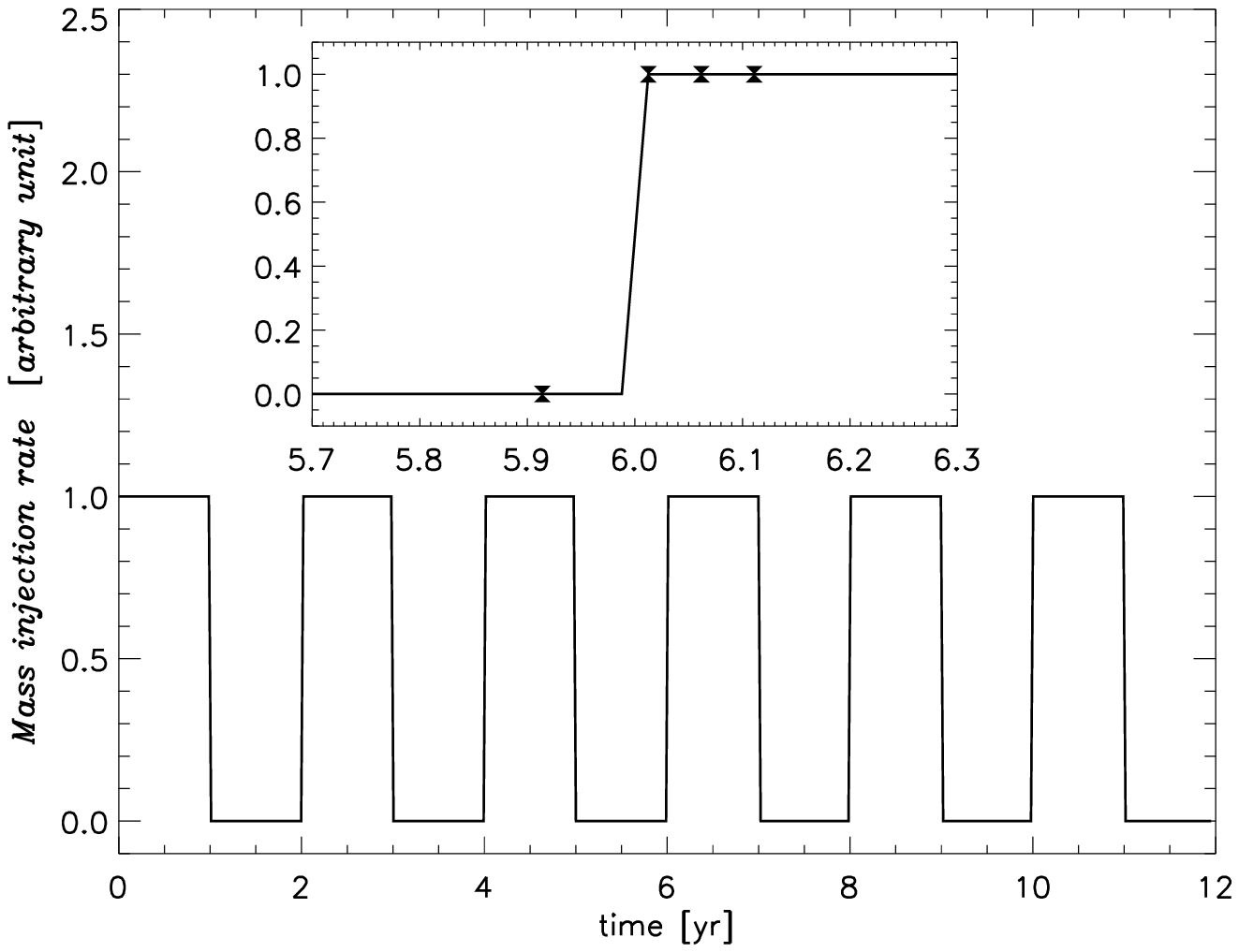} 
\hspace*{0.5 cm}
\includegraphics[width=3.4in]{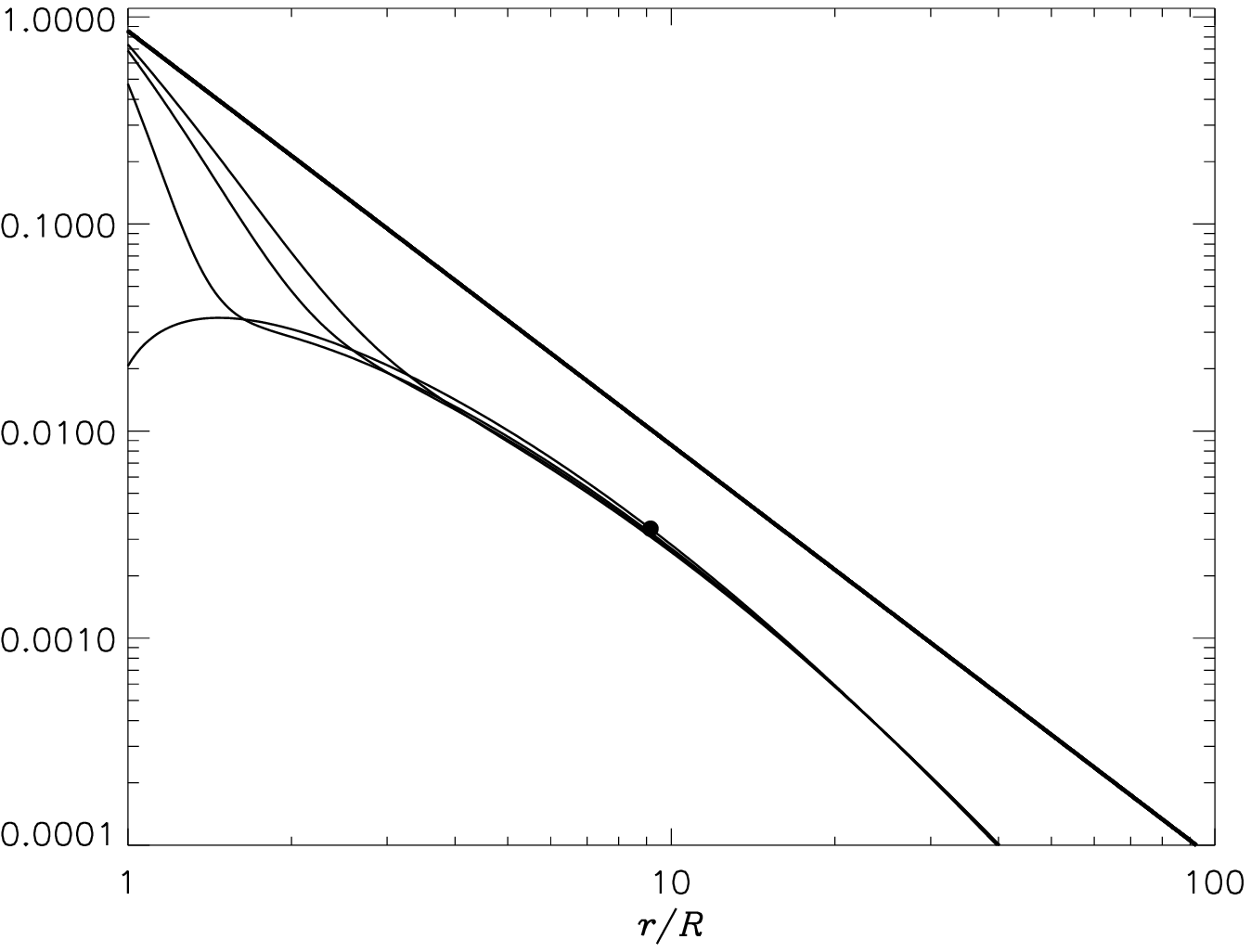} 
\includegraphics[width=3.4in]{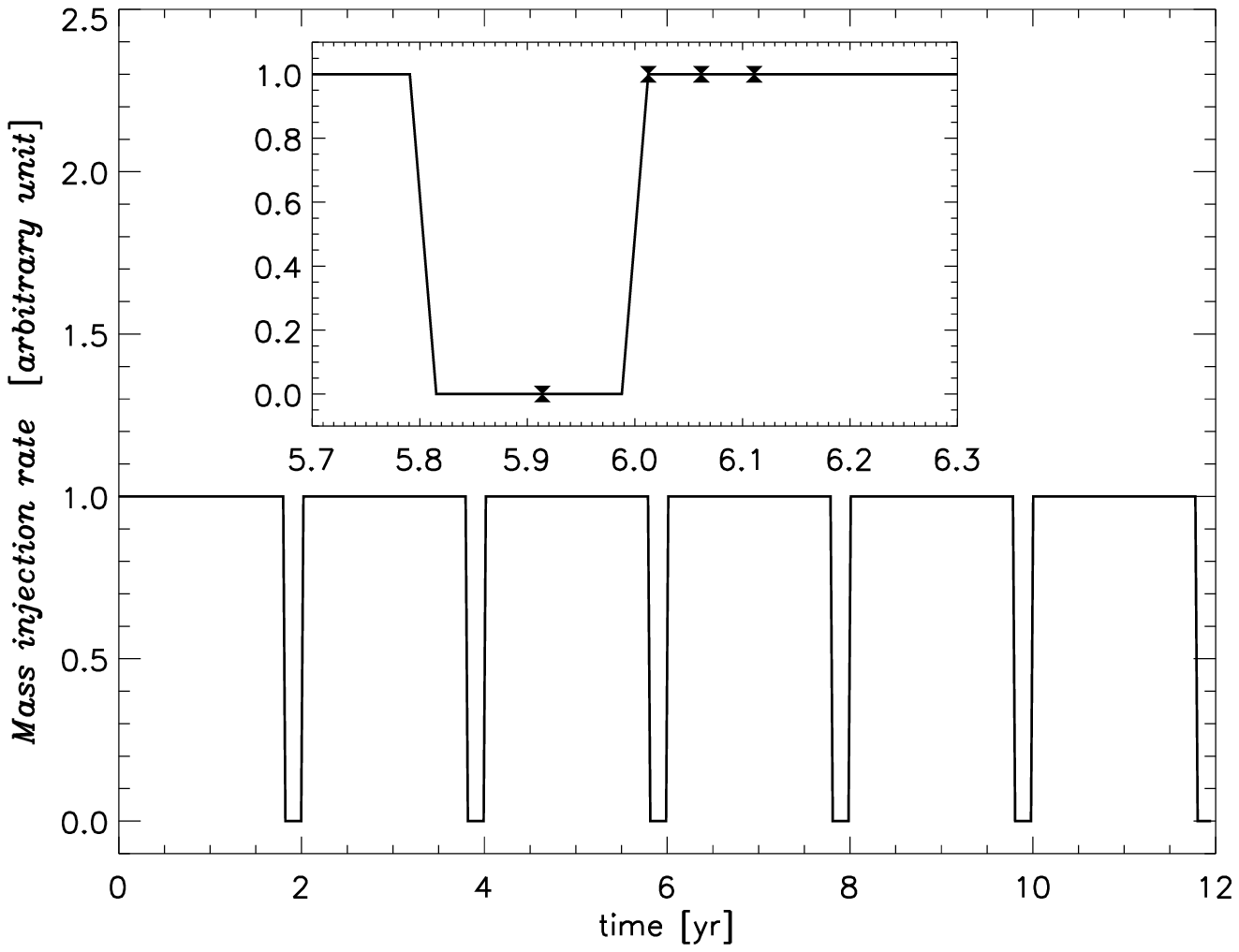}
\hspace*{0.5 cm} 
\includegraphics[width=3.4in]{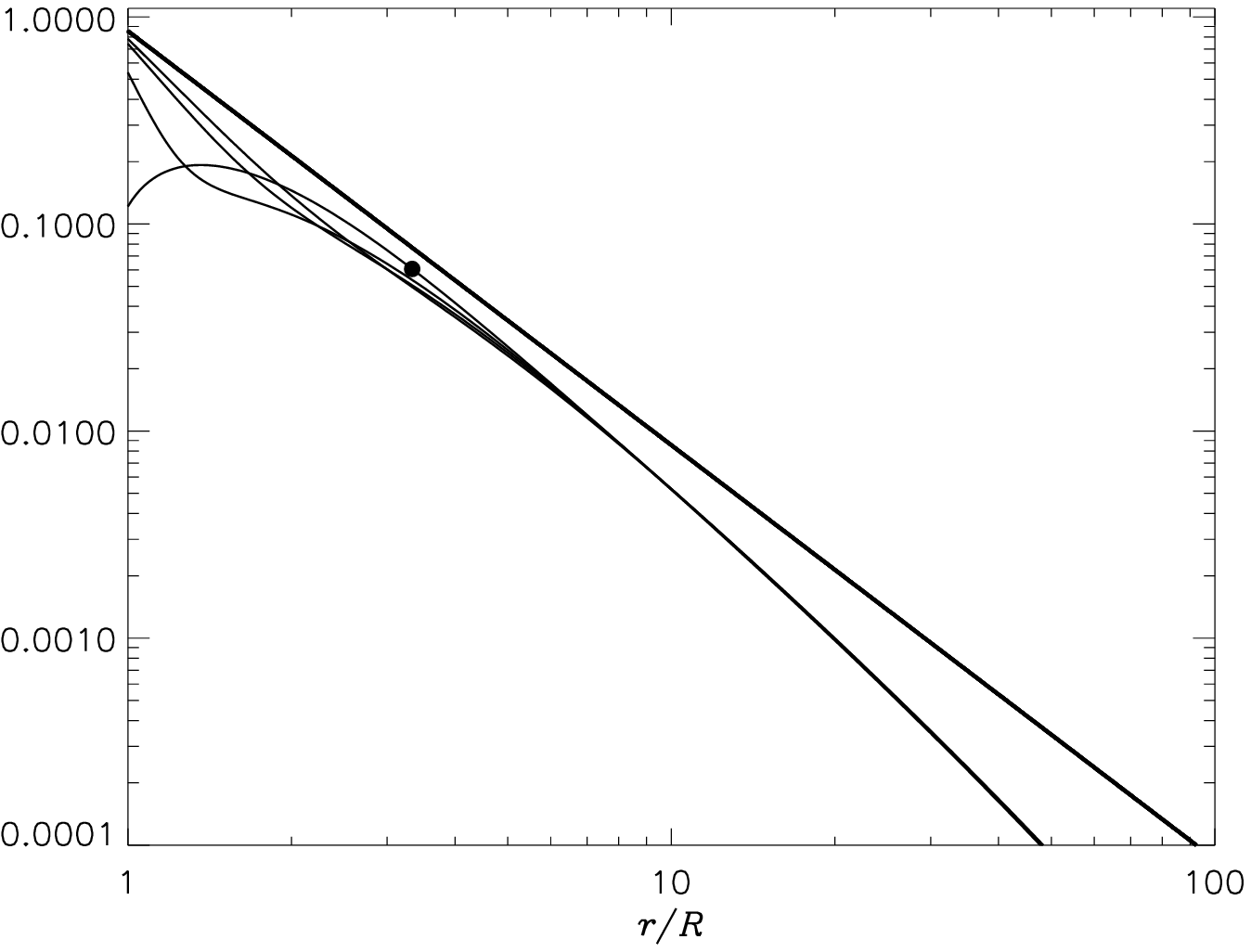} 

 \caption{\textit{Left}: Periodic mass injection rate with a 2 year period for different DCs (top: 10\%, middle: 50\%, bottom: 90\%). \textit{Right}: Corresponding surface density profile for $\alpha$=0.5 at chosen epochs. Bottom to top: 5.9, 6.01, 6.05 and 6.1 years.  Black dots indicates the stagnation point.}
   \label{scenarios}
\end{center}
\end{figure*}

Having studied the limiting case where the $\tau_{\rm in} \gg \tau_{\rm d}$, we now proceed to study cases for which $\tau_{\rm in} \sim \tau_{\rm d}$. To study how a disk forms and decays under the action of a variable mass loss, we studied some test-cases of periodic injection rates. In those models, we successively turned the injection rate on and off for different periods of time and computed the corresponding surface densities. Some examples are shown in Figure~\ref{scenarios}. For such  $\tau_{\rm in} \sim \tau_{\rm d}$ cases, we expect that the injection rate variation will generate a combination of the two previously described fundamental cases of disk build-up and decay. To investigate those periodic scenarios we used three parameters: the period, the $\alpha$ viscosity parameter and the duty cycle (hereafter DC, the time spent in an active state as a fraction of the total time).

%\rouge{ATSUO suggests to remove DC= 10\% and 90\% }

On Figure~\ref{scenarios}, we can see the effect on the surface density of a periodic material supply in a disk for different DCs. This succession of decay and build-up phases results in a complex temporal behaviour of the density profiles. For the same epochs, the curves are very specific for each DC because they depend on the previous dynamical history of the disk. However and as a general statement, we can say that for short timescales (i.e., within a given cycle) there is again a two-component structure: 

\vspace*{0.5cm}

\begin{itemize}
\item{the inner part of the disk that oscillates ; its surface density is consequently flapping up and down every period in fast reaction to the injection rate variations, and} 
\item{an outer part of the disk that is less affected by this intermittent mass injection but undergoes delayed, much slower variations.} 
\end{itemize}
%while the outer part of the disk is not quickly affected by this intermittent mass loss rate, the inner part of the disk reacts faster because of the radial dependency of the diffusion time and its surface density exhibits an oscillating profile.

 \begin{figure*}[t]
% \vspace*{-2.0 cm}
\begin{center}
\includegraphics[width=3.5in]{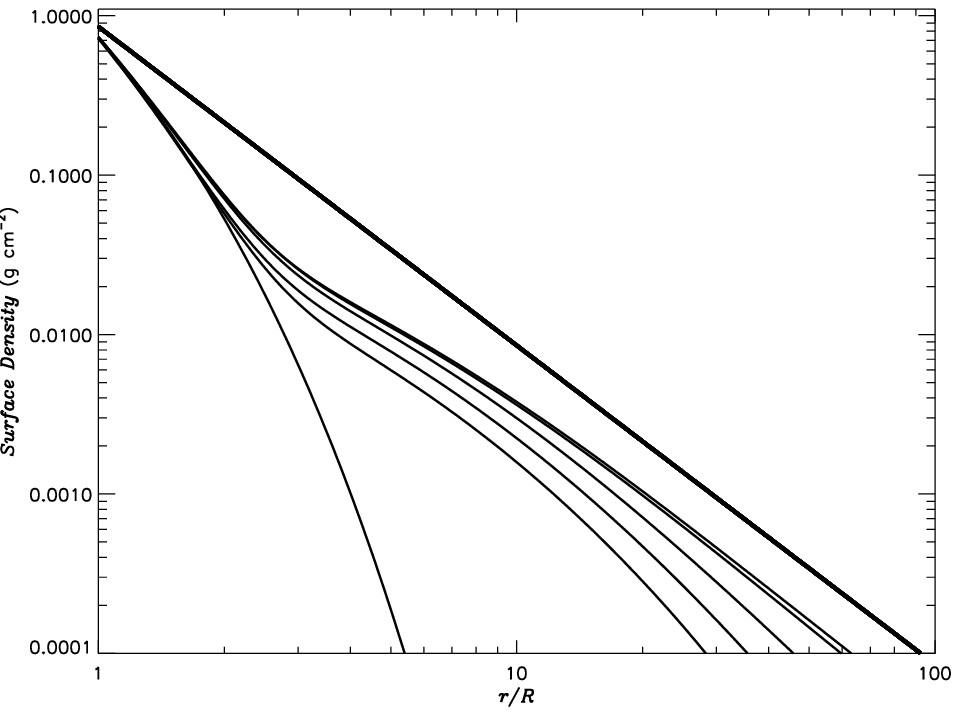}
\includegraphics[width=3.5in]{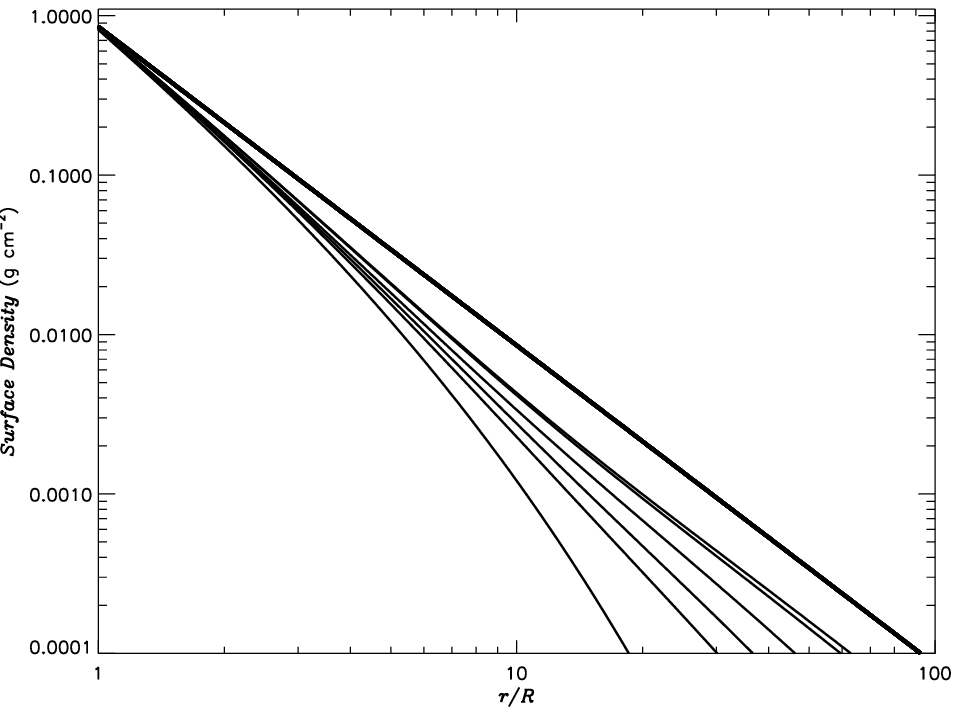}

\includegraphics[width=3.5in]{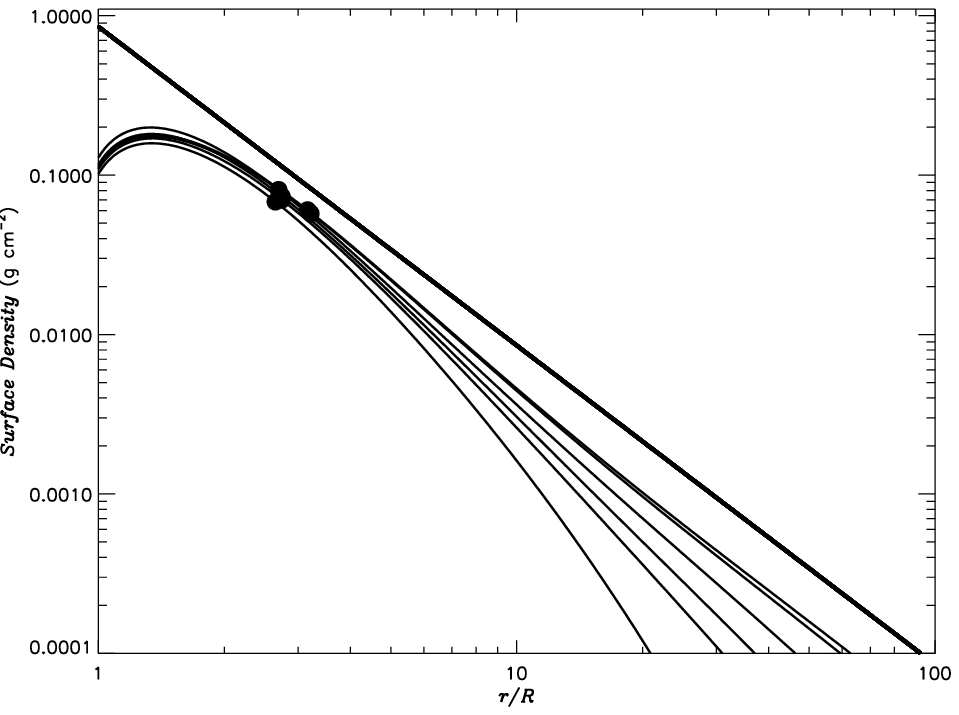}
\includegraphics[width=3.5in]{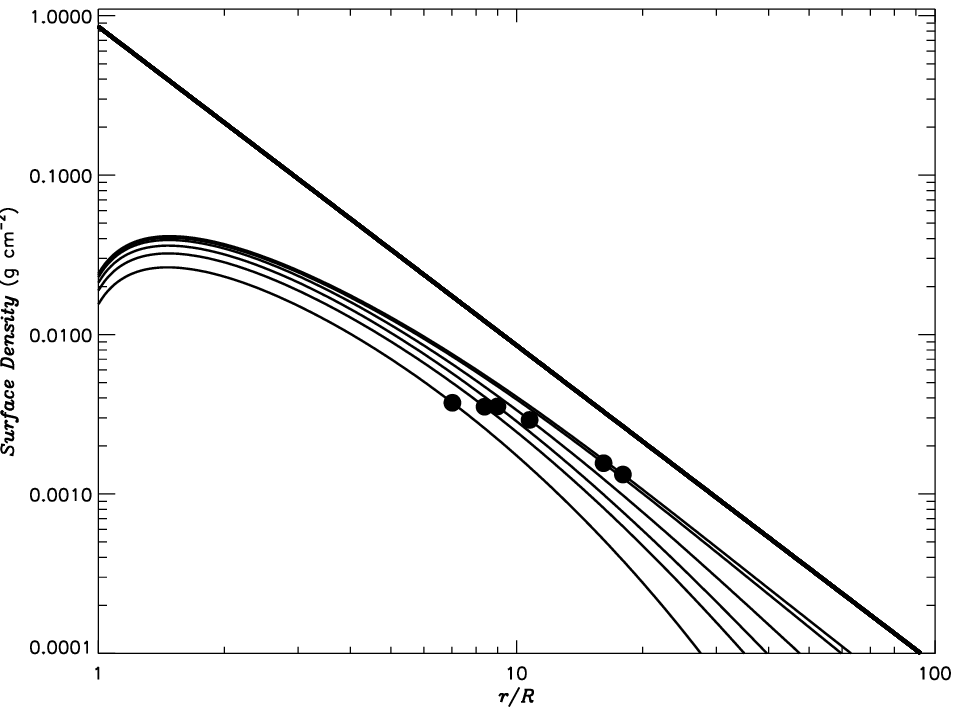}
% \caption{Left: Surface density profile for a periodic mass injection rate with a 2 year period. $\alpha$=0.5 and DC = 90\% shown at the same phases of 3 different cycles (the 2nd, 3rd and 51st). }
   \caption{Surface density profile for a periodic mass injection rate with a 2 year period, $\alpha$=0.5 and DC = 50\%. The four panels show four different phases at 6 different cycles: 1st, 2nd, 3rd, 6th, 26th and 51st. Top left: phase 0.05, top right: phase 0.45, bottom left: phase 0.55, bottom right: phase 0.95. Black dots indicates the stagnation point.}
   \label{scenarios2}
\end{center}
\end{figure*}

%\rouge{Atsuo thinks that figure is not mandatory and could be explained shorty in a text}

In order to illustrate the long-term changes of the disk,  it is interesting to look at the surface densities computed for the same phase along different cycles. Figure~\ref{scenarios2} shows the surface density for four different phases (0.05, 0.45, 0.55 and 0.95) at several cycles. In general, cycle to cycle changes of the inner disk (say, within 10 $R_\star$) are small. This is to be expected because the inner part responds quickly to changes of the mass injection rates. The outer part, however shows a steady growth in density as the disk is slowly filled up by the mass supplied in the previous cycles.  

\begin{figure}[t]
% \vspace*{-2.0 cm}
\begin{center}
\includegraphics[width=3.7in]{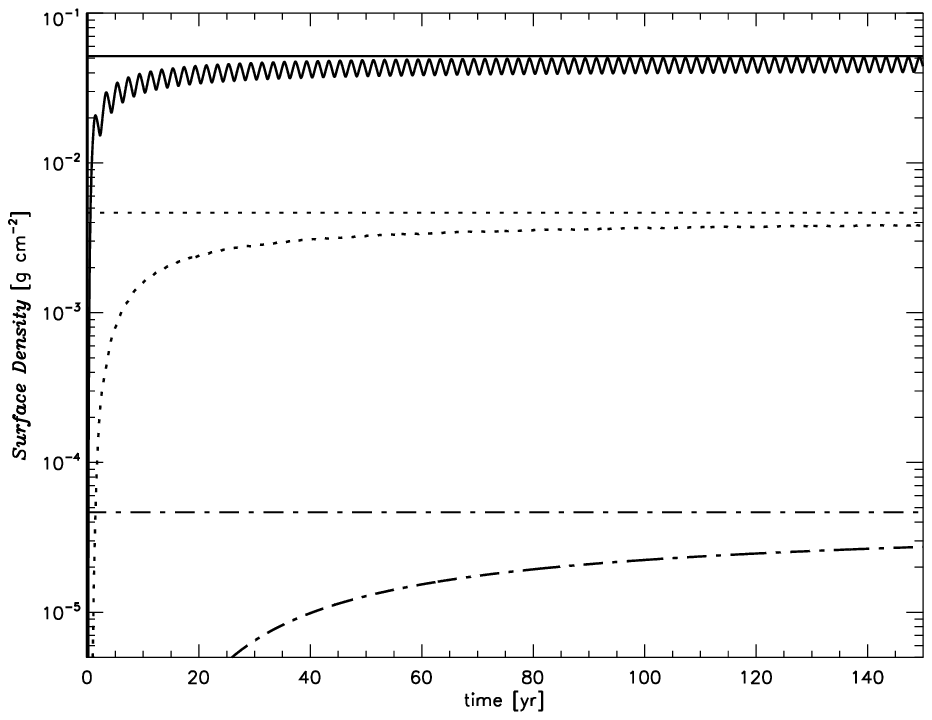} 
 \caption{Temporal evolution of the surface density at three different disk locations: 3 $R_\star$ (top in solid line), 10 $R_\star$ (middle in dotted line) and 100 $R_\star$ (bottom in dash-dotted line).
The corresponding  periodic mass injection rate has a 2 year period, $\alpha$=0.5 and DC = 50\%. The horizontal lines correspond to the limit value presented in \S~\ref{growth}.  The horizontal lines correspond to the limit value at that location ($\Sigma_0*r^{-2}$) times the DC.}
 \label{3rad}
\end{center}
\end{figure}

Given that the density is slowly but steadily increasing in the outer part,
one can wonder whether the density of a disk undergoing a periodic injection rate will ever reach a limit value in
its outermost reaches. Figure~\ref{3rad} shows the temporal variation of the
surface density at three different radii in order to answer to this question. It is demonstrated that indeed the
disk reaches a limiting value which is $\mathrm{DC} \times \Sigma_0(R_\star/r)^{2}$.

%\rouge{limit value * alpha dire que la valuer limit est la valeur de dm01 multiplie par DC because en moyenne Mdot = Mdot_dm01*DC}

\begin{figure}[t]
% \vspace*{-2.0 cm}
\begin{center}
\includegraphics[width=3.5in]{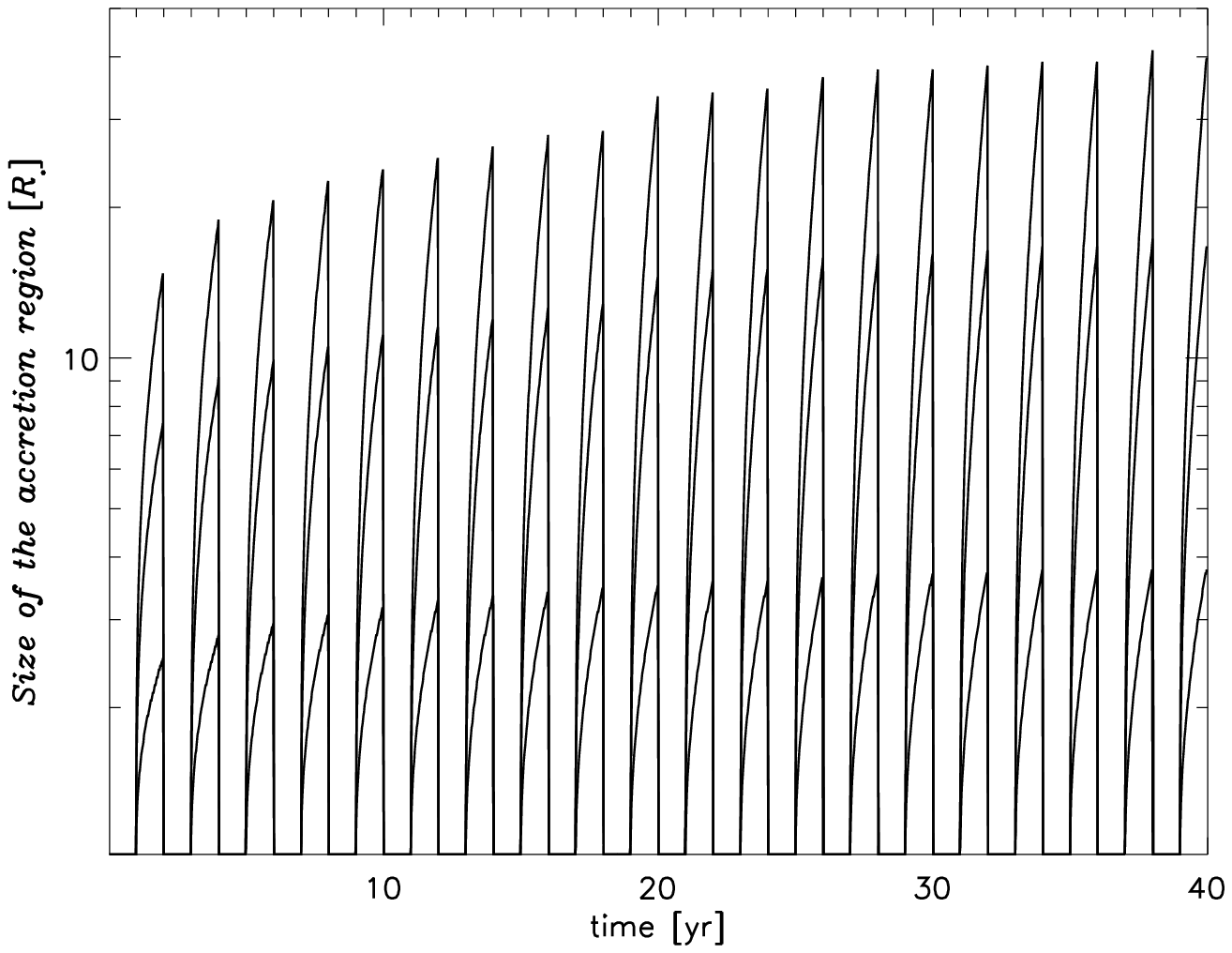} 
 \caption{Temporal evolution of the accretion region size for different values of $\alpha$ (top to bottom: 1.0, 0.5 and 0.1). Shown are results for a period of 2 years and a DC of 50\%.}
  \label{flap}
\end{center}
\end{figure}

It is interesting to note the density oscillation at $3\,R_\star$ in Figure~\ref{3rad}, a behavior that is not observed at larger radii.
This happens because for small radii the disk alternates between decretion and accretion, whereas the outer disk never experiences accretion, only decretion. This is quantified in Figure~\ref{flap}, which shows the position of the stagnation point as a function of time for three values of $\alpha$.

%\rouge{ATSUO suggested to remove that accretion region size figure.}

As mass injection stops, the accretion region increases in size until mass injection starts again. The higher the $\alpha$ values, the higher the viscous torque, the further the accretion region expands. The maximum extension of the accretion region stabilizes over time to reach 40 $R_\star$ ($\alpha =1.0$) after $\sim$ 40 years. To summarize, we find that long periods, low duty cycles (equivalent to short time outbursts) and high $\alpha$ values are needed to build the largest accretion regions.

In this subsection we saw how a disk builds in reaction to a periodic scenario. The evolution of the surface density radial profile depends strongly on the periodicity parameters  (DC and period) and also on $\alpha$. Contrary to the simpler scenarios explored in Sect.~\ref{refcase}, here the surface density depends on the details of the previous dynamical history of the system.

%Depending on its size, the oscillating region generates variability in different photometric band (see next section).  The observation of those photometric observables can thus reveal such oscillating structures in the surface density profile and lead to the determination of a decretion mass scenario.
%\rouge{mass budget ? }

\subsection{Episodic mass injections}
Even if periodicity is observed in lightcurves, they are also sometimes just random \citep{2002A&A...393..887M}. As an additional class of dynamical scenarios for which $\tau_{\rm in} \sim \tau_{\rm d}$, we investigate cases showing of a sudden increase in the mass injection rate (i.e. an outburst).

 \begin{figure}[h!]
\begin{center}
 \includegraphics[width=3.5in]{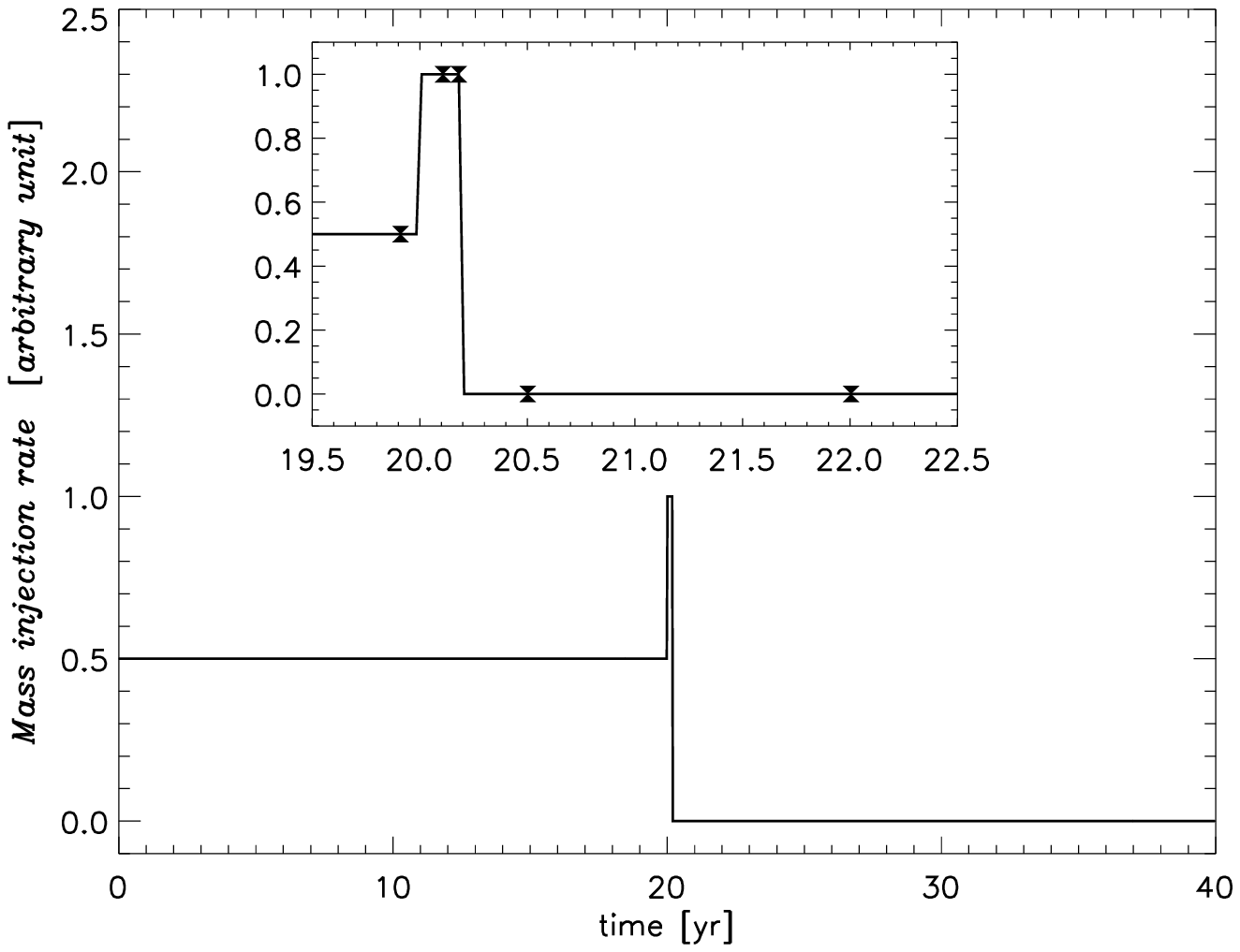} 
  \includegraphics[width=3.5in]{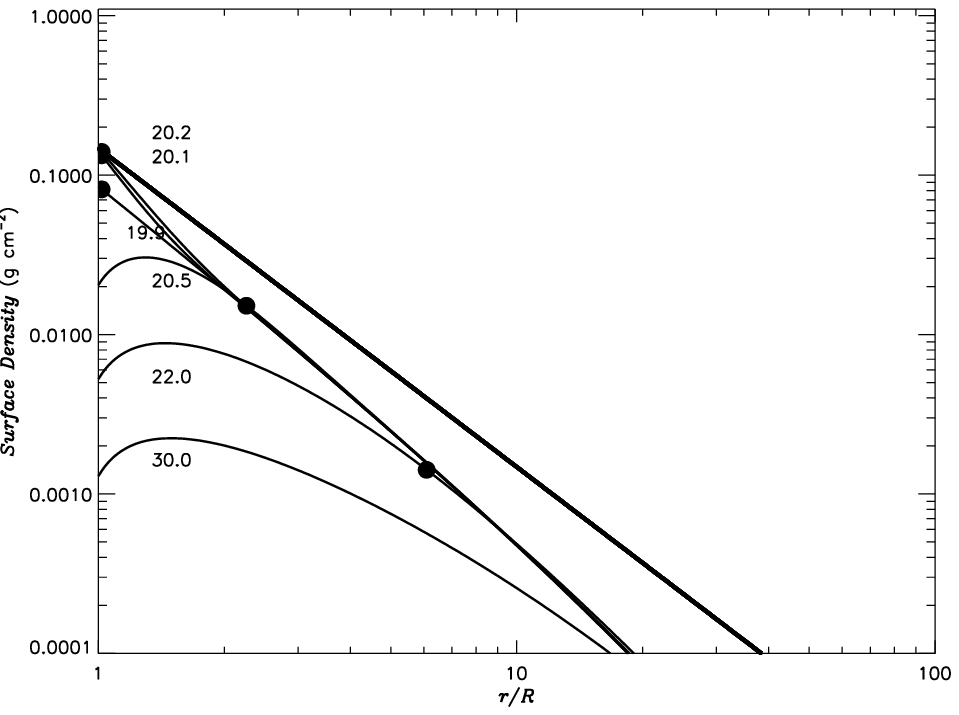} 
 \caption{Top:  Mass injection history of an outburst-like model. Bottom: Corresponding surface density profile for representative epochs, as indicated. For this model, $\alpha=0.5$. }
  \label{dm05}
\end{center}
\end{figure}
 
% In this model, the injection rate is set constant for 20 years until an outburst injects 2 times more mass during 0.2 year. Then the injection rate is set to zero. \rouge{The surface density of this model exhibits a profile comparable to one cycle of the periodic scenarios. Starting from a profile that gets closer to the limit value, the surface density increases all of sudden in the first radii and overcomes slightly the limit value corresponding to the initial mass loss rate. Then, accretion starts at 20.2 years making this density decrease. At further distances from the star, the disk decretes and density increases as in the build-up phase. Then once the accretion region gets large enough, these outer radii also have their density vanishing.}
A representative model is shown in Figure~\ref{dm05}. Here, a constant mass injection for 20 years is followed by an outburst with twice the mass injection rate lasting 0.2 year. After the outburst, the injection rate was set to zero. The surface density during the outburst reveals an important property of outburst-like scenarios: the outburst affects mostly the inner part of the disk, and how far out the surface density is affected depends on the length and strength of the outburst, as well as on the $\alpha$ parameter.

\vspace{0.5cm}

\section{Predictions of the dynamical properties of the  photometric observables}
\label{prediction}

In the previous section, we described the evolution of the disk structure for different dynamical scenarios. This part of this paper is devoted to {\sc hdust} predictions of the variability of photometric observables. A large variety of observables can be derived from the {\sc hdust} simulations. To structure the analysis, we choose to present in this paper only the most important photometric results. Polarimetric, spectroscopic and interferometric results will be analyzed in future publications. 

\begin{figure}[t]
% \vspace*{-2.0 cm}
\begin{center}
 \includegraphics[width=3.5in]{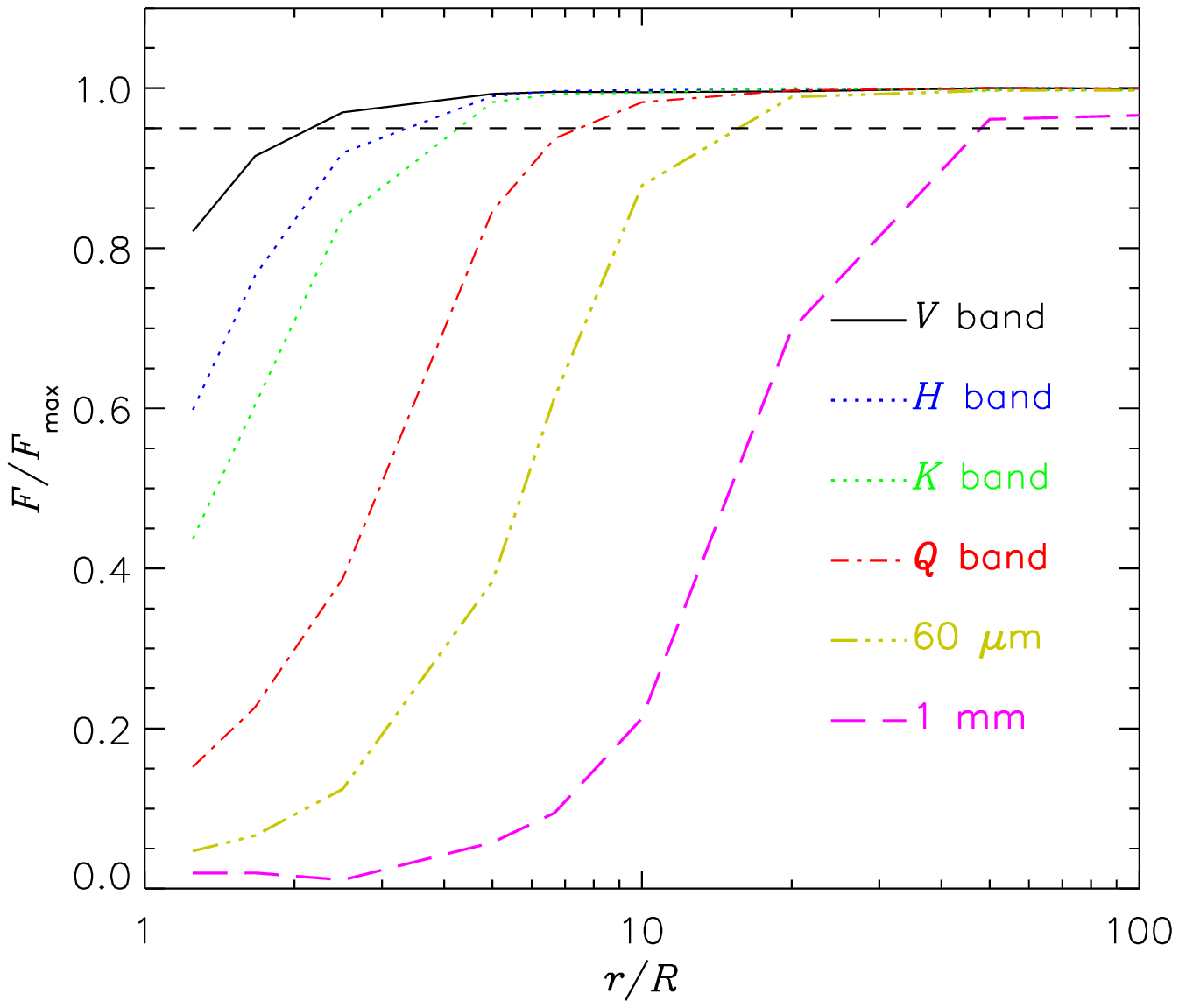} 
 %\vspace*{-1.0 cm}
 \caption{The formation loci of different observables in the continuum emission. The calculations assume the parameters listed in Tables~1 and 2.  The results were derived for an inclination angle of 30 \degr. Plotted is the ratio between the observed flux to the maximum flux $F_{\rm max}$, defined as the flux of a model with a disk outer radius of 1000 $R_\star$, as a function of the distance from the star.}
  \label{car10}
\end{center}
\end{figure}

Photometry is a powerful tool to use when observing variable stars because
fluxes at different wavelengths allow probing the disk at different radii
\citep[e.g., Figure~6 of][]{2006A&A...456.1027D}. As we want to follow the
outward growth and dissipation of the disk, it is critical to understand from
which part of the disk comes most of the emitted fluxes at different bands.
Figure~\ref{car10} shows the normalized radial flux distribution of a disk
around a Be star at various wavelengths. At a given band, the intersection of
the flux curve with the horizontal dashed line marks the position in the disk
whence about 95\% of the continuum excess comes. For instance, we see that the
$V$-band excess is formed very close to the star, within about 2 stellar
radii, whereas the excess at 1 mm originates from a much larger area of the
disk. To understand this plot, the disk can be regarded as a pseudo-photosphere around the star, whose effective size increases with wavelength. Note that these predictions for the formation loci are for an inclination angle $i=30\degr$. As we will see below, $i$ bears important effects on the emergent flux. Finally, Figure~\ref{car10} is in broad agreement with interferometric measurements of disk sizes for near-IR and mid-IR wavelengths \citep[e.g.][]{2007ApJ...654..527G,2009A&A...505..687M}. However, the extent of the $N$-band continuum emitting region seems systematically smaller than expected for example in \cite{2005A&A...435..275C} (see Figures 6 and 7), but in agreement with observations.

%\citep[e.g.][]{2009A&A...505..687M
%  ˆ vŽrifier}. 
%The sizes in the lines are however higher because the formation locus is located further away in the disk \citep{carciofi11}. This was also confirmed observationally \citep[e.g. for $\gamma$ Cas][]{2006AJ....131.2710T,2007ApJ...654..527G}.

Based on the previously described dynamical scenarios, we computed various photometric observables at different wavelengths with {\sc hdust} in order to study the mass redistribution process at different locations in the disk. In the following, we selected some results for the $V$-band, $K$-band and millimetric domains.

\subsection{$V$-band photometry}
\label{sec_Vband}
\begin{figure}[t]
% \vspace*{-2.0 cm}
\begin{center}
 \includegraphics[width=3.5in]{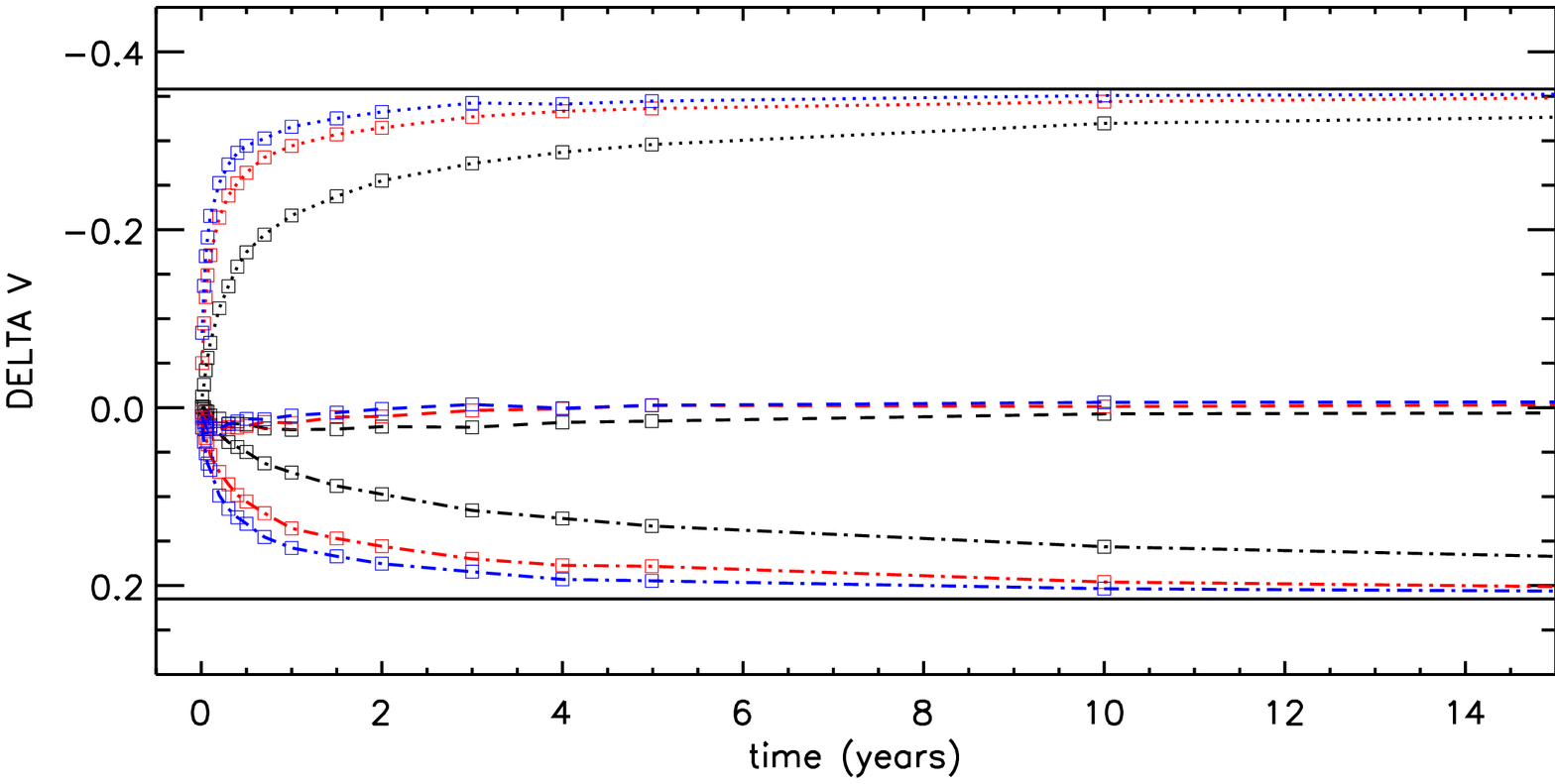}
 \includegraphics[width=3.5in]{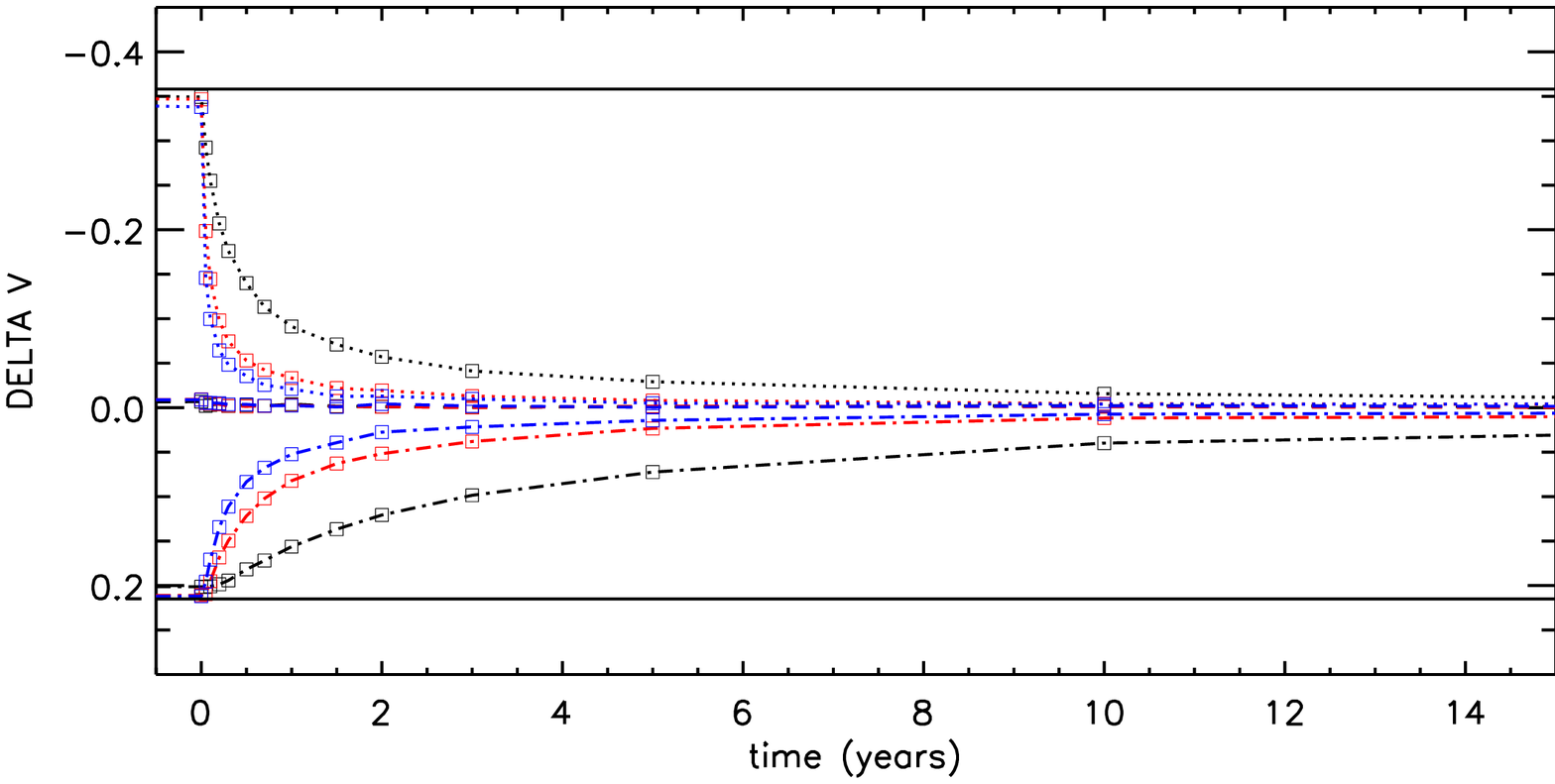} 
 \caption{$V$-band lightcurves associated with disk build-up (top panel) and dissipation (bottom panel). The dotted, dashed  and dot-dashed lines represent the lightcurves for inclination angles of 0\degr (face-on), 70\degr and 90\degr (edge-on), respectively. The black, red and blue colors represent models for $\alpha$=0.1, 0.5, and 1.0, respectively. The solid black lines indicate the asymptotic value of $\Delta V$.}
   \label{dV}
\end{center}
\end{figure}

The $V$-band magnitude in the disk is controlled by two processes: gas emission and absorption of the stellar radiation. From Figure~\ref{car10}, we expect that only the densest inner part of the disk (within $r \lesssim 2\,R_\star$) will affect the flux at this wavelength. Also, these two mechanisms will have a different impact on the observables depending on the inclination angle.  At low inclination angles, the disk is seen face-on, and little absorption of the stellar radiation is expected. At high inclination angles, the disk is seen edge-on, and absorption plays a stronger role.
Figure~\ref{dV} shows examples of $V$-band lightcurves computed for the dynamical scenarios described in \S~\ref{dissip} and \S~\ref{growth} for three different inclination angles. We plot $\Delta V$, the difference between the total flux and the photospheric flux in the $V$ band.

%The $\Delta$ $V$ will thus follow an opposite evolution of the mass injection rate: as the disks vanishes, less absorption occurs and the $V$ flux increases (V magnitude decreases).

% Mettre diff valeurs de alpha pour le premier graphe ou le deuxieme ?? Revoir commentaire en accord avec ca.

\begin{figure}[t]
% \vspace*{-2.0 cm}
\begin{center}
 \includegraphics[width=3.5in]{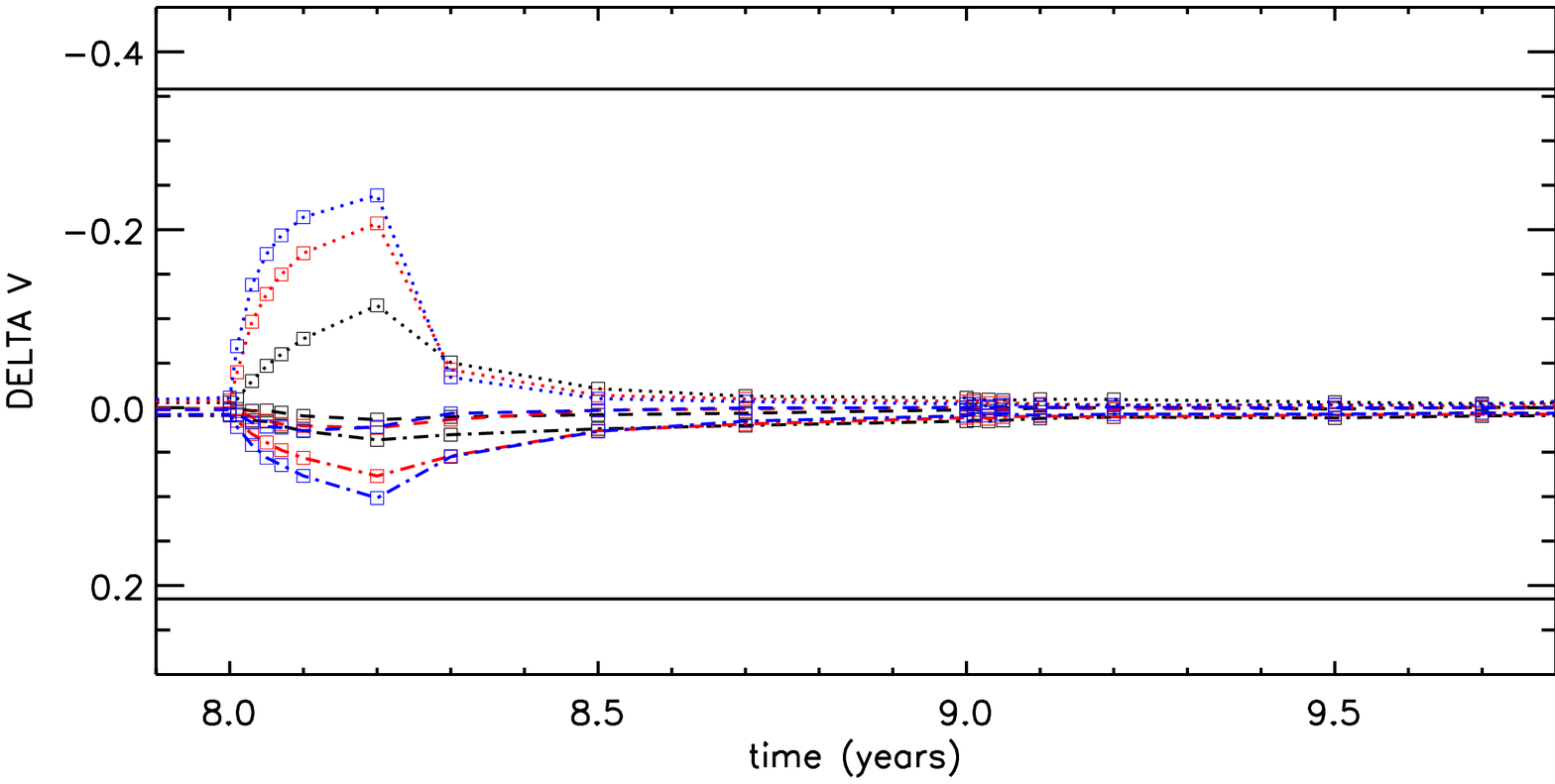}
 \includegraphics[width=3.5in]{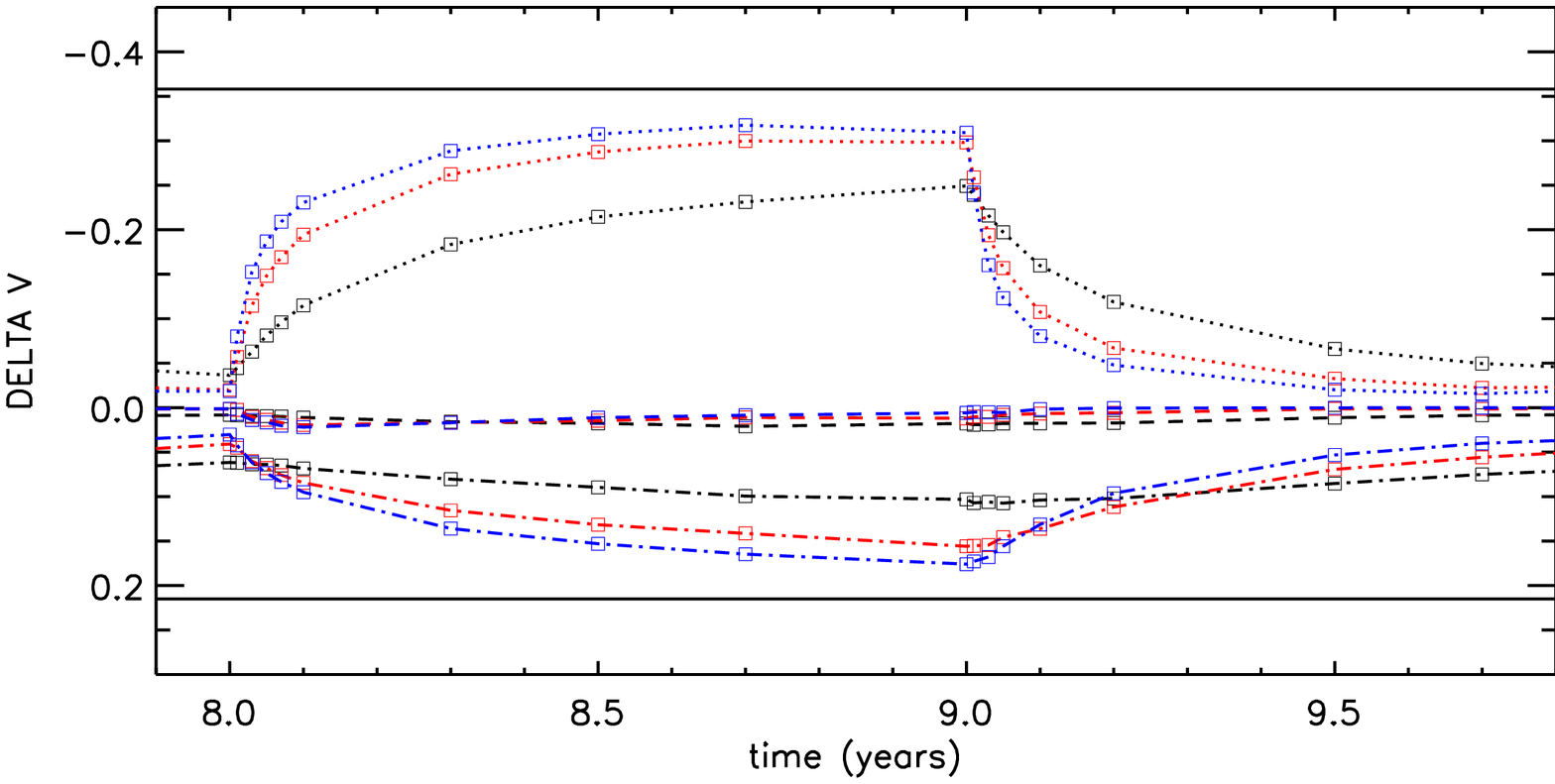} 
 \includegraphics[width=3.5in]{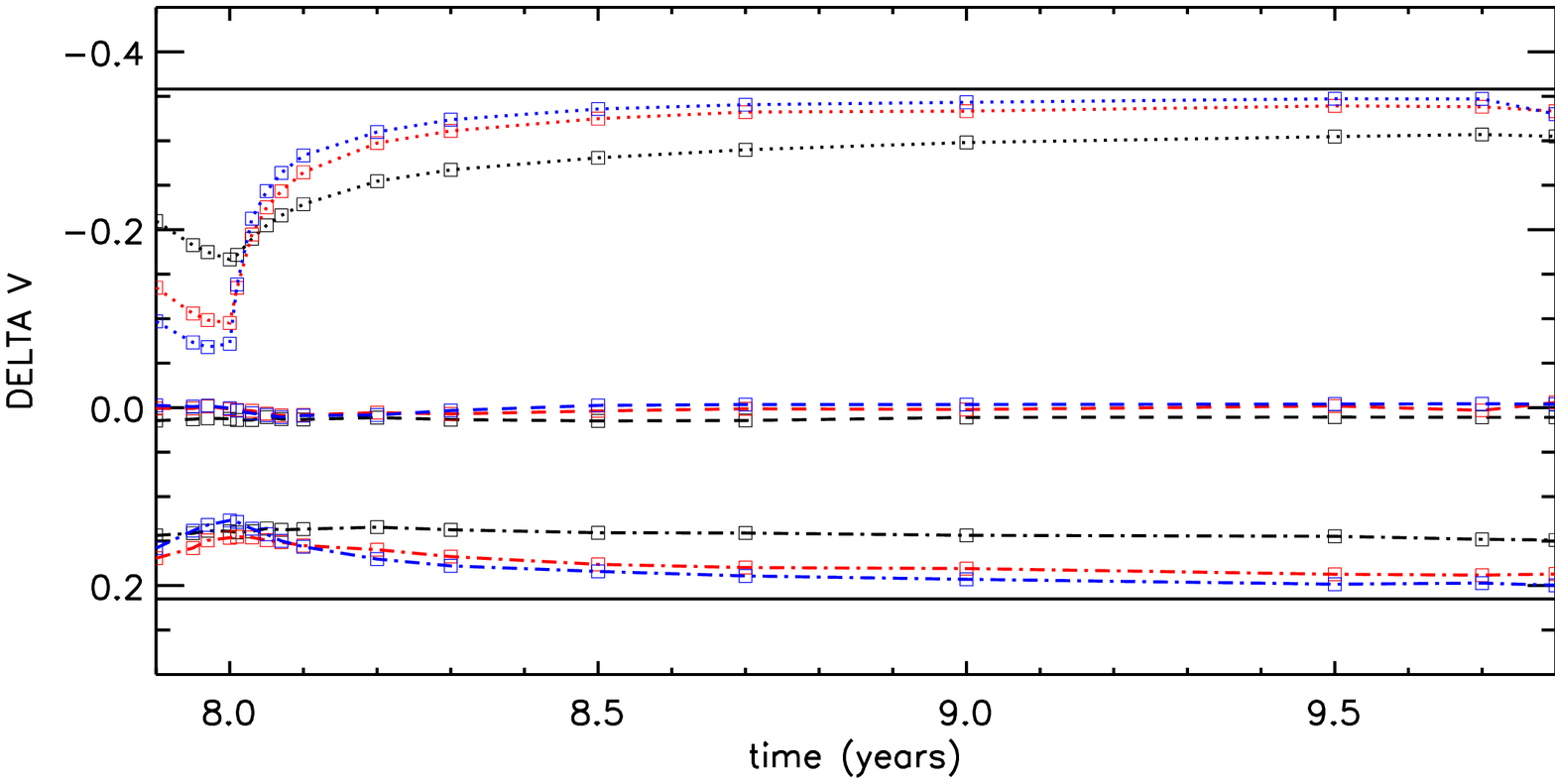} 
 \caption{Same as Figure~\ref{dV} for the periodic scenarios of Figure~\ref{scenarios} (top: DC=10\%, middle: DC=50\%, bottom: DC=90\%).}
   \label{dV2}
\end{center}
\end{figure}

For the build-up case, we can check the above statements on the inclination angle. For $i=0\degr$, for instance, the disk develops a strong $V$ excess of -0.4 magnitude. For $i=90\degr$, the disk causes a flux decrement of about 0.2 magnitude, as a result of absorption of photospheric light.
A balance between emission and absorption is met for an inclination angle of about 70\degr. This fact is wavelength-dependent as we will see that the balance is met at different values of $i$ for other spectral bands.

For a given $i$, the timescale for the $V$ evolution depends strongly on $\alpha$. As seen in \S\ref{refcase}, the higher the $\alpha$, the faster the disk will reach the limit density, so the faster the $V$-band variations will be. We should state here that the asymptotic values reached by $\Delta V$ depend on the density in the inner part of the disk and thus on the particular choice of $\Sigma_{0}$. Table~\ref{QS_Vband} summarizes how long  $\Delta V$  takes to reach 95\% of its asymptotic value depending on $\alpha$ and the inclination angle of the disk. As $V$ is mostly affected by the inner disk, those values are not sensitive to different disk sizes as long as $R_{d} \gtrsim 2\,R_\star$.

Figure~\ref{dV} illustrates another important prediction of the VDDM: the timescales for disk growth are shorter than the timescales for disk dissipation.  
This is due to the fact that while at disk growth, the timescales are set by the matter redistribution within a couple of stellar radii only. At disk dissipation, the timescales are controlled by the reaccretion from a much larger area of the disk.
This prediction is confirmed by observations \citep{car11b}.

%For the reference case scenario (disk build-up with a constant mass injection rate), we can check the above statements on the inclination angle and note that a balance of emission/absorption of $V$-band light is met for an inclination angle of about 70 \degr. This fact is wavelength-dependent, we will see it is different for other spectral bands. The evolution of the $V$-band magnitude depends strongly on the inclination angle and also on $\alpha$ values. As stated in \S\ref{refcase}, the higher the $\alpha$ value, the faster the disk will reach the limit density and get denser, so the faster the variations in $V$-band will be. We should state here that the asymptotic values that the $V$-band magnitude takes depend on the density in the inner part of the disk and thus on the particular choice of $\Sigma_{0}$. Table~\ref{QS_Vband} summarizes how long the $V$ magnitude takes to reach 95\% of its limit value depending on $\alpha$ and the inclination angle of the disk. As the $V$-band magnitude is mainly caused in the inner disk, those values are not sensitive to different disk sizes as long as  $R_{d} \geq 2 R_\star$. %Exploring different disk maximum sizes, we find it plays a secondary role and changes the  $\Delta$V values only by a 0.01 mag at high inclination angle because of a higher absorption.

  \begin{deluxetable}{ccccccc}
\tablecolumns{7}
\tablecaption{Time in years required for $V$ to reach 95\% of its asymptotic value.\label{QS_Vband}}
\tablehead{\colhead{$\alpha$ / Angle (deg)}  & \colhead{0} & \colhead{30} & \colhead{70} & \colhead{80} & \colhead{85} & \colhead{90}}
\startdata
0.1 & 27.8 & 29.3 & $<$ 0.1 & 26.0 & 71.0 &90.0 \\
0.3 & 11.4  & 12.7  & $<$ 0.1  & 9.7 & 23.6  & 24.9\\  
0.5 &  7.1 &  7.8  & $<$ 0.1 & 7.9 & 12.0 & 14.0 \\
0.7 & 4.1 & 4.3 & $<$ 0.1 & 4.4 & 9.7 & 11.4 \\  
1.0 & 2.7  & 2.8  &  $<$ 0.1  & 3.4 & 7.1 & 8.1 \\
\enddata
\end{deluxetable}

  \begin{deluxetable}{ccccccc}
\tablecolumns{7}
\tablecaption{Time in years required for $K$ to reach 95\% of its asymptotic value.\label{QS_Kband}.}
\tablehead{\colhead{$\alpha$ / Angle (deg)}  & \colhead{0} & \colhead{30} & \colhead{70} & \colhead{80} & \colhead{85} & \colhead{90}}
\startdata
0.1 & 26.3 & 28.5 & 92. &  $>$ 100. &  $>$ 100. &  $<$ 0.1   \\    %0.5
0.3 & 9.8 & 11.2 & 24.2 & 33.5 & 95.0 &  $<$ 0.1 \\       % $>$ 100.
0.5 & 6.0 & 6.7  & 16.6 & 31.3 & 46.6 &  $<$ 0.1  \\		%  8.4
0.7 & 4.0 & 4.4 & 9.9 & 20.2 & 27.6 &  $<$ 0.1 \\	 	% 0.15
1.0 & 2.9  & 3.3 & 8.4 & 12.1 &19.7  &  $<$ 0.1 \\		%  0.3 
\enddata
\end{deluxetable}

  \begin{deluxetable}{ccccccc}
\tablecolumns{7}
\tablecaption{Time in years required for the $mm$ magnitude to reach 95\% of its asymptotic value..\label{QS_subband}}
\tablehead{\colhead{$\alpha$ / Angle (deg)}  & \colhead{0} & \colhead{30} & \colhead{70} & \colhead{80} & \colhead{85} & \colhead{90}}
\startdata
0.1 & 26.2 & 26.7 & 28.4 & 33.4 & 38.3 & 40.1 \\
0.3 & 9.4 & 9.3  & 9.7 & 12.9 & 15.4 & 16.6 \\
0.5 & 5.8 &  4.6 & 6.4  & 7.4 & 8.1 & 8.4\\
0.7 & 3.9 & 2.6 & 4.2 & 5.3 & 6.2 & 6.5 \\
1.0 & 2.5  & 2.7  & 2.9  &  3.5 & 3.8  & 4.0 \\
\enddata
\end{deluxetable}

Figure~\ref{dV2} shows the predictions for the scenarios with periodic mass injections. 
Contrary to the case of a steady disk growth in which the inner disk forms without perturbation and steadily reaches the limiting density, the inner disk has not the time to fully develop in the case of periodic scenarios. This results in lower amplitudes for the variations in $\Delta V$. Clearly, this effect is more pronounced from smaller DCs. 

The diverse morphology of the $V$-band lightcurves shows that they are quite specific of the mass decretion scenario, in this case the DC and cycle length. They are also quite dependent on the $\alpha$ parameter. 
Therefore, follow-up observations of the $V$-band magnitude represent a powerful tool to infer these parameters. Its diagnostic potential was recently demonstrated by \citet{car11b} who used the lightcurve of the Be star 28\,CMa to measure, for the first time, the viscosity parameter in a Be star. 
It is important to keep in mind, however, that $V$-band variations are rather insensitive to the outer disk parts. 

If  lightcurves are of great importance to study Be stars, color-color and color-magnitude diagrams can also bring fundamental elements to the analysis because they are characteristic of different radiative transfer effects occurring in the disk (absorption, emission, etc) at different radii. The disk temperature is typically 60\% of  $T_{\rm eff} $ \citep{car06}, so the spectral shape of the disk emission is redder than the photospheric spectrum. Therefore, colors are a good tracer of the amount of stellar radiation that is reprocessed towards longer wavelengths by the disk. 

\begin{figure*}[t]
% \vspace*{-2.0 cm}
\begin{center}
 \includegraphics[width=2.0in]{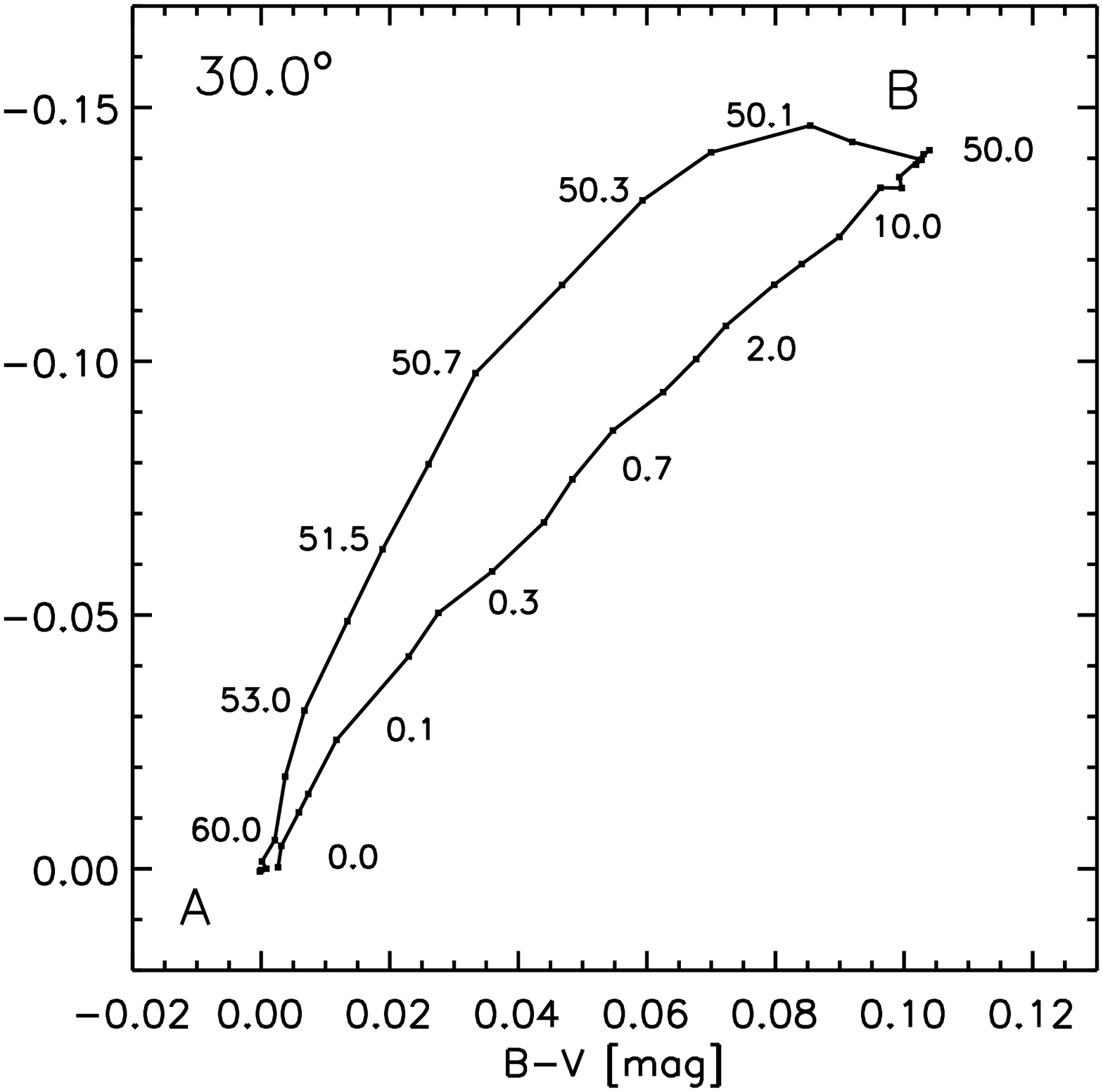} 
\hspace*{1cm}
   \includegraphics[width=2.0in]{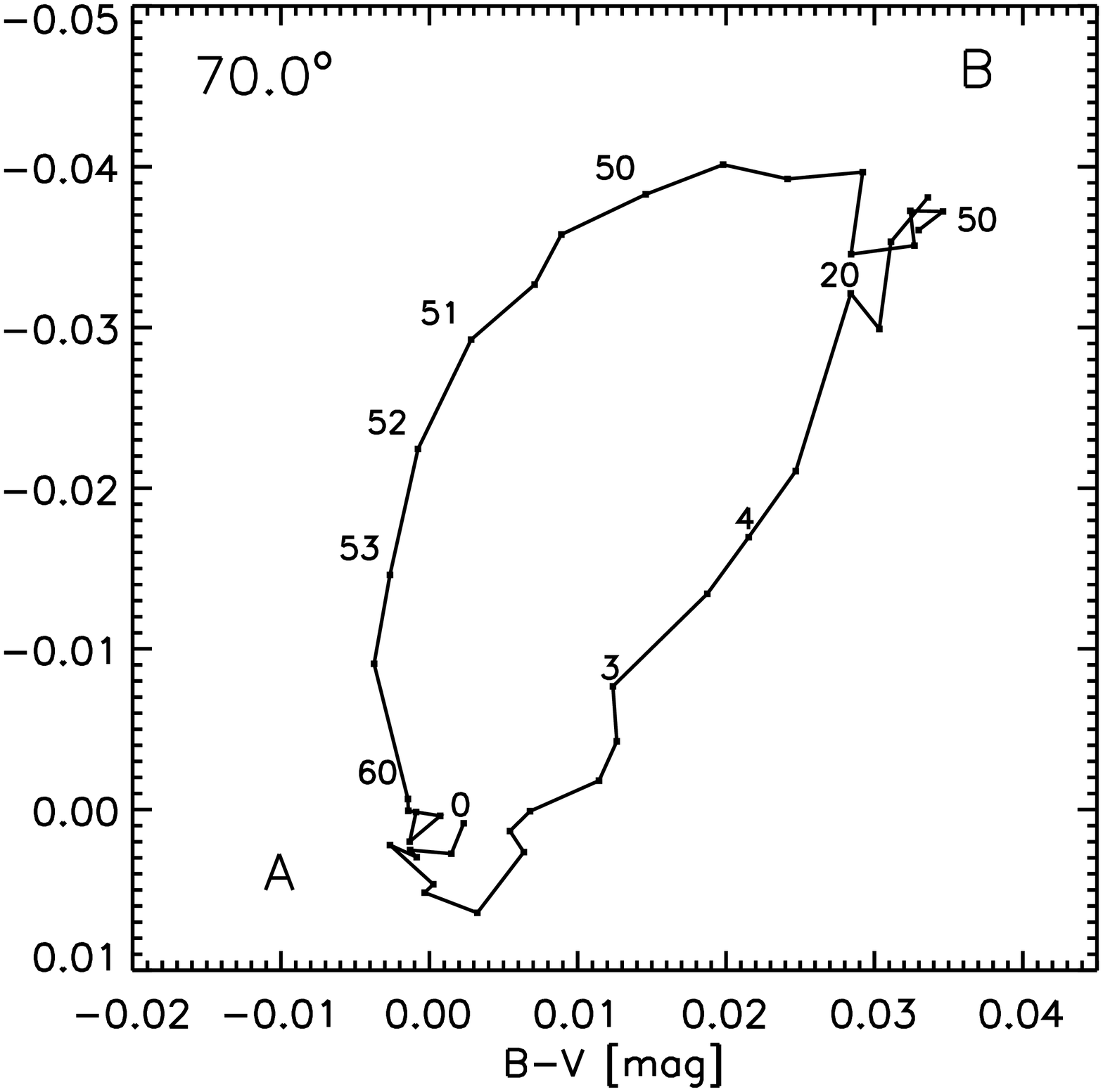} 
\hspace*{1cm}
      \includegraphics[width=2.0in]{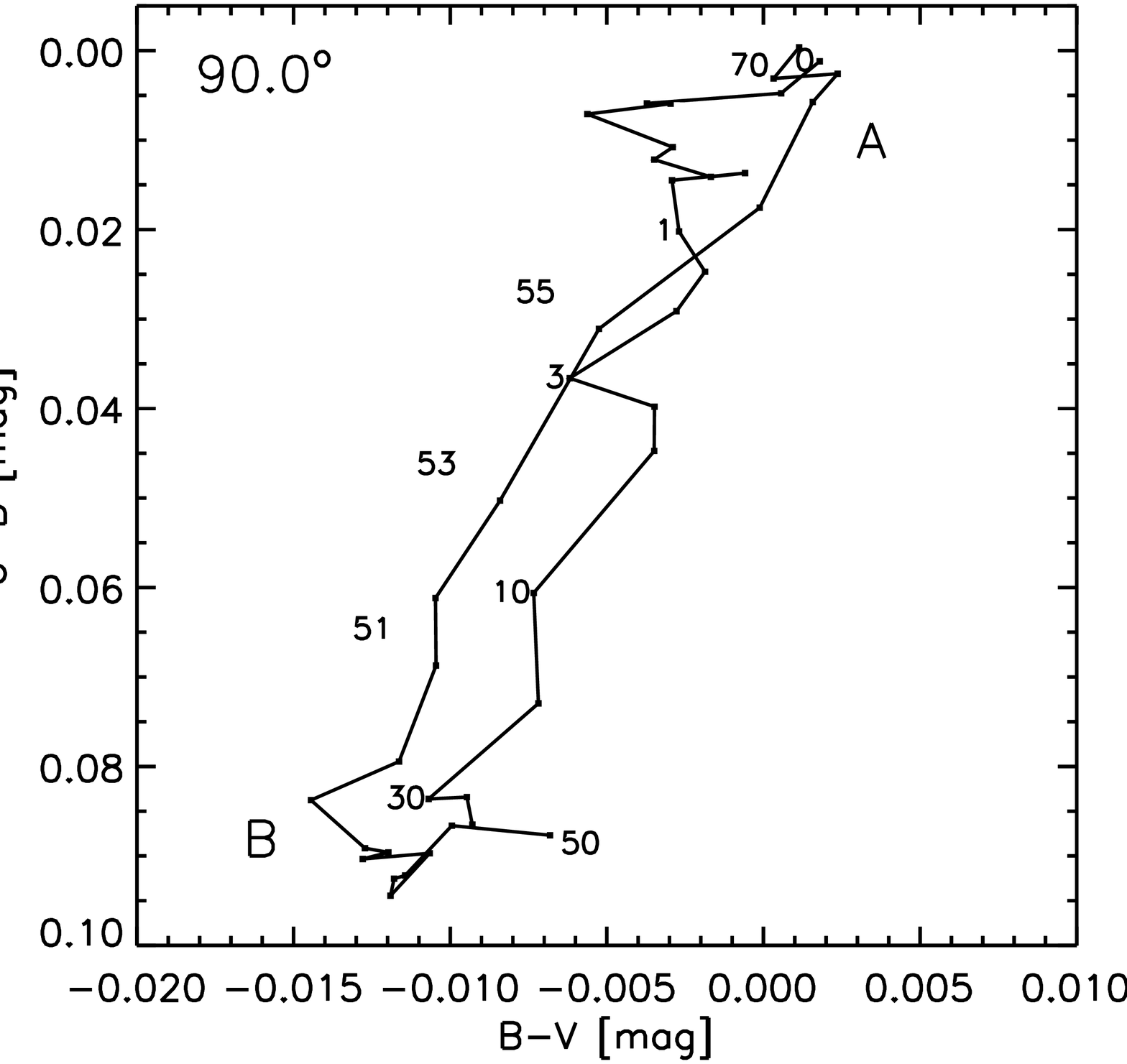} 
       \includegraphics[width=2.in]{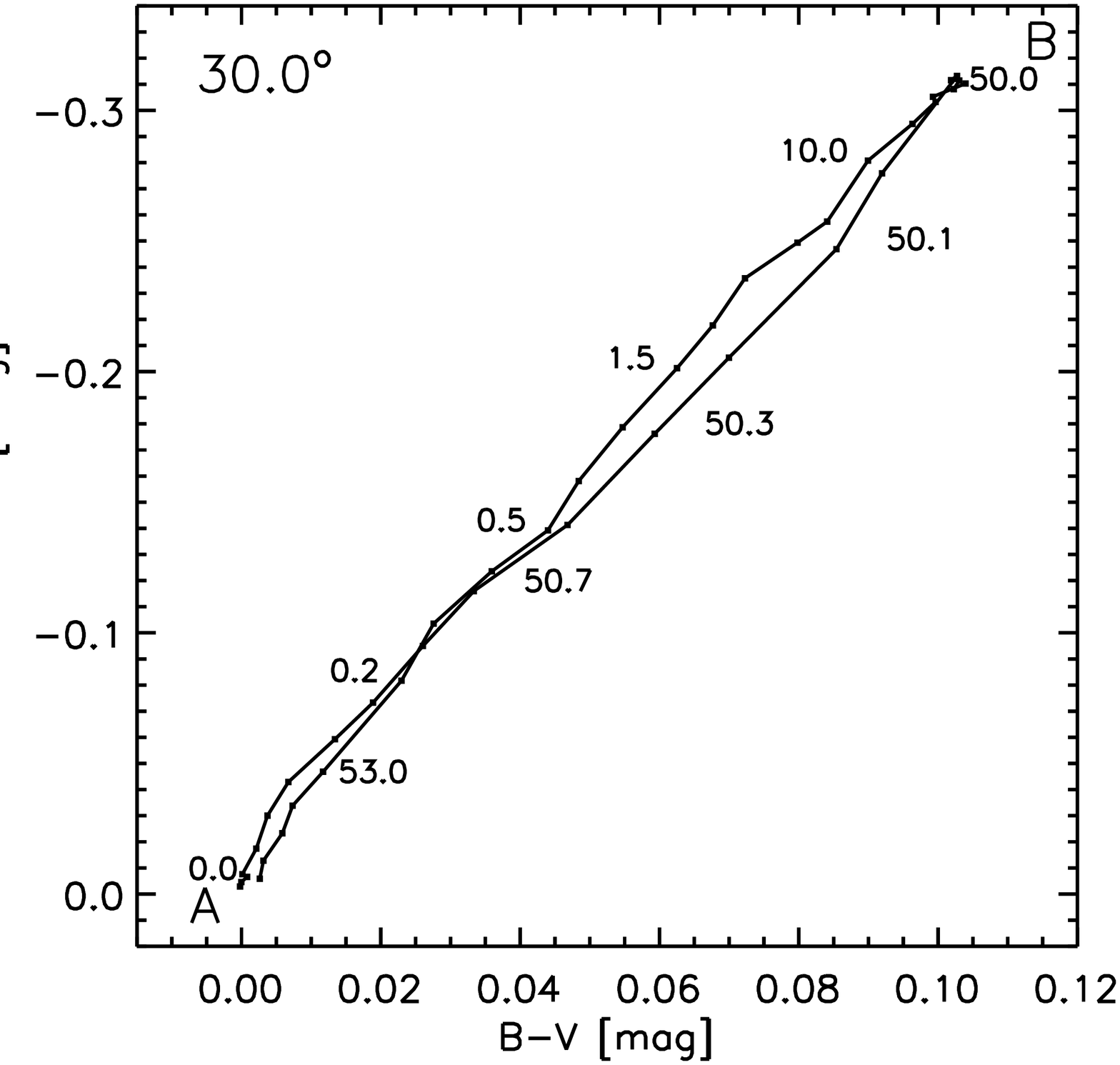} 
\hspace*{1cm} 
\includegraphics[width=2.in]{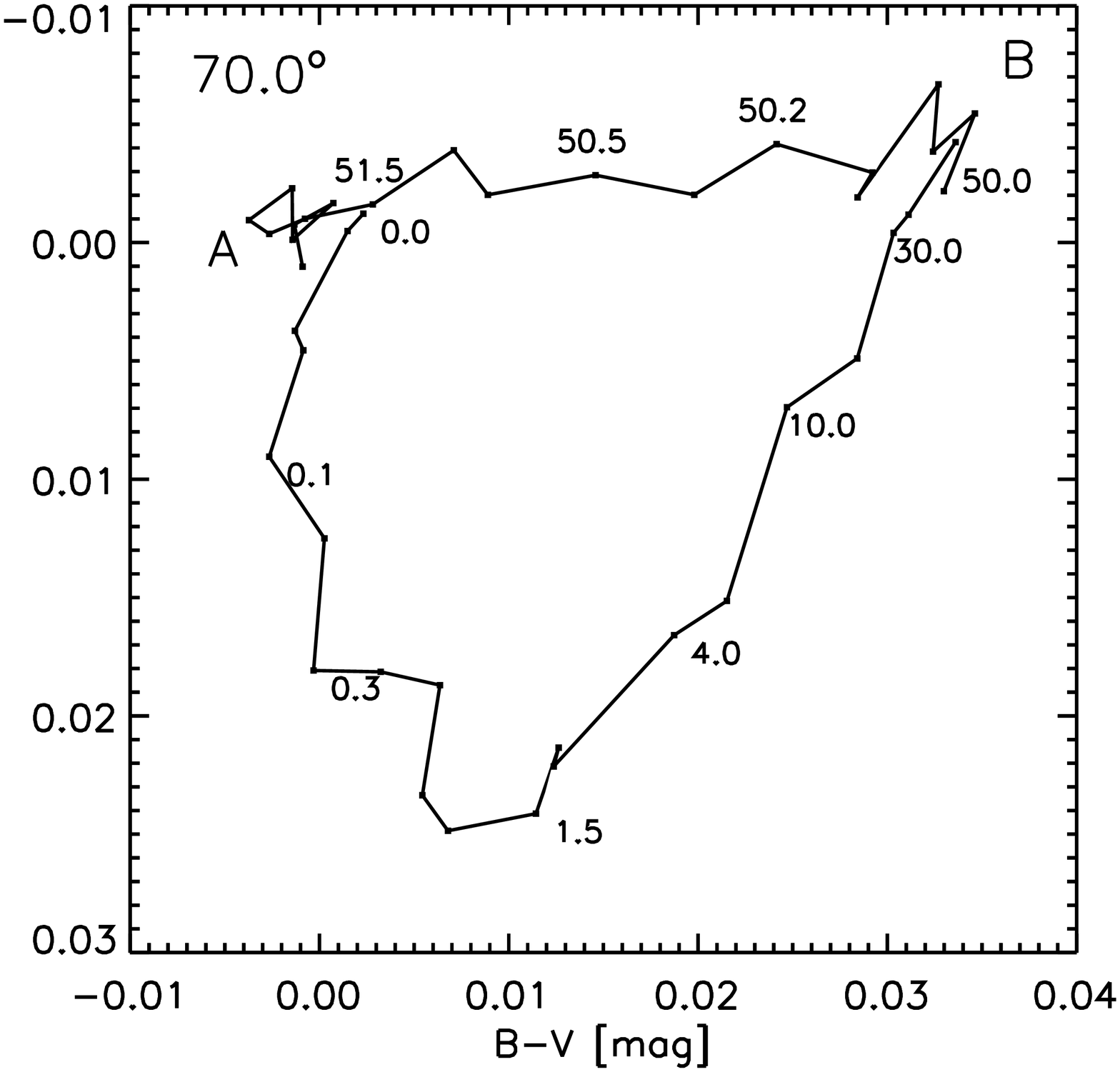} 
\hspace*{1cm}
\includegraphics[width=2.in]{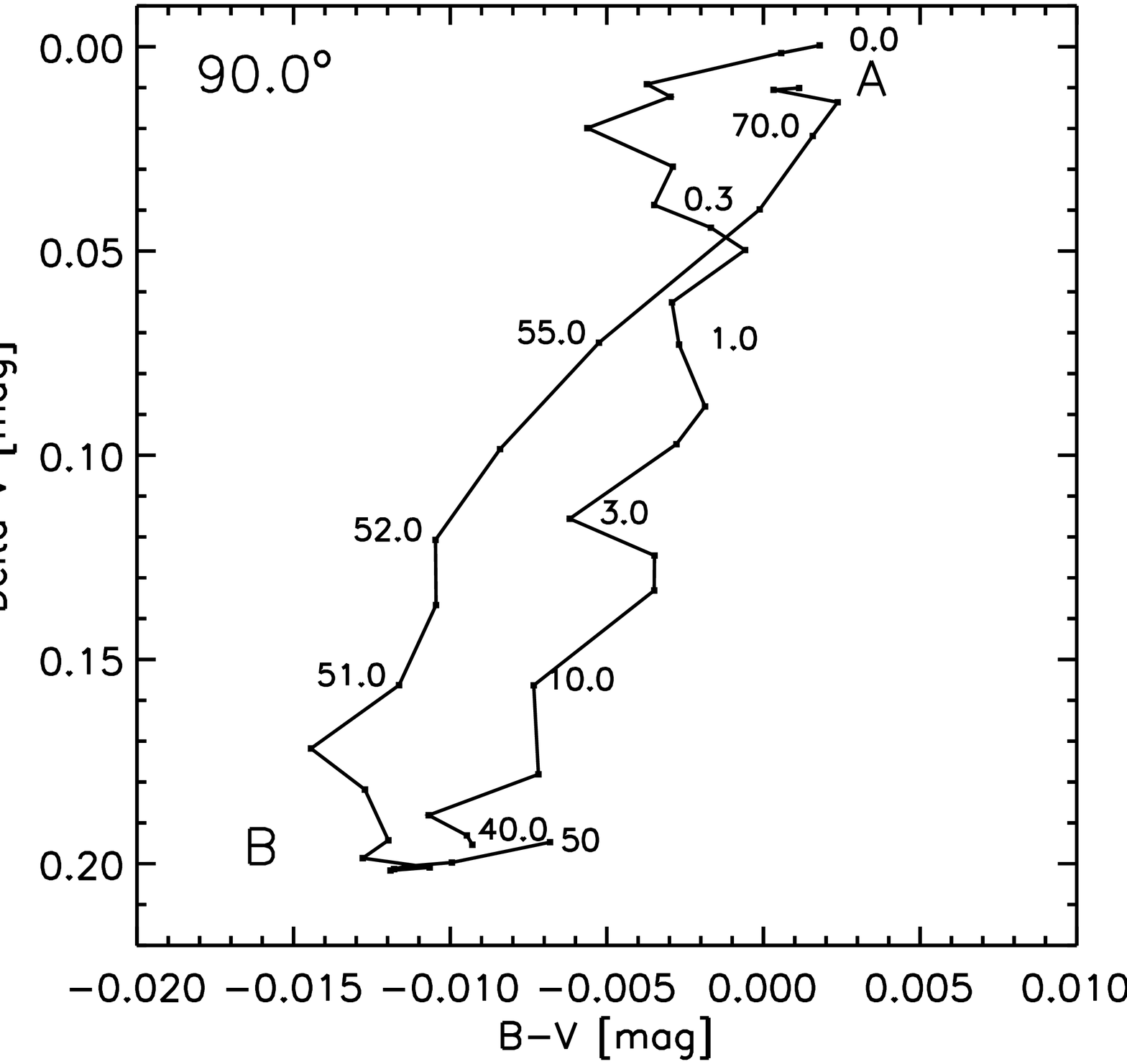} 
      
 \caption{Top panels : $\ub/ \bv$ diagrams for a 50 year long disk build-up followed by a 50 year long dissipation. Each panel shows the results for a different inclination angle. Epochs are indicated in years. The path A to B represents the build up phase and the B to A the dissipation phase. Bottom panels : corresponding $\Delta V/ \bv$ diagrams.}
%mod03dm03a0.70
   \label{harman1}
\end{center}
\end{figure*}

In Figure~\ref{harman1} we show examples of $\ub/\bv$ color-color and $\bv/V$ color-magnitude diagrams, illustrative of disk growth and dissipation.  
Starting from zone A (no-disk and a small color decrement) the system follows a reddening path towards an asymptotic value (located in zone B). When mass loss is turned off, the system follows a blueing path back to A that is \emph{different than the previous path}, therefore forming a loop in the color-color diagram. The different paths from A to B and from B to A come from the fact, discussed above, that the disk grows at a faster rate than it dissipates.

% Colors indicate 3 different $\alpha$ values (black: 0.1, blue: 0.5  and red: 1.0).

%All the diagrams shown on Figure~\ref{harman1} exhibit some loop-shaped curves. Starting from the zone A (no-disk), the color-color diagram follows a path to reach an asymptotic value when the disks is fully built (zone B) and then decays following another bluer path to close the loop when the disk has almost or entirely disappeared (zone A).

%\rouge{Interpretations of these diagrams TBDiscussed.  why is B increasing when the disk builds up ? Shouldn t it be $V$ instead ?obs and theory in $\bv$ seems opposite to each other for shell stars is really something that need a discussion.}

The building up and decay phases (and by extension periodic injection rates) show clear loop-like signatures in the color-color $\ub/\bv$ and color-magnitude diagrams. As the morphology of those loops depends strongly on the inclination angle, those diagrams represent a tool to estimate this parameter on top of the decretion history at play.
Photon noise is particularly visible on the 90 degree figures since the photometric signal is much lower than for the other inclination angles where absorption plays a lesser role.

\subsection{$K$-band photometry}
\label{kphot}

The excess flux in the IR comes mostly from a part located up to several stellar radii from the star ($\sim6\,R_\star$ in $K$-band, see Figure~\ref{car10}). This is also an interesting region because this is where we observe the highest density variations for a periodic decretion scenario (\S~\ref{period}), so we expect clear signatures in the $K$-band lightcurves. We mention here that a variability of a different sort in the disk structure is to be expected due to thermal effects that are not simulated in this work. Those effects have been studied in \cite{2008ApJ...684.1374C} and \cite{2008MNRAS.386.1922J}. Changes in density can be important in the inner disk \citep[e.g. see Figure 4 in][]{2008ApJ...684.1374C}. Their consequences upon short-wavelength observables remain to be investigated and constitute an important perspective for this work.

\begin{figure}[t!]
% \vspace*{-2.0 cm}
\begin{center}
 \includegraphics[width=3.5in]{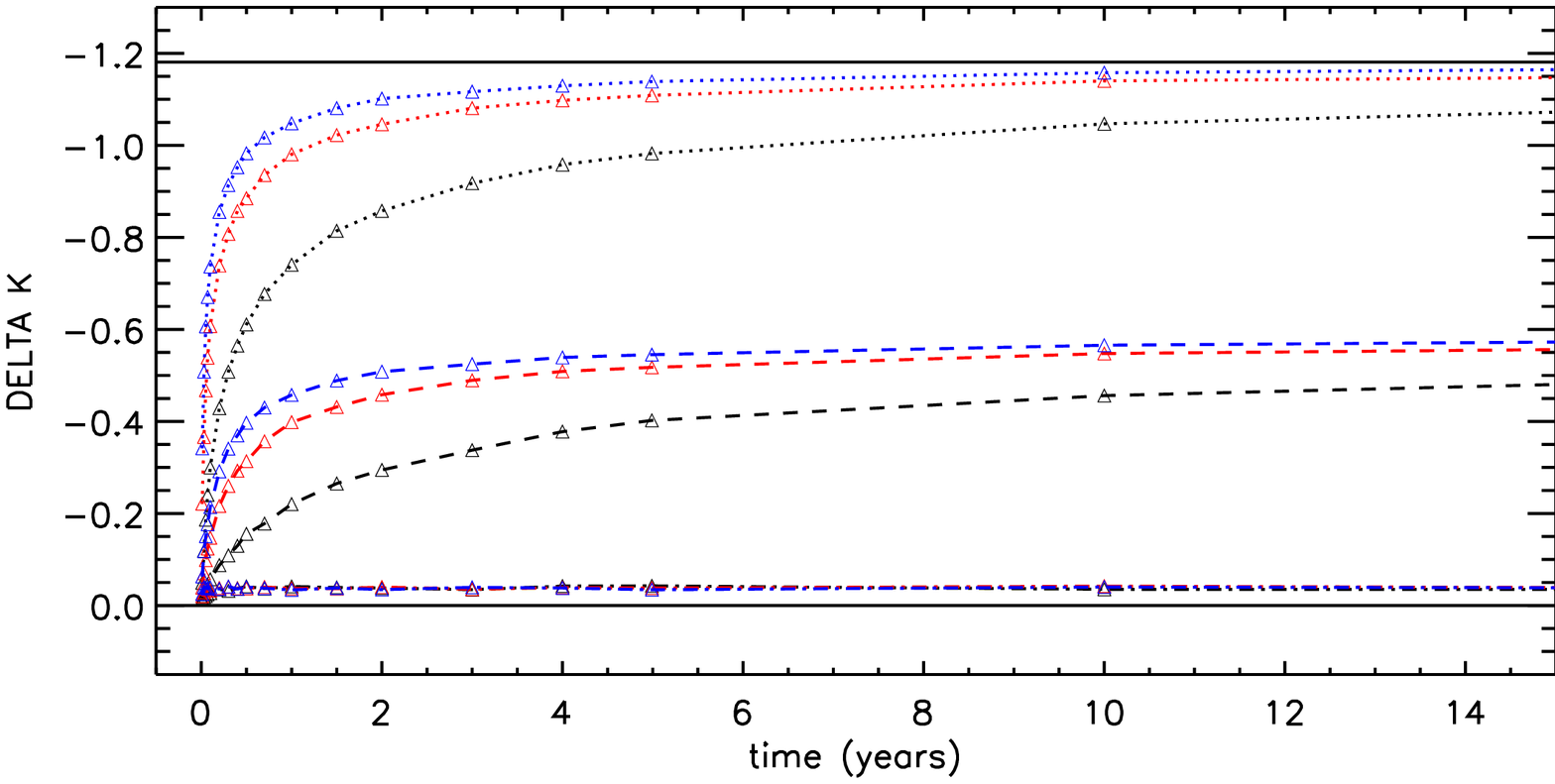} 
 \includegraphics[width=3.5in]{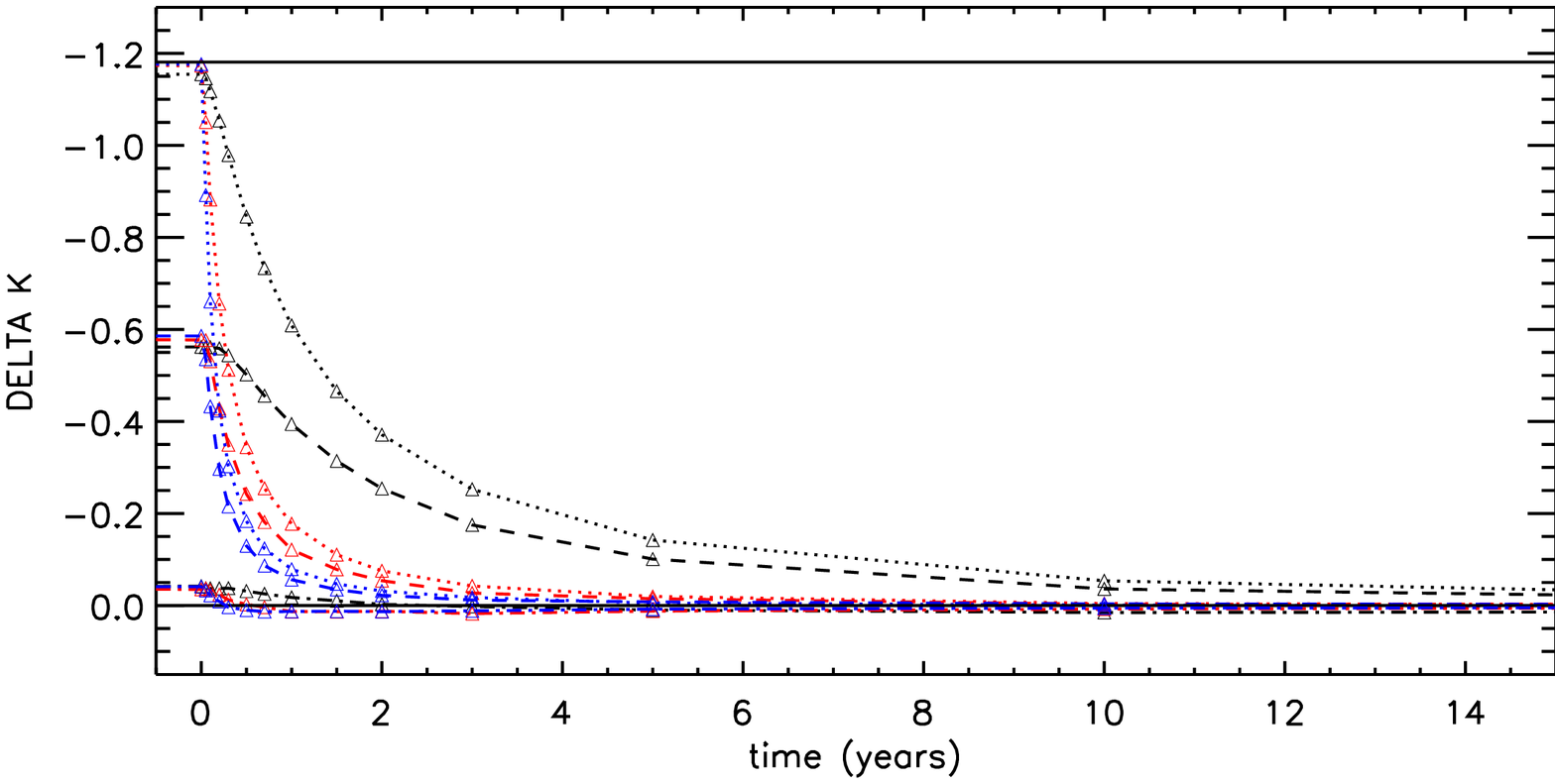}  
 \includegraphics[width=3.5in]{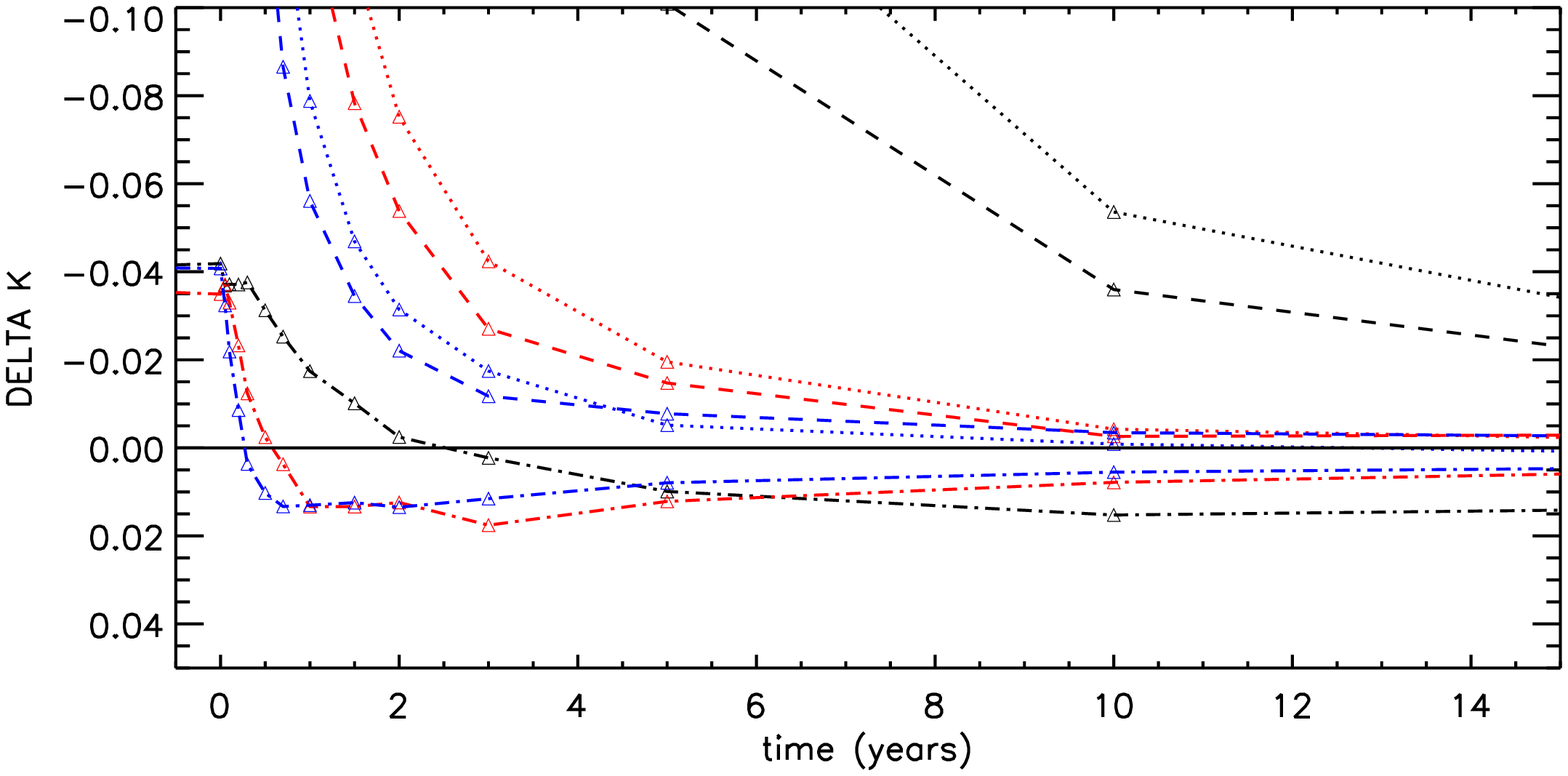}  
 \caption{Upper two panels: same as Figure~\ref{dV}, but for the $K$ band. Lower panel: a zoom into the dissipation curve allows to see that $\Delta K$ becomes positive for i=90\degr.}
   \label{dK}  
\end{center}
\end{figure}

\begin{figure}[t!]
% \vspace*{-2.0 cm}
\begin{center}
 \includegraphics[width=3.5in]{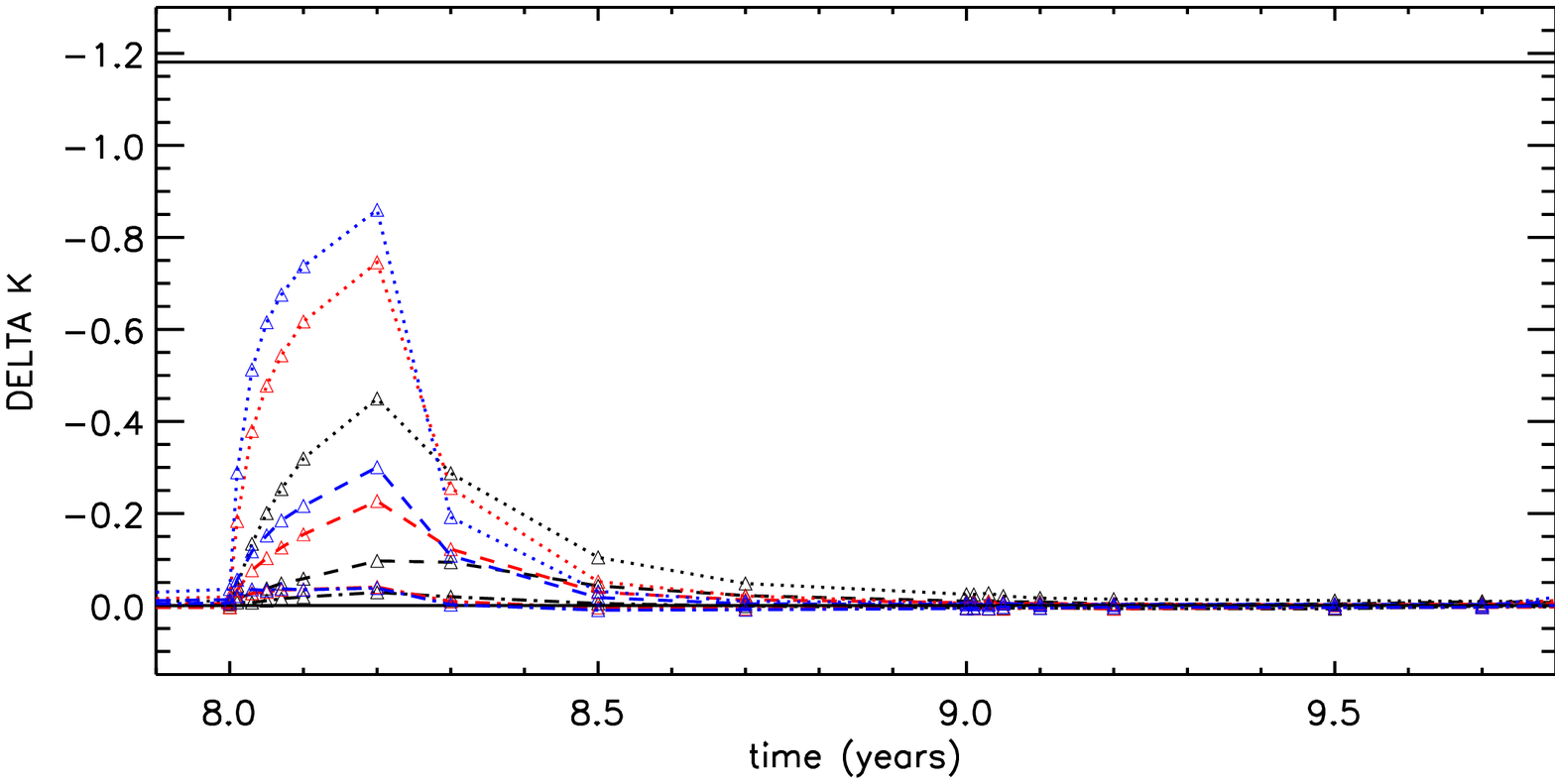} 
 \includegraphics[width=3.5in]{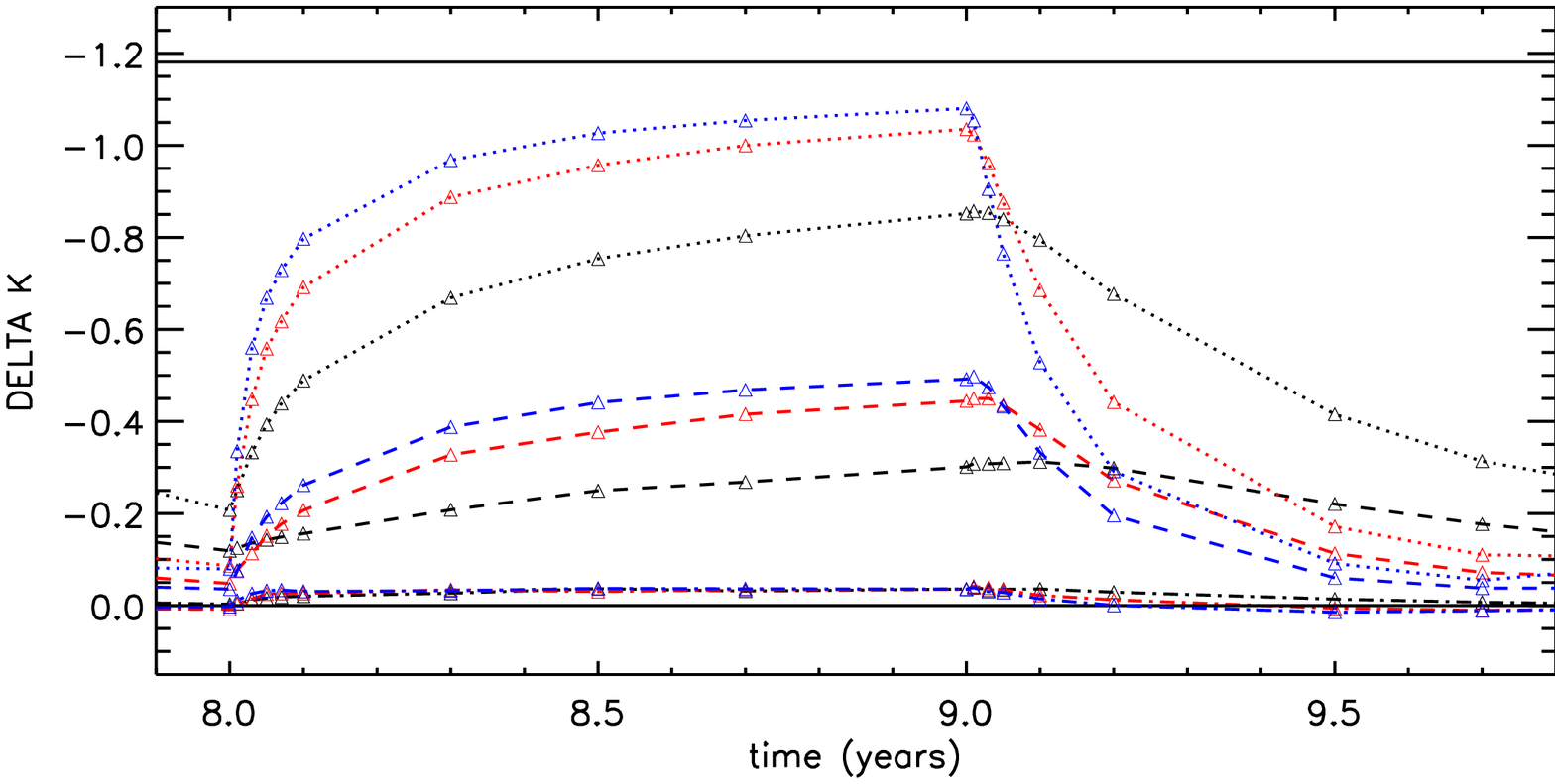}
 \includegraphics[width=3.5in]{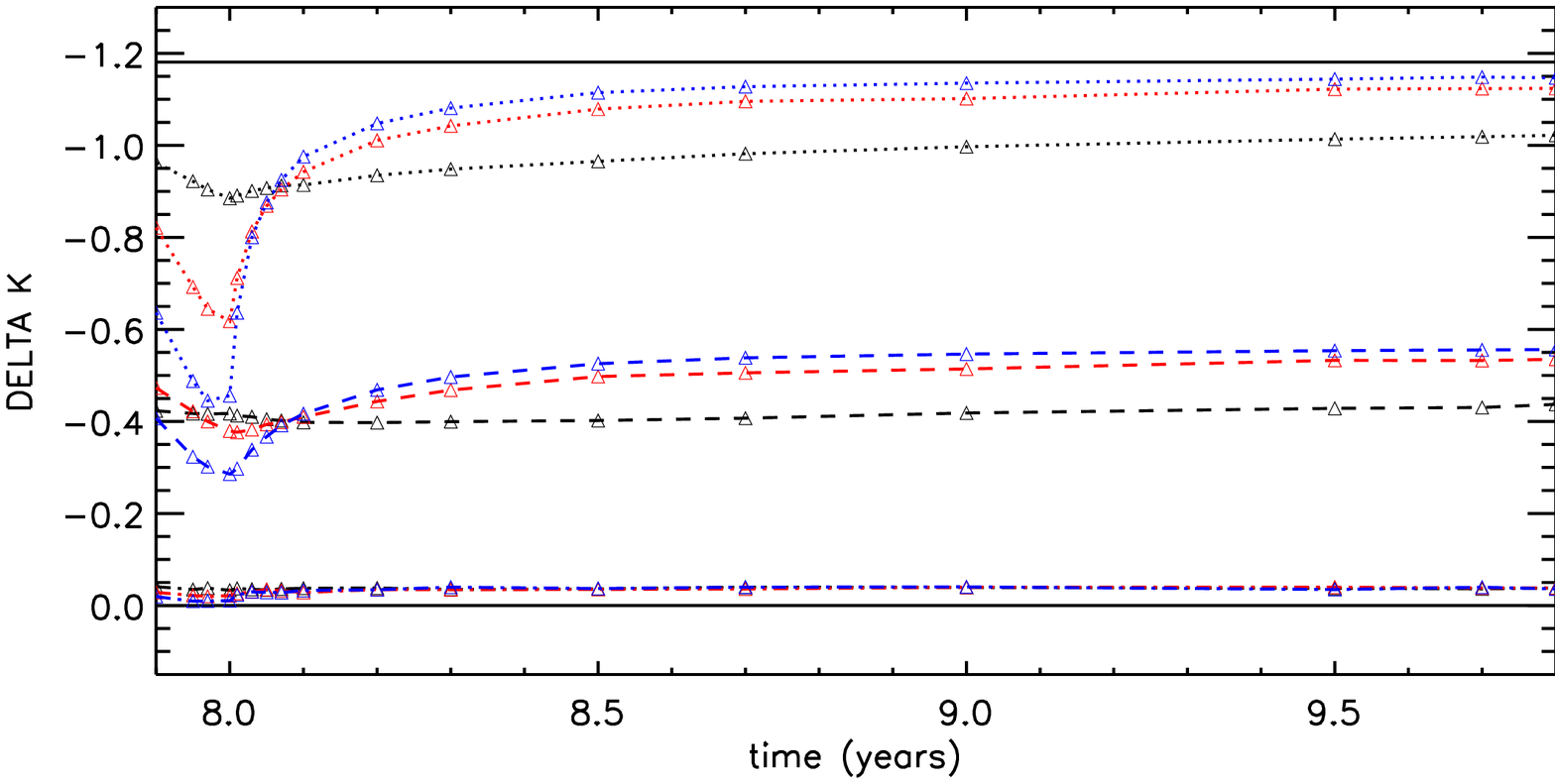}  
 \caption{$K$-band lightcurve for the periodic scenario. Legend is the same as for Figure~\ref{dV}.}
   \label{dK2}
\end{center}
\end{figure}

Figure~\ref{dK} shows the influence of the inclination angle and the $\alpha$ parameter on the $K$-band flux for the build-up and dissipation phases. Examples of lightcurves for the periodic scenarios are shown in Figure~\ref{dK2}.
While the behavior is similar to the $V$-band case, there are three important differences:
\begin{enumerate}
\item The asymptotic value of the excess reached for $i=0\degr$ is much larger than for the $V$-band. This is simply a result of the fact that the size of the pseudo-photosphere grows with wavelength.
\item Both the growth and decline rates of the $K$-band lightcurves are slower than in the $V$-band.
\item  The inclination for which emission and absorption balance each other is close to 90\degr. This means that, for systems seen at that such an inclination, even if the disk is experiencing long growth and dissipation phases, they will be barely detectable in this band.
\end{enumerate}

%Like for the $V$-band, the behaviour of the K band can be described in terms of a competition between absorption and re-emission. On the top panel of Figure~\ref{dK}, the build-up phase is very dependent of the inclination angle. While the $K$-band excess can exceed one magnitude for low inclination angles, we find this value to be lower than 0.05 mag at 90 \degr. This can be explained by the high absorption of the $K$-band-emitting region from the outer radii that get filled up with decreted mass in the equatorial plane. For lower angles, no or little absorption is present along the line of sight, so the $K$-band excess grows with the density in the first 7 radii. The influence of $\alpha$ is the same as described for the $V$-band lightcurve.

% During the dissipation phase (middle panel of Figure~\ref{dK}), the same processes invoked in the build-up phase take place but in a reversed way. The $K$-band emission decreases with the density at low inclination angles with an $\alpha$ dependence. For 90 \degr, an interesting phenomenon happens (lower panel of Figure~\ref{dK}): as the inner disk empties, the $K$-band emission from the disk decreases and at the same time, the outer parts still absorb the radiation both from the star and from the inner disk. Consequently, the $K$-band excess become positive until the absorbing part of the outer disk totally disappears and allow to see the full stellar $K$-band radiation.

From the above, we see that the combination of $V$ and $K$ bands has a strong diagnostic potential. For instance, it probably allows for constraints to be put on the inclination angle. More importantly, the fact that each lightcurve is a diagnostic of the mass redistribution timescale at different radii, combining $V$ and $K$ lightcurves opens the intriguing possibility of detecting possible variations of the $\alpha$ parameter with distance from the star. For example, a mismatch between the growth and/or decline rates of the model and observed lightcurve could be indicative of a radial variation of $\alpha$.

In the dissipation lightcurves, an interesting phenomenon happens for $i=90\degr$  (lower panel of Figure~\ref{dK}): as the inner disk empties, the $K$-band emission of the disk decreases and, at the same time, the outer parts still absorb radiation both from the star and from the inner disk. Consequently, $\Delta K$ becomes positive and then slowly goes to zero as the outer disk disappears.

%$K$-band lightcurves for some periodic scenarios are shown in Figure~\ref{dK2}. For 0 and 70 \degr of inclination angle, the variations in K-magnitude are highly visible with a peak-to-peak value up to about 1 mag. However, if the disk is seen edge-on, where the absorption is maximum along the line of sight, we find that a balance between emission and absorption is almost met. This means that even if the disk is experiencing strong dissipation and renewal phases, it will be barely visible with a few 0.01 mag peak-to-peak variations in this band. In a lower extent, the influence of the $\alpha$ parameter is also responsible for different amplitude of variations, the highest being obtained for high $\alpha$ values.  Like for the $V$-band, the magnitude in $K$-band is dependent of the disk state at each start of a new cycle. So for high DCs, the start and end excess values are all different for each inclination angle. 

%Finally, for disk smaller than the total emitting area  ($<$7-8 $R_\star$), we noticed lower amplitude variations of a few 0.01 mag in the $K$-band lightcurves than for bigger disk ($>$10 $R_\star$). This is in contrast with the $V$-band lightcurve which was not sensitive to the disk extent.

% decretion history,Angle, alpha, disque

\subsection{Millimetric photometry}

\begin{figure}[h!]
% \vspace*{-2.0 cm}
\begin{center}
 \includegraphics[width=3.5in]{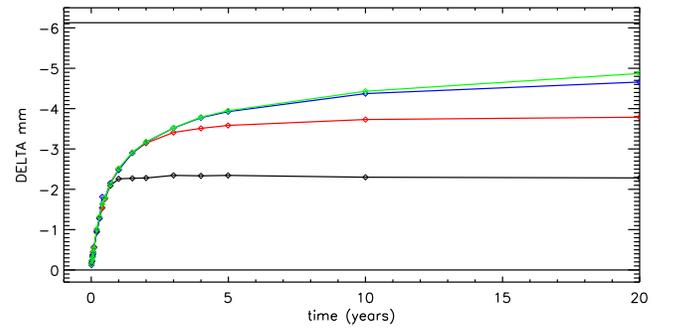} 
 \caption{Lightcurves at the $mm$ domain for a build-up phase under a constant injection rate ($i = 80 \degr$, $\alpha = 0.1$). The black, red, blue and green colors respectively represent the lightcurves for different disk sizes: 5, 10, 20 and 100 $R_\star$ respectively.}
   \label{dsub}
\end{center}
\end{figure}

The millimetric (mm) emission (at $\lambda$ = 1mm) comes from an extended outer part of the disk (mainly between 5 and 50$R_\star$). This implies that the $mm$ lightcurve is highly sensitive to the disk size. 
We thus explored four disk sizes in the range 5~---~$100\,R_\star$, to study the effect of disk truncation by a secondary star.
Figure~\ref{dsub} shows the $mm$ lightcurves during the build-up phase of these disks of different sizes. On this figure, we can see that the time required for the $mm$ magnitude to reach a stable value depends strongly on the disk size. The bigger the disk, the higher that value.

\begin{figure}[t!]
% \vspace*{-2.0 cm}
\begin{center}
 \includegraphics[width=3.5in]{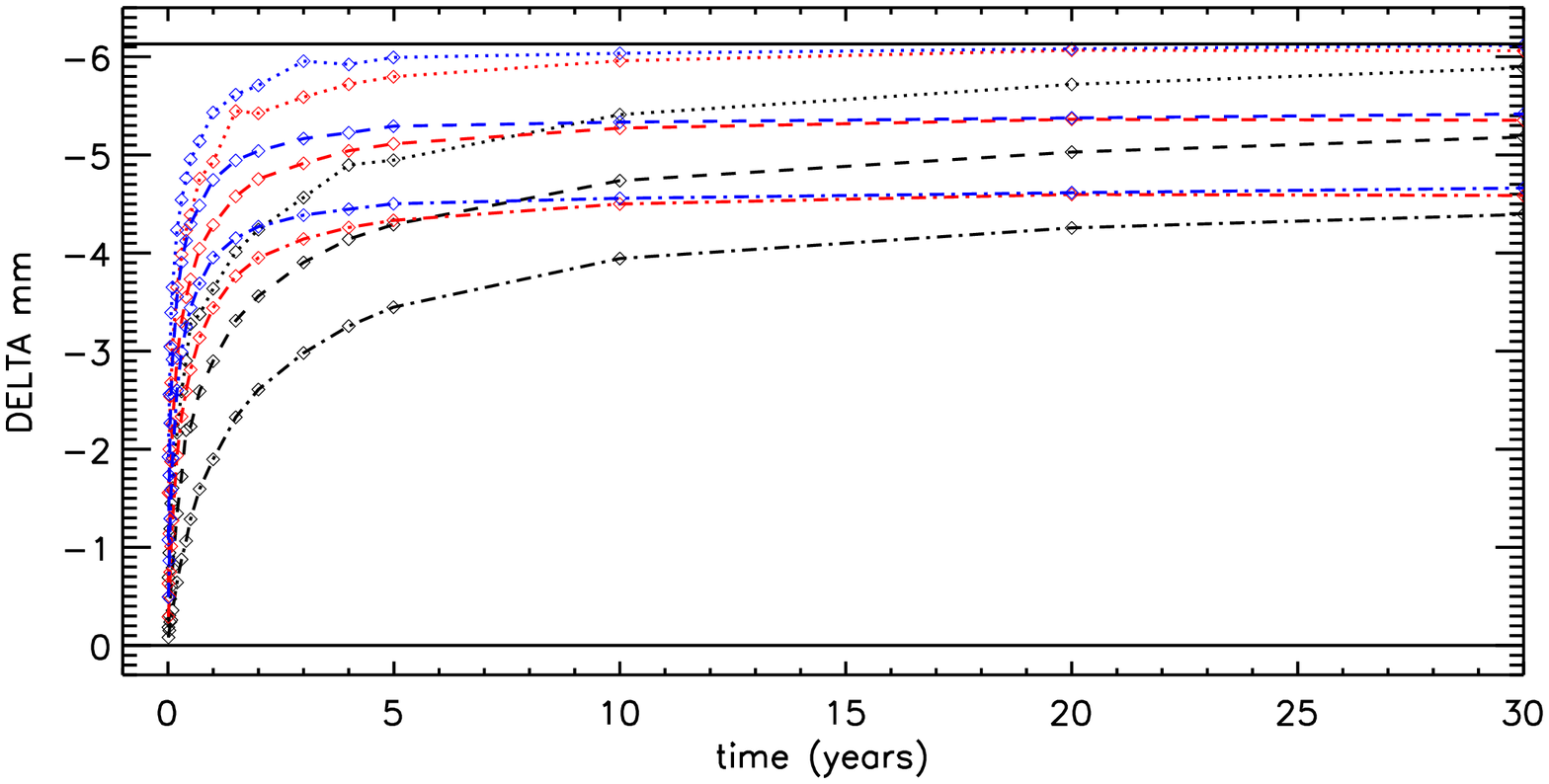} 
 \includegraphics[width=3.5in]{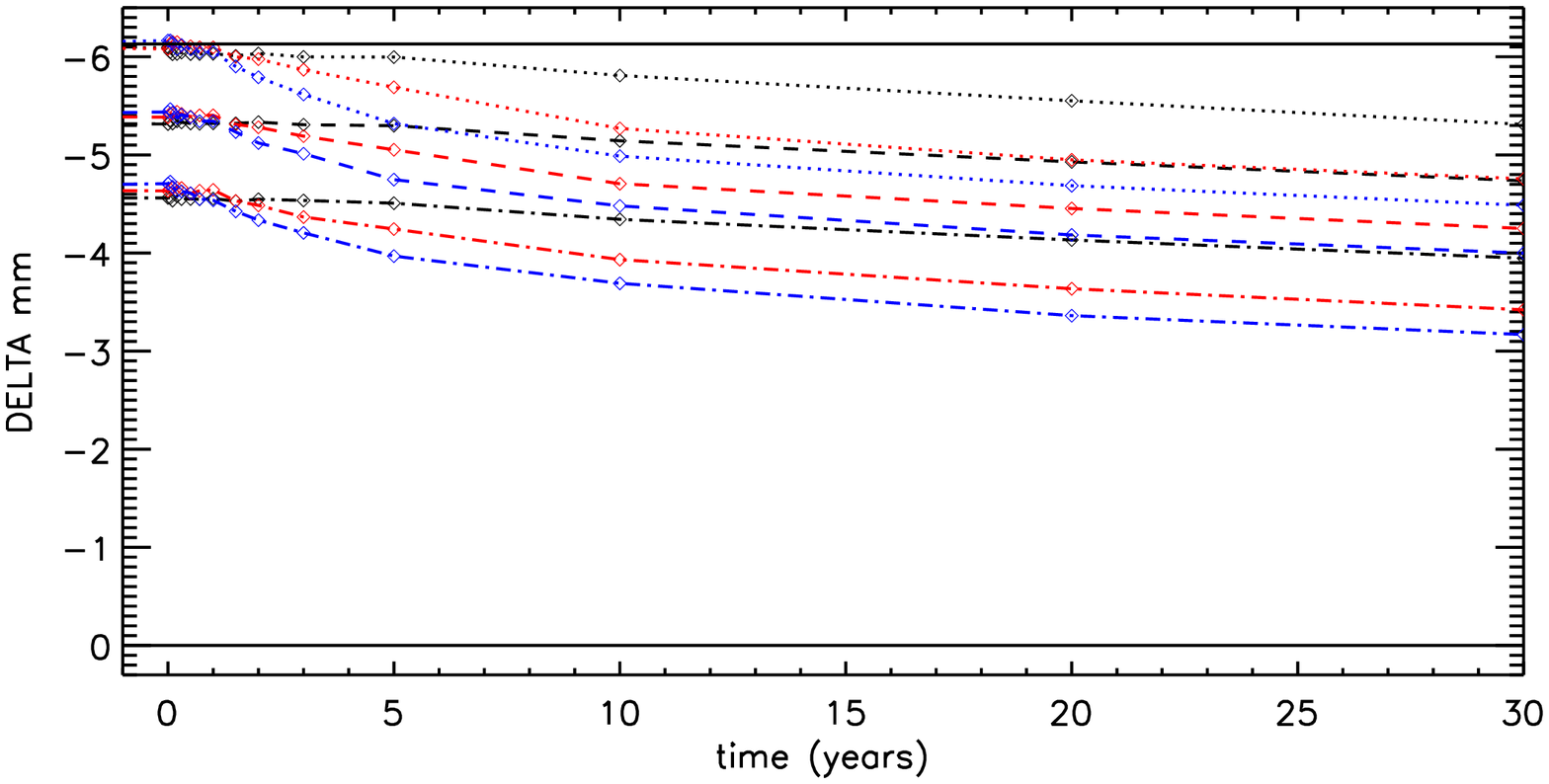}  
 \caption{Same as  Figure~\ref{dV}, but for a wavelength of 1 mm.}
   \label{dsub1}
\end{center}  
\end{figure}

%We should also state that $\sim$ 70\% of the total $mm$ flux for a 100 $R_\star$ size disk is met at 20 $R_\star$ (Figure~\ref{dsub1}).
 The magnitude variations at 1 mm computed for the build-up and decay phases are plotted in Figure~\ref{dsub1}.

As in the $V$ and $K$ bands, the $mm$ emission is produced quite fast after the mass decretion started. Table~\ref{QS_subband}  shows how fast the $mm$ disk excess reaches 95\% of its limit value. As a major difference with the $V$ and K band, during the decay phase, it takes a very long time before the emission decreases significantly: in 50 years, the excess at 0 \degr decreases by 1.5 magnitude for $\alpha$ = 0.1.

%\rouge{To ALEX : I don t get why the limit value differs for different alpha. I checked offset, they are ok. Could it be that the $mm$ magnitude isn t just about density ?  My guess is that we have something related to the disk size here. I should wait for the new mod04dm01 files (biggest disk) to be finished.} 

 \begin{figure}[t!]
% \vspace*{-2.0 cm}
\begin{center}
\includegraphics[width=3.5in]{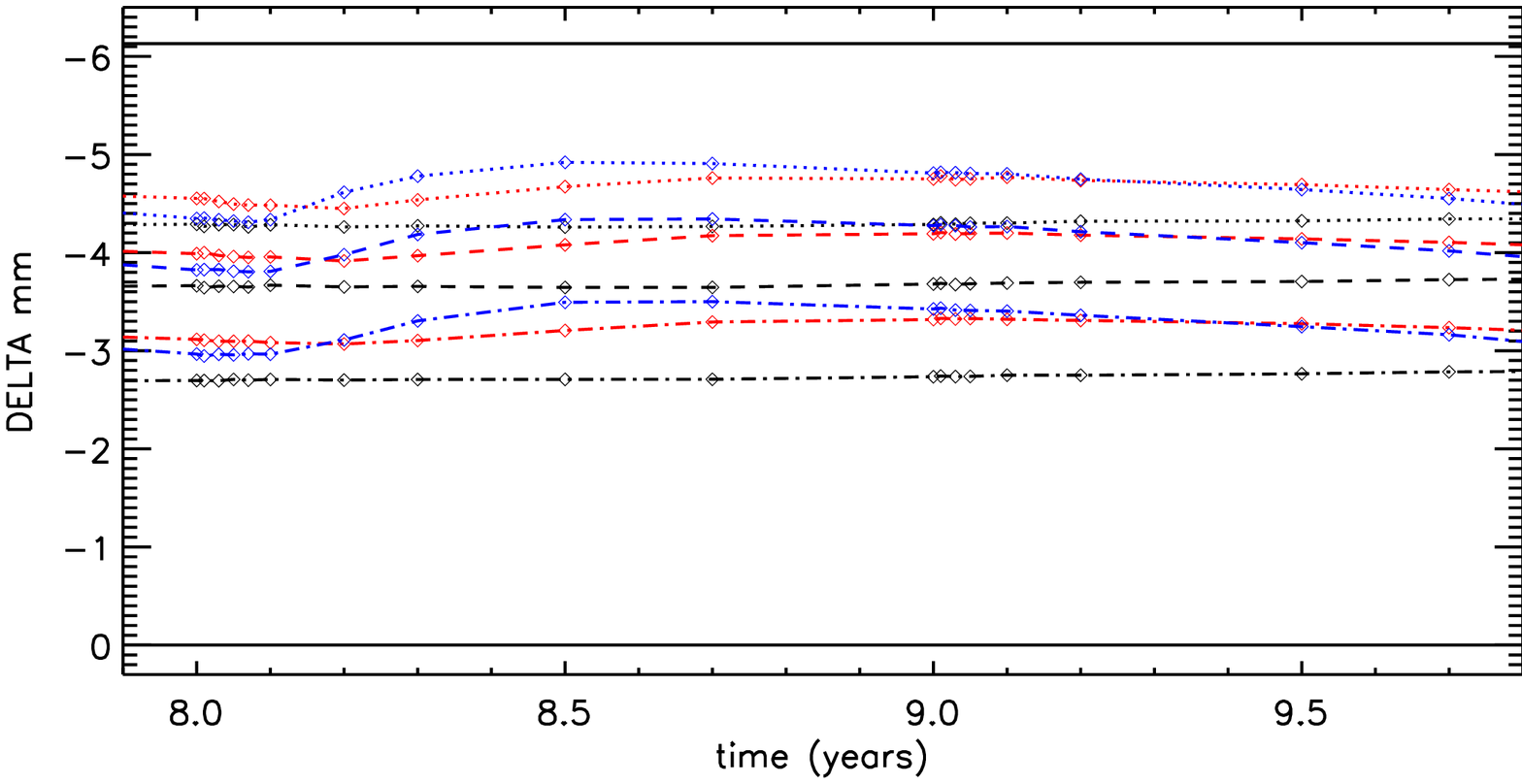}
\includegraphics[width=3.5in]{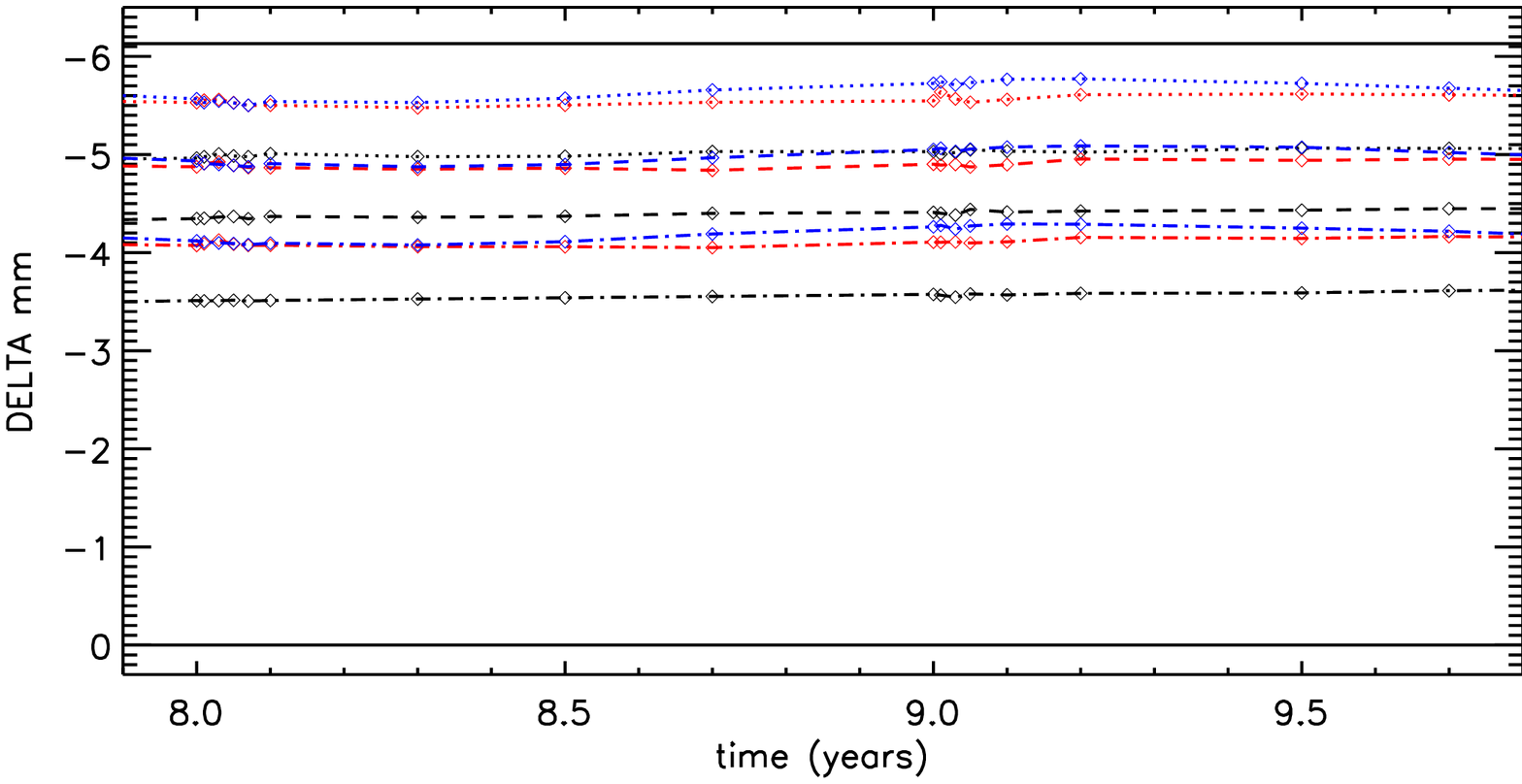}
\includegraphics[width=3.5in]{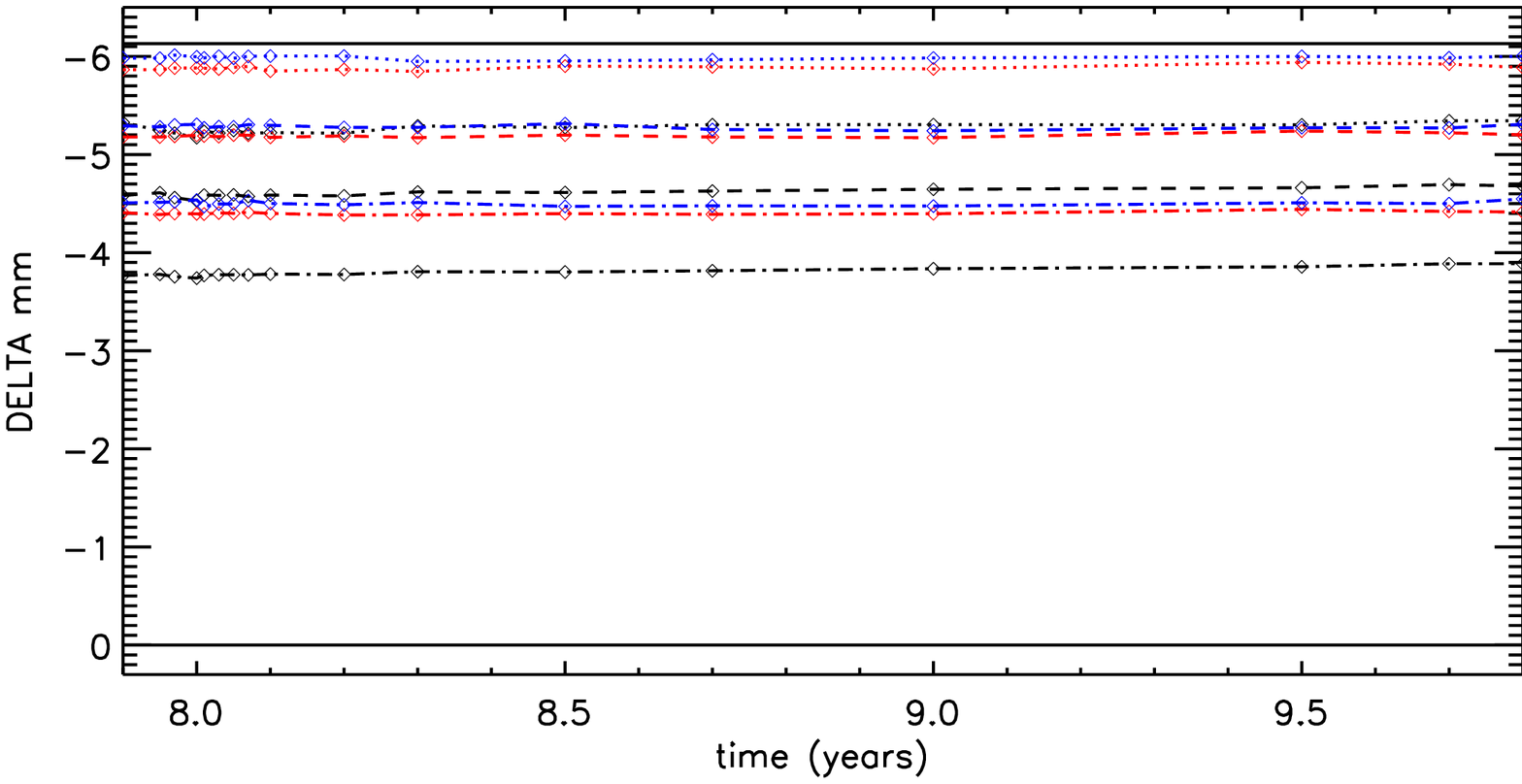}
 \caption{Mm-band lightcurves for the periodic scenario. Legend is the same as for Figure~\ref{dV2}. }
   \label{dsub2}
\end{center}
\end{figure}

%\rouge{Difference in the periodic scenarios lightcurves for other disk sizes ? Run dm06 for more important tendencies.}
Figure~\ref{dsub2} shows the $mm$ lightcurves during a periodic mass injection scenario for different angles and $\alpha$ parameters. The combined influence of the inclination angle, DC and $\alpha$  on the $mm$ lightcurves is not trivial. However, Figure~\ref{dsub2} clearly shows that for low $\alpha$ values, the periodic structure imposed by the decretion scenario is not visible in the lightcurves for high DCs (middle and lower panels of  Figure~\ref{dsub2}). The reason is that the area responsible for the $mm$ emission is so large that if the disk starts to vanish slowly from its inner parts for a few months and then gets filled up again, the resulting lack of density will propagate but will remain unnoticeable for the overall $mm$ emitting area. However for low DCs (upper panel of Figure~\ref{dsub2}), the effect of a periodic injection rate is visible because the mass supply injected into the disk is comparable to the overall mass of the $mm$ emitting region of the disk. However, this effect is more obvious for high $\alpha$ values where the density variations are more important because of large variations of the accretion region size (see discussion in \S~\ref{period}). 
Another interesting fact that we can observe on this figure is the temporal delay between the injection rate variation (turns on at 8 years) and the change in the lightcurve shape. Depending on the $\alpha$ value, the change of the disk state is visible in the lightcurves after 1 ($\alpha$ =1.0) or 2 ($\alpha$ = 0.5)  months.

We can summarize then that the $mm$ lightcurve is a good observing tool to infer a disk size but not to determine a short-term mass injection history neither to estimate unambiguously the inclination angle and the $\alpha$ parameter.

% dans l ordre d importance sur les valeurs : Taille du disque, angle, alpha, decretion history

\subsection{Photometric consequences for episodic models} 
\begin{figure}[h!]
  \begin{center}
\includegraphics[width=3.5in]{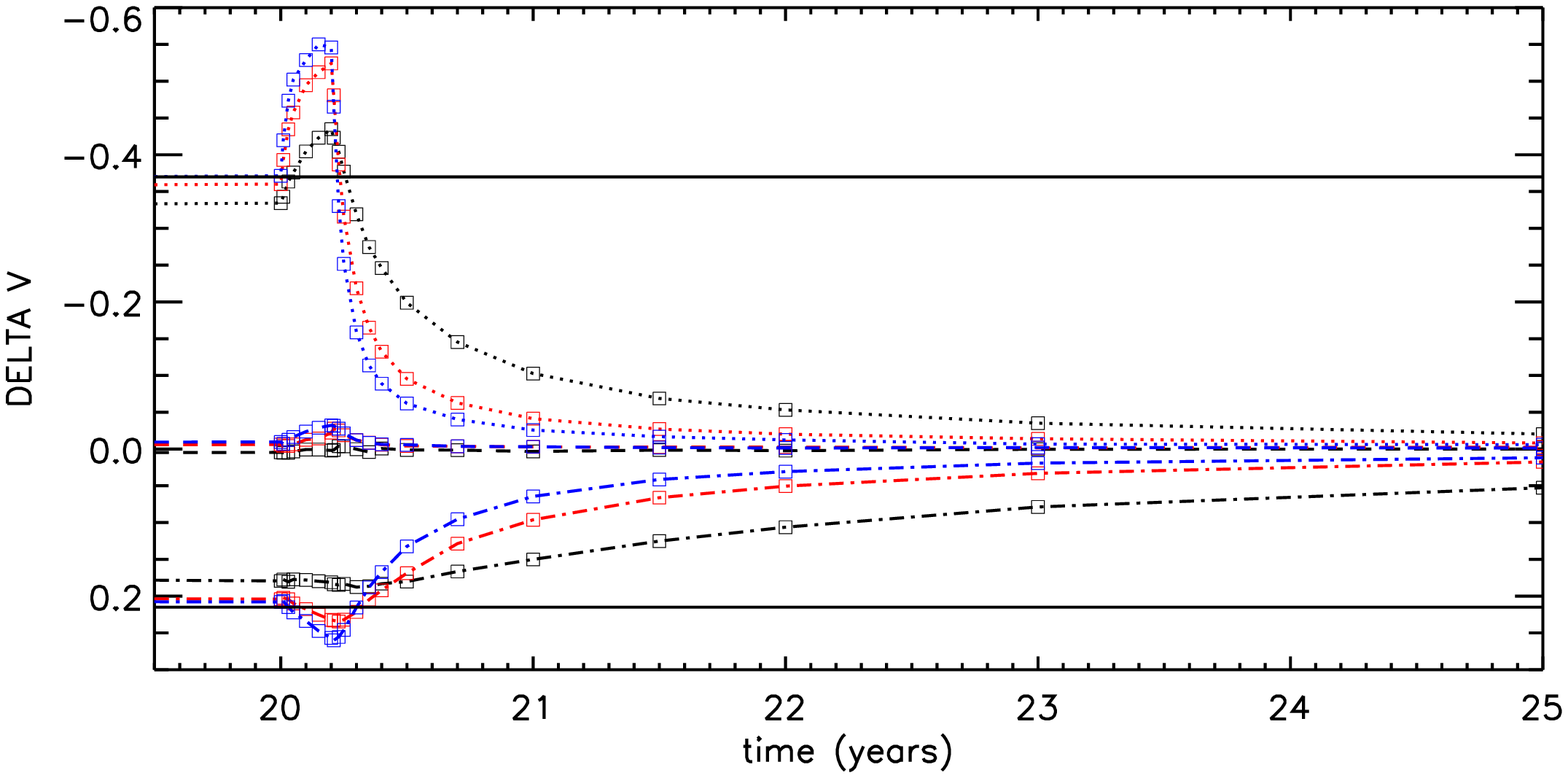}
\includegraphics[width=3.5in]{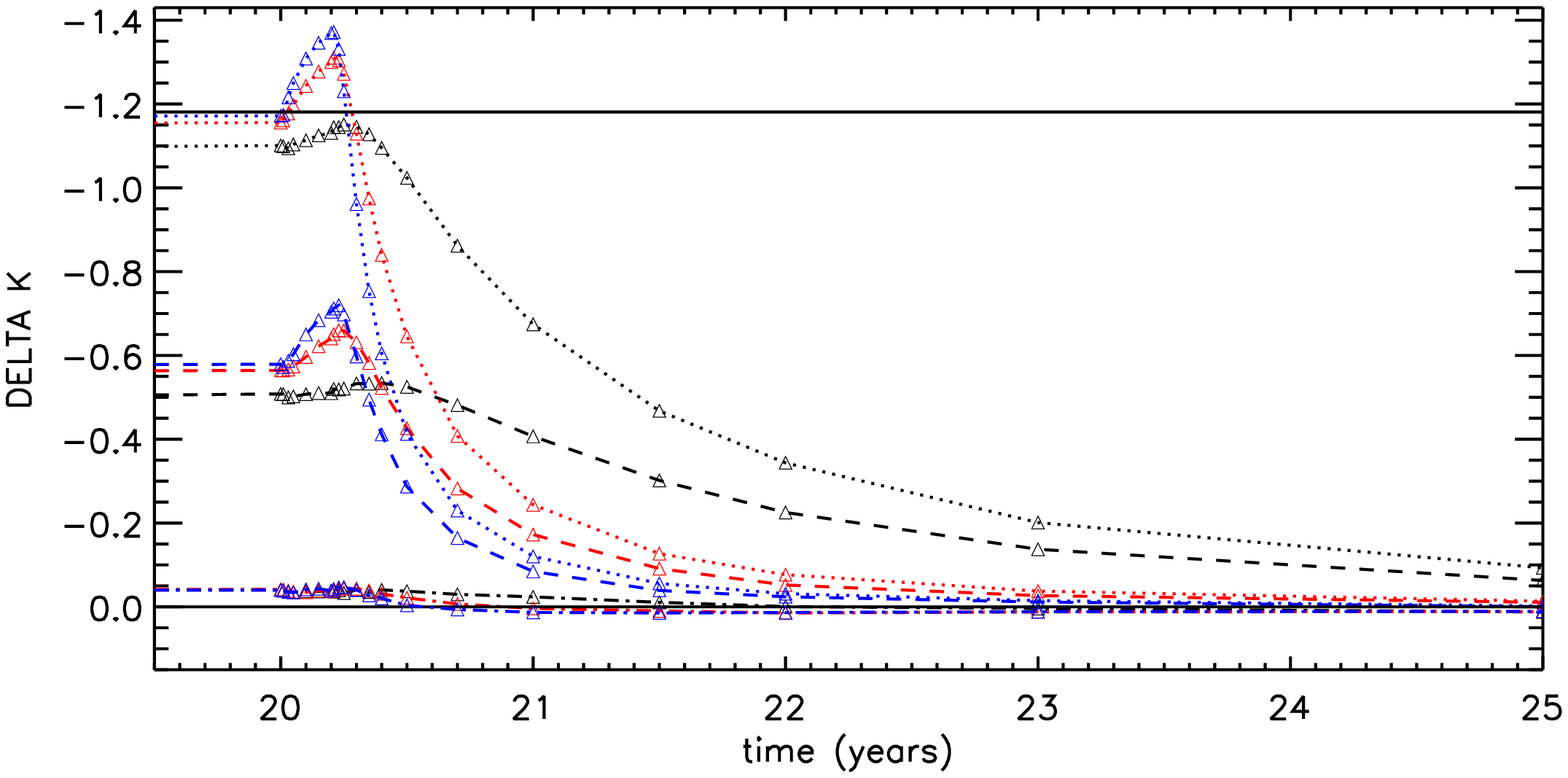}
\includegraphics[width=3.5in]{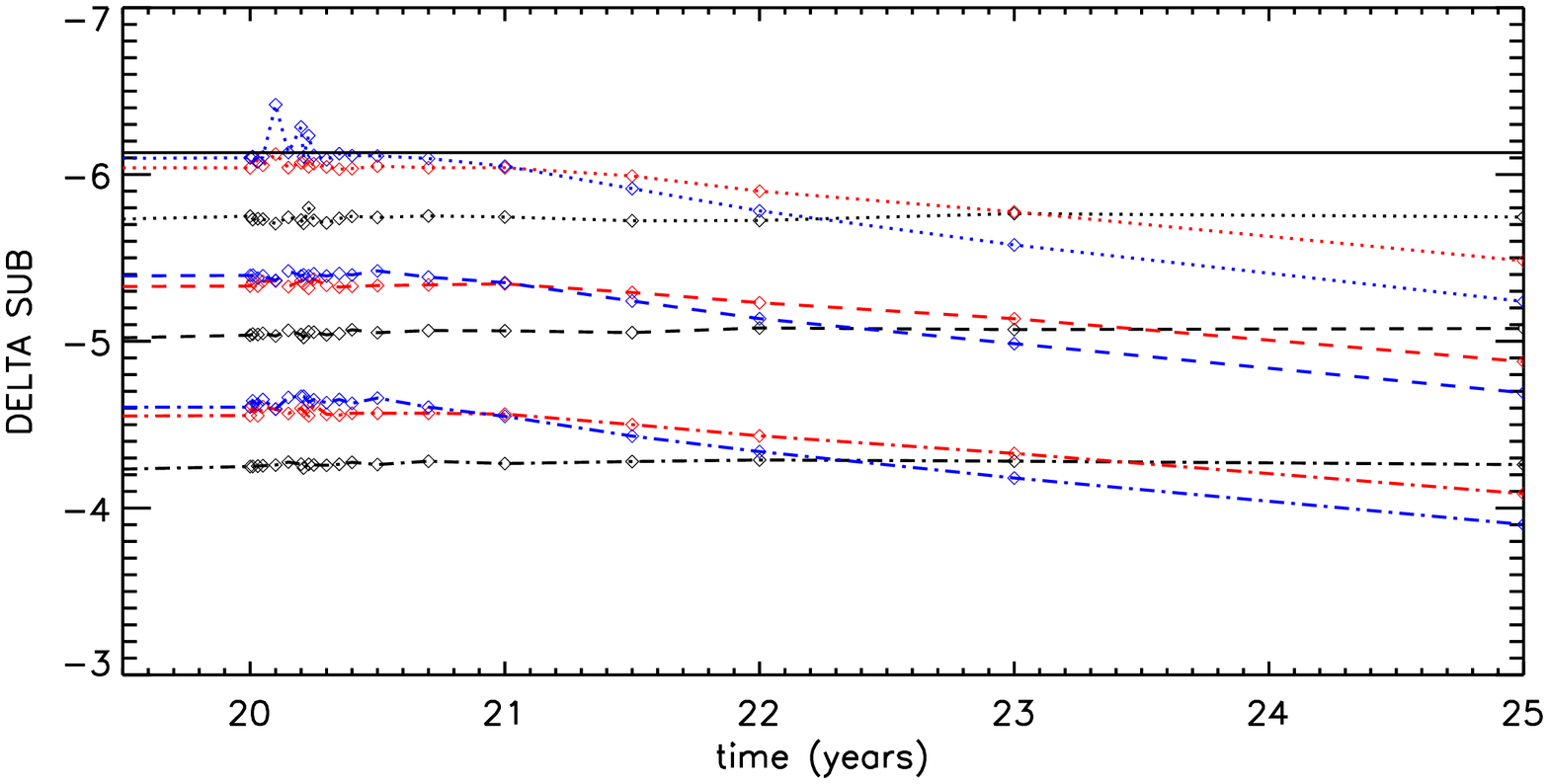} 
\caption{$V$-band, $K$-band and $mm$ lightcurves for the episodic scenario. Legend is the same as for Figure~\ref{dV}.}  \label{apert_phot} \end{center} \end{figure}

The lightcurves corresponding to the episodic model are presented in Figure~\ref{apert_phot}.  

%\rouge{The episodic model involves a 2 month outburst of stellar mass loss twice higher than the one that the disk underwent during its 20 first year of building. The resulting lightcurves of such an event can be explained with the same arguments than previously. We will not repeat them here. Nevertheless, it is interesting to compare the absolute excess values due to the outburst compared to the regular (with no outburst) build-up phase. The biggest difference is in the $V$-band where we see that an increase of the injection rate by a factor 2 results in a maximum magnitude value of -0.55 at 0 degree (-0.37 when no outburst). The outburst is then less obvious in the $K$-band. In the $mm$ lightcurves, it is almost invisible except for high $\alpha$ values (higher density variations) where the disk magnitude appears as a 0.3 high peak.}

The episodic model involves a 2 month outburst of stellar mass loss whose mass injection rate is twice the one the disk underwent during its 20 first years of building. The resulting lightcurves of such an event can be explained with the same arguments as previously. It is interesting to compare the absolute excess values due to the outburst with the excess at the end of the previous build-up phase. The biggest difference is in the $V$-band where we see that an increase of the injection rate by a factor 2 results in a maximum magnitude value of -0.55 at 0\degr. The outburst is a little less conspicuous in the $K$-band, and barely visible in the $mm$ lightcurves. This illustrates the important fact that the best indicator of the real time mass injection rate is photometry at short wavelengths.

\section{Discussion}
\label{discussion}

\begin{figure*}[t]
\begin{center}
\includegraphics[width=7in]{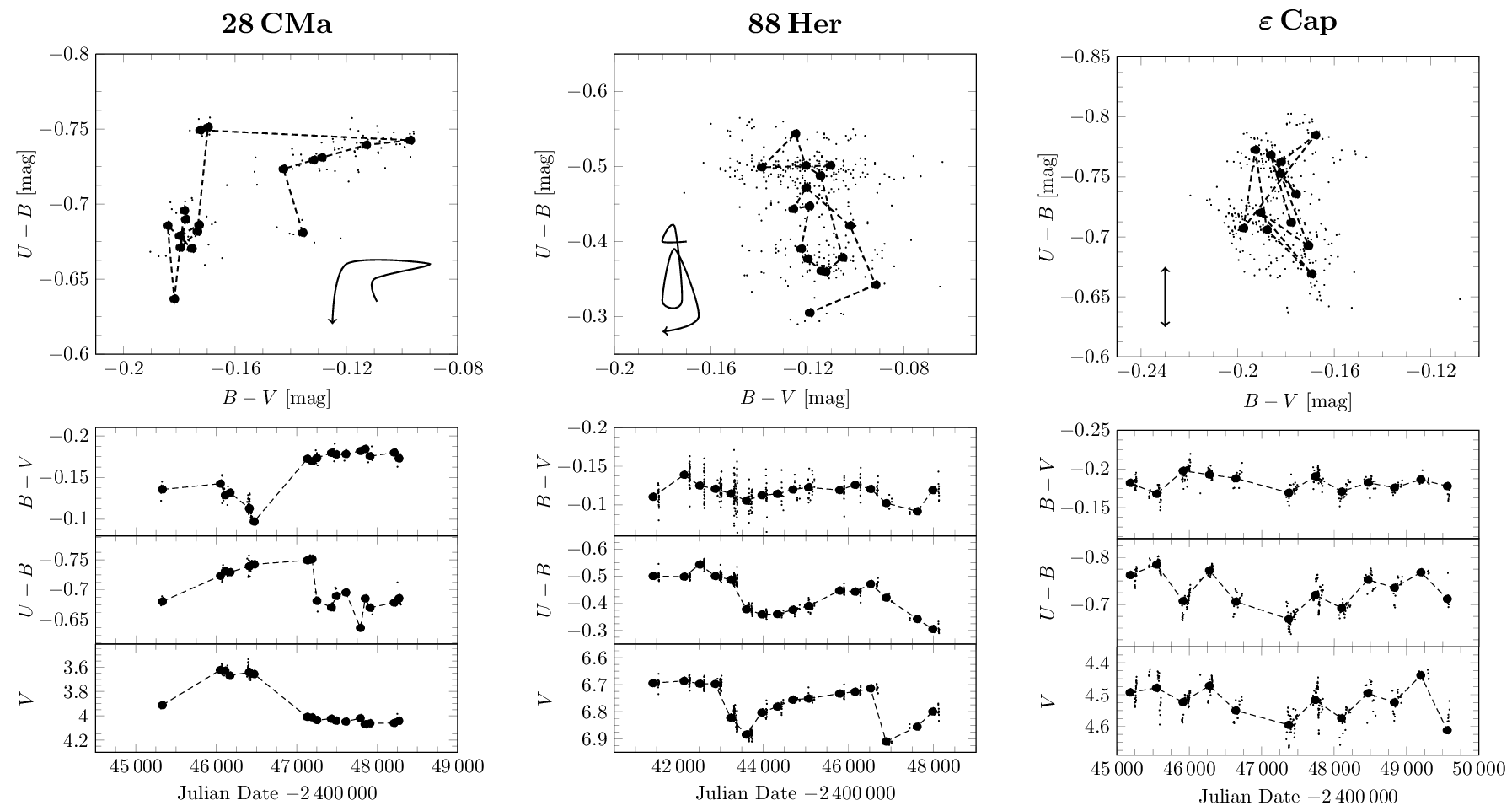}
%\plotone{Obs_col_col.ps}
\caption[]{Color and magnitude variations for three Be stars. The filled circles
  are bimonthly (28\,CMa) or annual (88\,Her, $\varepsilon$\,Cap) means. The
  line sketches in the respective uppermost panels indicate the direction of the
  path in time.}
\label{obs_col_col}
\end{center}
\end{figure*}

\subsection{Disk structure}
As outlined in \S~\ref{growth}, the density structure of the disk is often
parametrized in the form of a power law ($\rho (r) \propto r^{-n} $).  For a
well-developed disk, with a flaring parameter of 1.5, the expected value is
$n=3.5$. From visible interferometric observations of $\chi$ Oph,
\cite{2008ApJ...689..461T} found a best-fitting value of $n=2.5$, whereas
\cite{2008MNRAS.386.1922J} reports values of 4.2, 2.1 and 4.0 for $\kappa$
Dra, $\beta$ Psc and $\upsilon$ Cyg, respectively. \citet{1986A&A...162..121W}
report values in the range 2~--~3.5, based on an analysis of the IR SED of
several Be stars. Thus a wide range of values for $n$ is seen in the
literature.

We found \citep{2008ApJ...684.1374C} that non-isothermal viscous effects can be one possible explanation
for large scatter of the density radial exponent reported in the literature. However, another, and, in view of the perpetual variability of Be star disks, quite an attractive mechanism lies in the steepening of the index in decretion phases ($n>3.5$ in case decretion starts with an empty disk) and flattening of the index in re-accretion phases ($n<3.5$ in case decretion starts with a fully developed disk). Moreover, the index changes with distance from the star, and thus any determination of the index will be sensitive not only to the decretion history, but as well to the wavelength for which it is determined.

%Another important result concerning the disk structure is the rise of a local density maximum away of the stellar surface, only a short while after mass injection is turned off, in other words a ``ring''-like structure, though not with fully empty inner region. The loops seen on the color-color diagrams, with the return paths always being bluer, are a clear sign of this lower density regions in the inner disk wrt.\ further out. Such a disk structure has also also been reported observationally by \citet{2001A&A...379..257R}, though not based on photometric but on spectroscopic evidence.

%\addTRI{Effect (biases!) of photometric cancellation for e.g.\ photometricallyselected catalogs (like Mennickent SMC/LMC/bulge)? Or reserve that for the  actual paper on those light-curves? }

\subsection{Observed Galactic Be stars}

Data for some example stars were taken from published photometric
databases. The three best examples found were 28\,CMa, representing a low $i$
Be star with a decaying disk, 88\,Her, a shell star (i.e.\ at high $i$) which
has shown signs of two about one-year long mass transfer events during the
almost ten years of observations, and $\varepsilon$\,Cap, another shell star,
which alternated between a bright and faint state every other year for about
eight years of observation. For the discussion, we note that colors are bluer
when they are more negative. Data are shown in Figure~\ref{obs_col_col}.

The dataset of the almost pole-on Be star 28\,CMa has already been described
by \citet{2003A&A...402..253S}. The observations started in a phase of mass
transfer, reaching a state of a well developed disk around MJD=46\,000 in
which $\bv$ is reddest and $\ub$ is bluest. Then the disk started decaying
about a year later, i.e. mass transfer weakened or ceased. The color-color
path during decreasing brightness is bluer than during increasing brightness,
to finally reach a state in which $\bv$ is bluest and $\ub$ is reddest.
This agrees with \citet{1983HvaOB...7...55H}, as well as with the results
of \S~\ref{sec_Vband} and Figure~\ref{harman1}. Another star showing such
behaviour of secular color changes during disk built-up and decay is
$\kappa$\,Dra in the 1980s \citep{1994A&AS..107..403J}.

The shell stars 88\,Her and $\varepsilon$\,Cap show a different behaviour. The
data were taken from \citet{1997A&AS..125...75P} and
\citet{1994A&AS..108..237M}, respectively, and if necessary converted to the
UBV system using the relations given by \citet{2001A&A...369.1140H}. These are
bluer in \ub\ when having little circumstellar material (meaning when they
are bright due to lack of absorbing gas). As material is decreted, \ub\
clearly and \bv\ tendentially become redder. The \ub\ behaviour is
again in good agreement with \S~\ref{sec_Vband} and Figure~\ref{harman1},
while \bv\ is not. However, the amplitudes, both observed and predicted, in
\bv\ are much smaller than in \ub.\ As mass injection ceases, the star
88\,Her returns to the bright state on a bluer path, for $\varepsilon$\,Cap
this cannot be said with certainty. Also this is in agreement with the
modeling.

Most other stars in the photometric databases show a generally less clean
picture, with the notable exception of 48\,Lib. In that star \ub\ and
\bv\ are positively correlated, which is not seen in any of the model
computations. We note, that such a correlation is also given by
\citet{1983HvaOB...7...55H} as the normal behaviour of shell stars. However,
the colors not only correlate well with each other, but as well with the state
of the long-term $V/R$ variation \citep[see
  e.g.][]{1998A&A...330..631M,2000ASPC..214..460M}, while at the same time the
H$\alpha$ equivalent width and emission height was constant \citep[Figure~11
  of][]{1995A&A...300..163H}. This suggests that the color and magnitude
changes observed in 48\,Lib are not linked to variability of the mass
injection rate (and hence the radial density structure), but instead to the
azimuthal structure of the disk governed by the one-armed density wave, seen
at different aspect angles over the years. Since the \bv\ amplitude of
48\,Lib is twice as high as the ones of 88\,Her or $\varepsilon$\,Cap, it is
possible that a correlation as given by Harmanec is governed rather by
$V/R$-cycle related changes than by changes related to the mass injection.

\subsection{Comparison with previous models}
Already in \citeyear{1978ApJS...38..229P} \citeauthor{1978ApJS...38..229P}
published theroretical computations for brightness and color variations for a
set of steady-state disk models, i.e.\ assuming the stable equilibrium of a
fully developed disk. Even if the basic assumption of those models have been
revised considerably in the more than 30 years since, their findings
concerning the photometric behaviour remained the only available parameter
study until now (their Figures~29 to 34). In quantitative terms, their findings
do not differ strongly from the ones presented here for disks after a long
built-up phase, i.e.\ as described in \S~\ref{growth}.

\citeauthor{1978ApJS...38..229P} also find an inclination at which the
brightness change is zero. The inclination depends on wavelength, but the
value at which the cancellation of absorption vs.\ additional emission occurs
is generally smaller. For the $V$-band, \citeauthor{1978ApJS...38..229P} find
an invariant magnitude for an inclination of about 45 to 60 \degr, while
here it is 70.  At a wavelength of 2.7$\mu$m, an exactly equator-on disk still
reduces the brightness in the model of \citeauthor{1978ApJS...38..229P}, while
here it is found that longwards of about the $K$-band even a perfectly
equator-on seen disk will increase the brightness of the star. Only at
10$\mu$m both model predictions agree again qualitatively for an edge on disk,
in that they both brighten. Another important quantitative difference is that for
\citeauthor{1978ApJS...38..229P}, the (positive) magnitude difference in the
edge-on case is about twice the value of the (negative) magnitude difference
when seen pole-on ($\Delta V_{edge-on} = -2  \Delta V_{pole-on} $), whereas this ratio turns out to be reversed in this work ($\Delta V_{edge-on} = -0.5  \Delta V_{pole-on} )$.

In terms of color-color variations, for a comparable base density,
\citeauthor{1978ApJS...38..229P} derive somewhat smaller changes than here, but
it is important to note that these changes have the same sense: 
\begin{itemize}
\item For a disk seen face-on, $V$ is brighter, $\ub$ is bluer, while
  $\bv$ becomes redder (vs.\ a diskless star).
\item For a disk seen edge-on, $V$ is fainter, $\ub$ is redder, while
  $\bv$ becomes bluer (vs.\ a diskless star).
\end{itemize}

As already pointed out above, the second point is at variance with the observational correlation given by \citet{1983HvaOB...7...55H}, which, for stars with "inverse correlation between \ion{H}{1} emission and luminosity" states that the "fading in $V$ is accompanied by the reddening of both $\ub$ and $\bv$". Both \citeauthor{1978ApJS...38..229P}  and this work find a blueing of $\bv$ if the fading in $V$ is caused by an axisymmetric increase of disk density and mass. This lends some support to the hypothesis that the observed correlation might in fact not (fully) be governed by growth or decay of the disk.

The quantitative differences are most likely because the disk considered by
\citeauthor{1978ApJS...38..229P} was quite a simplified one compared to the
current understanding of circumstellar disks, e.g.\ it included no flaring and
was isothermal (and hotter) than the disk considered here. In qualitative
terms, however, we find good agreement with the previous results.

\section{Conclusions}
\label{conclusion}

The goal of this paper is to report on photometric predictions derived from the modeling of viscous decretion disks that undergo variable mass injection rates. We thus simulated the disk evolution under the influence of different dynamical scenarios. In order to understand the evolution of the disk structure in details, we looked at the temporal variations of some fundamental quantities derived from the surface density profiles. There are several timescales at play in the disk that helps to understand its evolution. To provide the reader with some reference numerical values, we made a quantified comparison between the viscous diffusion time and the time the disk surface density requires to reach its limit value. Coming also from the study of density profiles, we brought some theoretical arguments to explain the variety of power-law index values reported in the literature. We then presented $V$-band, $K$-band and $mm$ lightcurves at some epochs of the dynamical scenarios. The constant, periodic or episodic decretion scenarios generated characteristic variations on those photometric observables. We noted that the general behaviour of the $V$ and $K$ excesses is quite similar. These lightcurves vary accordingly to the inclination angle in a first extent, then to the $\alpha$ parameter. They consequently represent a good tool to estimate those parameters. Moreover, as they result from an emission produced in the first stellar radii, they are therefore a good way to infer, at a several days timescale, the mass loss history of the central star. We found that $mm$ lightcurves are more useful for disk size determination. However, there is some degeneracy, i.e.\ several sets of decretion scenarios and parameters could describe a same given short-term observed lightcurve. Consequently we stress that the more observables with a long time coverage, the easier to infer a dynamical scenario and physical parameters of the system. We finally compared our results to reported observations and results from previous modelings. In order to further test the $\alpha$-disk theory, we plan to compare our models with lightcurves from a representative sample of Be stars observed at different wavelengths.  Moreover, additional predictions on spectroscopic, polarimetric and interferometric observables will be presented in further publications.

%Viscous diffusion is improving, a non isothermal will be soon released.
%Link toward the website : fix this color scale pbm wit IDL.
%Further scenarios to investigate : more complex mass loss decretion, blob  puffing.
%Opening on interferometry and spectroscopy.

\begin{acknowledgements}
We acknowledge support from Fapesp grants  2009/07477-1 (XH), 2010/19029-0 (ACC), 2010/16037-2 (JEB) and CNPq grant 308985/2009-5 (ACC). This work has made use of the computing facilities of the Laboratory of Astroinformatics (IAG/USP, NAT/Unicsul), whose purchase was made possible by the Brazilian agency FAPESP (grant 2009/54006-4) and the INCT-A.

\end{acknowledgements}


%% in parentheses, after the keyword, but they will not be verified.



%\appendix

%\section{Appendix material}


\begin{thebibliography}{}

\bibitem[Balona et al.(1992)]{1992A&AS...92..533B} Balona, L.~A., Cuypers, J., \& Marang, F.\ 1992, \aaps, 92, 533 

\bibitem[Bjorkman(1997)]{1997LNP...497..239B} Bjorkman, J.~E.\ 1997, Stellar Atmospheres: Theory and Observations, 497, 239 

\bibitem[Beichman et al.(1988)]{1988iras....1.....B} Beichman, C.~A., 
Neugebauer, G., Habing, H.~J., Clegg, P.~E., 
\& Chester, T.~J.\ 1988, Infrared astronomical satellite (IRAS) catalogs and atlases.~Volume 1: Explanatory supplement, 1,  

\bibitem[Bjorkman \& Carciofi(2005)]{2005ASPC..337...75B} Bjorkman, J.~E., \& Carciofi, A.~C.\ 2005, The Nature and Evolution of Disks Around Hot Stars, 337, 75 

\bibitem[Carciofi et al.(2006)]{car06b} Carciofi, A.~C., Miroshnichenko, A.~S., Kusakin, A.~V., et al.\ 2006, \apj, 652, 1617 

\bibitem[Carciofi \& Bjorkman(2006)]{car06} Carciofi, A.C. \& Bjorkman, J.E. 2006, ApJ, 639, 1081

\bibitem[Carciofi \& Bjorkman(2008)]{2008ApJ...684.1374C} Carciofi, A.~C., \& Bjorkman, J.~E.\ 2008, \apj, 684, 1374 

\bibitem[Carciofi et al.(2009)]{car09} Carciofi, A.~C., Okazaki, A.~T., Le Bouquin, J.-B., Sö{v}tefl, S., Rivinius, T., Baade, D., Bjorkman, J.~E., \& Hummel, C.~A.\ 2009, \aap, 504, 915 

\bibitem[Carciofi(2011)]{carciofi11} Carciofi, A.~C.\ 2011, IAU Symposium, 272, 325 

\bibitem[Carciofi et al.(2012)]{car11b} Carciofi, A.~C., Bjorkman, J.~E., Otero, S., et al. \ 2012, ApJ, 744, L15

\bibitem[Chesneau et al.(2005)]{2005A&A...435..275C} Chesneau, O., Meilland, A., Rivinius, T., et al.\ 2005, \aap, 435, 275 

\bibitem[Cranmer(2005)]{2005ApJ...634..585C} Cranmer, S.~R.\ 2005, \apj, 634, 585 

\bibitem[Cuypers et al.(1989)]{1989A&AS...81..151C} Cuypers, J., Balona, L.~A., \& Marang, F.\ 1989, \aaps, 81, 151 

\bibitem[Delaa et al.(2011)]{2011A&A...529A..87D} Delaa, O., Stee, P., Meilland, A., et al.\ 2011, \aap, 529, A87 

\bibitem[Draper et al.(2011)]{2011ApJ...728L..40D} Draper, Z.~H., Wisniewski, J.~P., Bjorkman, K.~S., Haubois, X., Carciofi, A.~C., Bjorkman, J.~E., Meade, M.~R., \& Okazaki, A.\ 2011, \apjl, 728, L40 

\bibitem[Gehrz, Hackwell, \& Jones(1974)]{geh74}
 Gehrz R.D., Hackwell J.A., Jones T.W., 1974, \apj, 191, 675

\bibitem[Gies et al.(2007)]{2007ApJ...654..527G} Gies, D.~R., Bagnuolo, W.~G., Jr., Baines, E.~K., et al.\ 2007, \apj, 654, 527 

\bibitem[Hanuschik(1986)]{1986A&A...166..185H} Hanuschik, R.~W.\ 1986, \aap, 166, 185 

\bibitem[Hanuschik et 
al.(1993)]{1993A&A...274..356H} Hanuschik, R.~W., Dachs, J., Baudzus, M., \& Thimm, G.\ 1993, \aap, 274, 356 

\bibitem[{{Hanuschik} {et~al.}(1995){Hanuschik}, {Hummel}, {Dietle}, \&
  {Sutorius}}]{1995A&A...300..163H}
{Hanuschik}, R.~W., {Hummel}, W., {Dietle}, O., \& {Sutorius}, E. 1995, \aap,
  300, 163

\bibitem[Haubois et al. (2010)]{iau272} Haubois, X., Carciofi A.~C., , Okazaki, A.~T, Bjorkman, J.~E., Proceeding OB stars Paris 2010

\bibitem[Harmanec(1983)]{1983HvaOB...7...55H} Harmanec, P.\ 1983, Hvar Observatory Bulletin, 7, 55

\bibitem[{{Harmanec} \& {Bo{\v z}i{\'c}}(2001)}]{2001A&A...369.1140H}
{Harmanec}, P. \& {Bo{\v z}i{\'c}}, H. 2001, \aap, 369, 1140

\bibitem[Hubert et al.(2000)]{2000ASPC..214..348H} Hubert, A.~M., Floquet, M., \& Zorec, J.\ 2000, IAU Colloq.~175: The Be Phenomenon in Early-Type Stars, 214, 348 

\bibitem[Jones et al.(2008)]{2008MNRAS.386.1922J} Jones, C.~E., Sigut, T.~A.~A., \& Porter, J.~M.\ 2008, \mnras, 386, 1922 

\bibitem[Jones et al.(2011)]{2011AJ....141..150J} Jones, C.~E., Tycner, C., \& Smith, A.~D.\ 2011, \aj, 141, 150

\bibitem[{{Juza} {et~al.}(1994){Juza}, {Harmanec}, {Bozic}, {Pavlovski},  {Ziznovsky}, {Tarasov}, {Horn}, \& {Koubsky}}]{1994A&AS..107..403J}{Juza}, K., {Harmanec}, P., {Bozic}, H., {Pavlovski}, K., {Ziznovsky}, J., {Tarasov}, A.~E., {Horn}, J., \& {Koubsky}, P. 1994, \aaps, 107, 403

\bibitem[Kraus et al.(2011)]{2011arXiv1109.3447K} Kraus, S., Monnier, J.~D., Che, X., et al.\ 2011, arXiv:1109.3447 

\bibitem[Kurucz(1994)]{Kur94} Kurucz, R.\ 1994, Solar abundance model atmospheres for 0,1,2,4,8 km/s.~Kurucz CD-ROM No.~19.~ Cambridge, Mass.: Smithsonian Astrophysical Observatory, 1994., 19

\bibitem[Lee et al.(1991)]{1991MNRAS.250..432L} Lee, U., Osaki, Y., \& Saio, H.\ 1991, \mnras, 250, 432 

\bibitem[Lynden-Bell \& Pringle(1974)]{1974MNRAS.168..603L} Lynden-Bell, D., \& Pringle, J.~E.\ 1974, \mnras, 168, 603 

\bibitem[{{McDavid} {et~al.}(2000){McDavid}, {Bjorkman}, {Bjorkman}, \&
  {Okazaki}}]{2000ASPC..214..460M}
{McDavid}, D., {Bjorkman}, K.~S., {Bjorkman}, J.~E., \& {Okazaki}, A.~T. 2000,
  in Astronomical Society of the Pacific Conference Series, Vol. 214, IAU
  Colloq. 175: The Be Phenomenon in Early-Type Stars, ed. {M.~A.~Smith,
  H.~F.~Henrichs, \& J.~Fabregat}, 460

\bibitem[Meilland et al.(2006)]{2006A&A...455..953M} Meilland, A., Stee, P., Zorec, J., \& Kanaan, S.\ 2006, \aap, 455, 953 

\bibitem[Meilland et al.(2007)]{2007A&A...464...73M} Meilland, A., et al.\ 2007, \aap, 464, 73 

\bibitem[Meilland et al.(2009)]{2009A&A...505..687M} Meilland, A., Stee, P., Chesneau, O., \& Jones, C.\ 2009, \aap, 505, 687 

\bibitem[Meilland et al.(2011)]{2011arXivMeilland} Meilland, A., Millour, F., Kanaan, S., et al.\ 2011, arXiv:1111.2487 
 
\bibitem[{{Mennickent} {et~al.}(1998){Mennickent}, {Sterken}, \&
  {Vogt}}]{1998A&A...330..631M}
{Mennickent}, R.~E., {Sterken}, C., \& {Vogt}, N. 1998, \aap, 330, 631

\bibitem[Mennickent et al.(1994)]{1994A&AS..108..237M} Mennickent, R.~E., Vogt, N., \& Sterken, C.\ 1994, \aaps, 108, 237 

\bibitem[Mennickent et al.(2002)]{2002A&A...393..887M} Mennickent, R.~E., Pietrzy{\'n}ski, G., Gieren, W., \& Szewczyk, O.\ 2002, \aap, 393, 887 

\bibitem[Okazaki et al.(2002)]{2002MNRAS.337..967O} Okazaki, A.~T., Bate, M.~R., Ogilvie, G.~I., \& Pringle, J.~E.\ 2002, \mnras, 337, 967 

\bibitem[Okazaki(2007)]{2007ASPC..361..230O} Okazaki, A.~T.\ 2007, Active OB-Stars: Laboratories for Stellare and Circumstellar Physics, 361, 230 

\bibitem[Oudmaijer \& Parr(2010)]{2010MNRAS.405.2439O} Oudmaijer, R.~D., \& Parr, A.~M.\ 2010, \mnras, 405, 2439 

\bibitem[Owocki(2006)]{2006ASPC..355..219O} Owocki, S.\ 2006, Stars with the B[e] Phenomenon, 355, 219 

\bibitem[{{Pavlovski} {et~al.}(1997){Pavlovski}, {Harmanec}, {Bozic}, {Koubsky}, {Hadrava}, {Kriiz}, {Ruzic}, \& {Stefl}}]{1997A&AS..125...75P}{Pavlovski}, K., {Harmanec}, P., {Bozic}, H., {Koubsky}, P., {Hadrava}, P., {Kriiz}, S., {Ruzic}, Z., \& {Stefl}, S. 1997, \aaps, 125, 75

\bibitem[{{Poeckert} \& {Marlborough}(1978)}]{1978ApJS...38..229P}
{Poeckert}, R. \& {Marlborough}, J.~M. 1978, \apjs, 38, 229

\bibitem[Porter(1999)]{1999A&A...348..512P} Porter, J.~M.\ 1999, \aap, 348, 512 

\bibitem[Porter \& Rivinius(2003)]{2003PASP..115.1153P} Porter, J.~M., \& Rivinius, T.\ 2003, \pasp, 115, 1153 

\bibitem[Pringle(1981)]{pri81} Pringle, J.E.\ 1981, ARAA, 19, 137

\bibitem[Quirrenbach et al.(1997)]{1997ApJ...479..477Q} Quirrenbach, A., Bjorkman, K.~S., Bjorkman, J.~E., et al.\ 1997, \apj, 479, 477 

\bibitem[Rivinius et al.(2001)]{2001A&A...379..257R} Rivinius, T., Baade, D., {\v S}tefl, S., \& Maintz, M.\ 2001, \aap, 379, 257 

\bibitem[Sabogal et al.(2008)]{2008A&A...478..659S} Sabogal, B.~E., Mennickent, R.~E., Pietrzy{\'n}ski, G., et al.\ 2008, \aap, 478, 659 

\bibitem[Shakura \& Sunyaev(1973)]{1973A&A....24..337S} Shakura, N.~I., \& Sunyaev, R.~A.\ 1973, \aap, 24, 337 

\bibitem[Sigut \& Jones(2007)]{2007ApJ...668..481S} Sigut, T.~A.~A., \& Jones, C.~E.\ 2007, \apj, 668, 481 

\bibitem[{{{\v S}tefl} {et~al.}(2003){{\v S}tefl}, {Baade}, {Rivinius},{Otero}, {Stahl}, {Budovi{\v c}ov{\'a}}, {Kaufer}, \&{Maintz}}]{2003A&A...402..253S}{{\v S}tefl}, S., {Baade}, D.,{Rivinius}, T., {Otero}, S., {Stahl}, O.,  {Budovi{\v c}ov{\'a}}, A., {Kaufer}, A., \& {Maintz}, M. 2003, \aap, 402, 253

\bibitem[Sterken et al.(1996)]{1996A&A...311..579S} Sterken, C., Vogt, N., \& Mennickent, R.~E.\ 1996, \aap, 311, 579 

\bibitem[Tycner et al.(2006)]{2006AJ....131.2710T} Tycner, C., Gilbreath, G.~C., Zavala, R.~T., et al.\ 2006, \aj, 131, 2710 

\bibitem[Tycner et al.(2008)]{2008ApJ...689..461T} Tycner, C., Jones, C.~E., Sigut, T.~A.~A., Schmitt, H.~R., Benson, J.~A., Hutter, D.~J., \& Zavala, R.~T.\ 2008, \apj, 689, 461 

\bibitem[Stagg(1987)]{1987MNRAS.227..213S} Stagg, C.\ 1987, \mnras, 227, 213

\bibitem[{\v S}tefl et al.(2009)]{ste09} {\v S}tefl, S., Rivinius, T., Carciofi, A.~C., et al.\ 2009, \aap, 504, 929 

\bibitem[{\v S}tefl et al.(2011)]{2011IAUS..272..430S} {\v S}tefl, S., Carciofi, A.~C., Baade, D., et al.\ 2011, IAU Symposium, 272, 430 

\bibitem[Waters(1986)]{1986A&A...162..121W} Waters, L.~B.~F.~M.\ 1986, \aap, 162, 121

\bibitem[Waters et al.(1988)]{1988A&A...198..200W} Waters, L.~B.~F.~M., van den Heuvel, E.~P.~J., Taylor, A.~R., Habets, G.~M.~H.~J., \& Persi, P.\ 1988, \aap, 198, 200 

\bibitem[de Wit et al.(2006)]{2006A&A...456.1027D} de Wit, W.~J., Lamers, H.~J.~G.~L.~M., Marquette, J.~B., \& Beaulieu, J.~P.\ 2006, \aap, 456, 1027 

\bibitem[Wisniewski et al.(2010)]{2010ApJ...709.1306W} Wisniewski, J.~P., Draper, Z.~H., Bjorkman, K.~S., et al.\ 2010, \apj, 709, 1306 

\bibitem[Zorec \& Briot(1997)]{1997A&A...318..443Z} Zorec, J., \& Briot, D.\ 1997, \aap, 318, 443 

\end{thebibliography}
\end{document}